%% file: ElsPaper.tex
\pdfoutput=1
\documentclass[3p]{elsarticle}
\biboptions{sort&compress}
\usepackage{graphicx}
\usepackage{subcaption}
\graphicspath{{./Figures/}}
\usepackage{tikz}
\usetikzlibrary{decorations.markings}
\usepackage{hyperref}
\hypersetup{hidelinks}
\usepackage{siunitx}
\include{ChemChrgNom}
\usepackage{epic}
\makeatletter
\newcommand{\vast}{\bBigg@{4}}
\newcommand{\Vast}{\bBigg@{5}}
\makeatother
\begin{document}
\begin{frontmatter}
\title{An efficient flamelet progress-variable method for modeling non-premixed flames in weak electric fields} 
\author[dmmm]{M. \textsc{Di Renzo}}
\ead{mario.direnzo@poliba.it}
\author[dmmm]{P. \textsc{De Palma}}
\ead{pietro.depalma@poliba.it}
\author[dmmm]{M. D. \textsc{de Tullio}}
\ead{marcodonato.detullio@poliba.it}
\author[dmmm]{G. \textsc{Pascazio}\corref{cor1}}
\ead{giuseppe.pascazio@poliba.it}
\address[dmmm]{\normalsize Dipartimento di Meccanica, Matematica e Management \& Centro di Eccellenza in Meccanica Computazionale, Politecnico di Bari, Bari, Italy}
\date{}
\cortext[cor1]{Corresponding author}
%
%
\begin{abstract}
	Combustion stabilization and enhancement of the flammability limits are mandatory objectives to improve nowadays combustion chambers.
	At this purpose, the use of an electric field in the flame region provides a solution which is, at the same time, easy to implement and effective to modify the flame structure.
	The present work describes an efficient flamelet progress-variable approach developed to model the fluid dynamics of flames immersed in an electric field.
	The main feature of this model is that it can use complex ionization mechanisms without increasing the computational cost of the simulation.
	The model is based on the assumption that the combustion process is not directly influenced by the electric field and has been tested using two chemi-ionization mechanisms of different complexity in order to examine its behavior with and without the presence of heavy anions in the mixture.
	Using a one- and two-dimensional numerical test cases, the present approach has been able to reproduce all the major aspects encountered when a flame is subject to an imposed electric field and the main effects of the different chemical mechanisms.
	Moreover, the proposed model is shown to produce a large reduction in the computational cost, being able to shorten the time needed to perform a simulation up to 40 times.
\end{abstract}
%
%
\end{frontmatter}
%
%
%
\input{./Sections/Intro.tex}
\input{./Sections/Model.tex}
\input{./Sections/Num_proc.tex}
\input{./Sections/Chem_analysis.tex}
\input{./Sections/1D_Verification.tex}
\input{./Sections/Speelman.tex}
\input{./Sections/Test_case.tex}
\input{./Sections/Num_setup.tex}
\input{./Sections/Results.tex}

\input{./Sections/Comp_cost.tex}
\input{./Sections/Conclusions.tex}
%
%
\section*{Acknowledgments}
This research has been supported by grant n. PON03PE\_00067\_6 APULIA SPACE.
The authors want also to thank the reviewer for his precious suggestions, which contributed to improve the clarity of the present work.
%
%
\section*{References}
\bibliographystyle{model1-num-names_mod}
\bibliography{local_bib,Mendeley}
%
%
%
%
\end{document}

%% file: ChemChrgNom.tex
\usepackage[acronym,nonumberlist,nowarn]{glossaries}
\glsdisablehyper
\makeglossaries
\newcommand*{\vect}[1]{\ensuremath{\mathbf{#1}}} 
\newcommand*{\tens}[1]{\ensuremath{\overline{\overline{#1}}}} 
\newcommand*{\tder}[1]{\ensuremath{\frac{\partial {#1}}{\partial \tim}}} 
\newcommand*{\xder}[1]{\ensuremath{\frac{\partial {#1}}{\partial \x}}} 
\newcommand{\isumsp}{\ensuremath{\sum_{i=1}^{\ns}}} 
\newcommand{\ksumsp}{\ensuremath{\sum_{k=1}^{\ns}}} 
\newcommand{\jsumre}{\ensuremath{\sum_{j=1}^{\nr}}} 
\newcommand{\divg}{\ensuremath{\nabla \cdot }} 
\newcommand{\lap}{\ensuremath{\nabla^2 }} 
\newcommand{\tim}{\ensuremath{t} } 
\newcommand{\x}{\ensuremath{x}} 
\newcommand{\mf}{\ensuremath{Y}} 
\newcommand{\xf}{\ensuremath{X}} 
\newcommand{\mm}{\ensuremath{\mathcal{M}}} 
\newcommand{\dif}{\ensuremath{\alpha}} 
\newcommand{\mob}{\ensuremath{k}} 
\newcommand{\Le}{\ensuremath{Le}} 
\newcommand{\pr}{\ensuremath{\dot{\omega}}} 
\newcommand{\ns}{\ensuremath{N_s}} 
\newcommand{\nr}{\ensuremath{N_r}} 
\newcommand{\np}{\ensuremath{n}} 
\newglossaryentry{imf}{
	name = {\ensuremath{\mf_i}},
	description = {I$^{th}$ species mass fraction [-]}
}
\newglossaryentry{idif}{
	name = {\ensuremath{\dif_i}},
	description = {I$^{th}$ species diffusivity [\si{\square\meter\per\second}]}
}
\newglossaryentry{ivdif}{
	name = {\ensuremath{\vect{V}_{i}}},
	description = {I$^{th}$ species diffusion velocity [\si{\meter\per\second}]}
}
\newglossaryentry{iLe}{
	name = {\ensuremath{\Le_i}},
	description = {I$^{th}$ species Lewis number [-]}
}
\newglossaryentry{imob}{
	name = {\ensuremath{\mob_i}},
	description = {I$^{th}$ species mobility [\si{\square\meter\per\second\per\volt}]}
}
\newglossaryentry{Si}{
	name = {\ensuremath{S_i}},
	description = {I$^{th}$ species number of elementary charges [-]}
}
\newglossaryentry{ipr}{
	name = {\ensuremath{\pr_i }},
	description = {I$^{th}$ species production rate [\si[per-mode=reciprocal]{\per\second}]}
}
\newglossaryentry{icp}{
	name = {\ensuremath{c_{p_i}}},
	description = {I$^{th}$ species constant pressure heat capacity [\si{\joule\per\kilo\gram\per\kelvin}]}
}
\newglossaryentry{xi}{
	name = {\ensuremath{\xf_i}},
	description = {I$^{th}$ species molar fraction [-]}
}
\newglossaryentry{ilam}{	
	name = {\ensuremath{\lambda_{i}}},
	description = {I$^{th}$ species thermal conductivity [\si{\watt\per\meter\per\second}]}
}
\newglossaryentry{ih}{	
	name = {\ensuremath{h_i}},
	description = {I$^{th}$ species specific enthalpy [\si{\joule\per\kilo\gram}]}
}
\newglossaryentry{imm}{	
	name = {\ensuremath{\mm_{i}}},
	description = {I$^{th}$ species molar mass [\si{\kilo\gram\per\mole}]}
}
\newglossaryentry{inu}{	
	name = {\ensuremath{\nu_{i}}},
	description = {I$^{th}$ species kinematic viscosity [\si{\square\meter\per\second}]}
}
\newglossaryentry{LJd}{	
	name = {\ensuremath{\sigma_{i}}},
	description = {I$^{th}$ species Lennard-Jones collision diameter [\si{\meter}]}
}
\newglossaryentry{LJp}{	
	name = {\ensuremath{\epsilon_{i}}},
	description = {I$^{th}$ species Lennard-Jones potential well depth [\si{\joule}]}
}
\newglossaryentry{ci11}{	
	name = {\ensuremath{\Omega^{(1,1)}_{ij}}},
	description = {I$^{th}$ species collision integral with j$^{th}$ species (based on Stockmayer potentials) [-]}
}
\newglossaryentry{ci22}{	
	name = {\ensuremath{\Omega^{(2,2)}_{i}}},
	description = {I$^{th}$ species collision integral [-]}
}
\newglossaryentry{Aj}{
	name = {\ensuremath{A_j}},
	description = {J$^{th}$ reaction pre-exponential coefficient [\si{\cubic\meter\per\mole\per\second}]}
}
\newglossaryentry{Eaj}{
	name = {\ensuremath{E_j}},
	description = {J$^{th}$ reaction specific activation energy [\si{\joule\per\mole}]}
}
\newglossaryentry{den}{
	name = {\ensuremath{\rho}},
	description = {Gas density [\si{\kilo\gram\per\cubic\meter}]}
}
\newglossaryentry{tdif}{
	name = {\ensuremath{\dif}},
	description = {Thermal diffusivity [\si{\square\meter\per\second}]}
}
\newglossaryentry{lam}{	
	name = {\ensuremath{\lambda}},
	description = {Mixture thermal conductivity [\si{\watt\per\meter\per\second}]}
}
\newglossaryentry{cp}{	
	name = {\ensuremath{c_p}},
	description = {Mixture constant pressure heat capacity ($\gls{cp} = \isumsp \gls{imf} \gls{icp}$) [\si{\joule\per\kilo\gram\per\kelvin}]}
}
\newglossaryentry{Et}{	
	name = {\ensuremath{\mathcal{E}}},
	description = {Specific total energy ($\frac{\gls{um}^2}{2} + \gls{Ei}$) [\si{\joule\per\kilo\gram}]}
}
\newglossaryentry{Ei}{	
	name = {\ensuremath{\mathcal{E}_i}},
	description = {Specific internal energy [\si{\joule\per\kilo\gram}]}
}
\newglossaryentry{h}{	
	name = {\ensuremath{h}},
	description = {Mixture specific enthalpy ($\frac{\gls{pre}}{\gls{den}} + \gls{Ei} = \isumsp \gls{imf} \gls{ih} $) [\si{\joule\per\kilo\gram}]}
}
\newglossaryentry{mmm}{	
	name = {\ensuremath{\mm}},
	description = {mixture molar mass [\si{\kilo\gram\per\mole}]}
}
\newglossaryentry{T}{	
	name = {\ensuremath{T}},
	description = {Mixture temperature [\si{\kelvin}]}
}
\newglossaryentry{nu}{	
	name = {\ensuremath{\nu}},
	description = {Mixture kinematic viscosity [\si{\square\meter\per\second}]}
}
\newglossaryentry{u}{
	name = {\vect{u}},
	description = {Flow velocity [\si{\meter\per\second}]}
}
\newglossaryentry{um}{
	name = {\ensuremath{u}},
	description = {Flow velocity magnitude [\si{\meter\per\second}]}
}
\newglossaryentry{pre}{	
	name = {\ensuremath{p}},
	description = {pressure [\si{\pascal}]}
}
\newglossaryentry{fm}{
	name = {\vect{f}},
	description = {General specific force field [\si{\newton\per\cubic\meter}]}
}
\newglossaryentry{StrT}{	
	name = {\tens{S}},
	description = {Strain tensor [\si[per-mode=reciprocal]{\per\second}]}
}
\newglossaryentry{sig}{	
	name = {\tens{\sigma}},
	description = {Viscous stress tensor [\si{\pascal}]}
}
\newglossaryentry{mf}{
	name = {\ensuremath{Z}},
	description = {Mixture fraction [-]}
}
\newglossaryentry{prv}{
	name = {\ensuremath{C}},
	description = {Progress variable [-]}
}
\newglossaryentry{cpr}{
	name = {\ensuremath{\pr_{\gls{prv}}}},
	description = {Progress variable production rate [\si[per-mode=reciprocal]{\per\second}]}
}
\newglossaryentry{zLe}{
	name = {\ensuremath{\Le_{\gls{mf}}}},
	description = {Mixture fraction Lewis number [-]}
}
\newglossaryentry{zdif}{
	name = {\ensuremath{\dif_{\gls{mf}}}},
	description = {Mixture fraction diffusivity [\si{\square\meter\per\second}]}
}
\newglossaryentry{cdif}{
	name = {\ensuremath{\dif_{\gls{prv}}}},
	description = {Progress variable diffusivity [\si{\square\meter\per\second}]}
}
\newglossaryentry{chi}{
	name = {\ensuremath{\chi}},
	description = {Scalar dissipation rate [\si[per-mode=reciprocal]{\per\second}]  ($\gls{chi} = 2 \gls{idif} | \nabla \gls{mf}|^2$)}
}
\newglossaryentry{pcrg}{
	name = {\ensuremath{P}},
	description = {Positive charge density [\si{\coulomb\per\kilo\gram}]}
}
\newglossaryentry{ncrg}{
	name = {\ensuremath{M}},
	description = {Negative charge density [\si{\coulomb\per\kilo\gram}]}
}
\newglossaryentry{ppr}{
	name = {\ensuremath{\pr_{\gls{pcrg}}}},
	description = {Positive charge density production rate [\si[per-mode=reciprocal]{\per\second}]}
}
\newglossaryentry{mpr}{
	name = {\ensuremath{\pr_{\gls{ncrg}}}},
	description = {Negative charge density production rate [\si[per-mode=reciprocal]{\per\second}]}
}
\newglossaryentry{pdif}{
	name = {\ensuremath{\dif_{\gls{pcrg}}}},
	description = {Positive charge density diffusivity [\si{\square\meter\per\second}]}
}
\newglossaryentry{mdif}{
	name = {\ensuremath{\dif_{\gls{ncrg}}}},
	description = {Negative charge density diffusivity [\si{\square\meter\per\second}]}
}
\newglossaryentry{pmob}{
	name = {\ensuremath{\mob_{\gls{pcrg}}}},
	description = {Positive charge density mobility [\si{\square\meter\per\second\per\volt}]}
}
\newglossaryentry{mmob}{
	name = {\ensuremath{\mob_{\gls{ncrg}}}},
	description = {Negative charge density mobility [\si{\square\meter\per\second\per\volt}]}
}
\newglossaryentry{ef}{
	name = {\vect{E}},
	description = {Electric field [\si{\volt\per\meter}]}
}
\newglossaryentry{v}{
	name = {\ensuremath{V}},
	description = {Electric potential [\si{\volt}]}
}
\newglossaryentry{epr}{	
	name = {\ensuremath{\epsilon_r}},
	description = {Relative electric permittivity [\si{\volt\per\coulomb\per\mole}]}
}
\newglossaryentry{R}{	
	name = {\ensuremath{\mathcal{R}}},
	description = {Universal gas constant [\SI{8.3144598}{\joule\per\kelvin\per\mole}]}
}
\newglossaryentry{ec}{	
	name = {\ensuremath{e}},
	description = {Elementary charge [\SI{1.60218d-19}{\coulomb}]}
}
\newglossaryentry{Na}{	
	name = {\ensuremath{N_a}},
	description = {Avogadro Number [\SI{6.02214d23}{\per\mole}]}
}
\newglossaryentry{kb}{	
	name = {\ensuremath{k_B}},
	description = {Boltzmann constant [\SI{1.38064852d-23}{\joule\per\kelvin}]}
}
\newglossaryentry{ep0}{	
	name = {\ensuremath{\epsilon_0}},
	description = {Free-space electric permittivity [\SI{8.8541878176d-12}{\volt\per\coulomb\per\mole}]}
}
\newglossaryentry{Lref}{	
	name = {\ensuremath{L_{ref}}},
	description = {Reference length [\si{\meter}]}
}
\newglossaryentry{Uref}{	
	name = {\ensuremath{\gls{um}_{ref}}},
	description = {Reference velocity [\si{\meter\per\second}]}
}
\newglossaryentry{denref}{	
	name = {\ensuremath{\gls{den}_{ref}}},
	description = {Reference density [\si{\kilo\gram\per\cubic\meter}]}
}
\newglossaryentry{nuref}{	
	name = {\ensuremath{\gls{nu}_{ref}}},
	description = {Reference kinematic viscosity [\si{\square\meter\per\second}]}
}
\newglossaryentry{efref}{	
	name = {\ensuremath{E_{ref}}},
	description = {Reference electric field [\si{\volt\per\meter}]}
}
\newglossaryentry{imfref}{
	name = {\ensuremath{\mf_{i,ref}}},
	description = {Reference mass fraction of I$^{th}$ species [-]}
}
\newglossaryentry{mmref}{	
	name = {\ensuremath{\mm_{ref}}},
	description = {Reference molar mass [\si{\kilo\gram\per\mole}]}
}
\newglossaryentry{difref}{	
	name = {\ensuremath{\dif_{ref}}},
	description = {Reference molecular diffusivity [\si{\square\meter\per\second}]}
}
\newglossaryentry{mobref}{	
	name = {\ensuremath{\mob_{ref}}},
	description = {Reference mobility [\si{\square\meter\per\second\per\volt}]}
}
\newglossaryentry{Ajref}{
	name = {\ensuremath{A_{j,ref}}},
	description = {Reference reaction pre-exponential coefficient [\si{\cubic\meter\per\mole\per\second}]}
}
\newglossaryentry{Eajref}{
	name = {\ensuremath{E_{j,ref}}},
	description = {Reference reaction specific activation energy [\si{\joule\per\mole}]}
}
\newglossaryentry{delTref}{
	name = {\ensuremath{\Delta \gls{T}_{ref}}},
	description = {Reference temperature delta [\si{\kelvin}]}
}
\newglossaryentry{cpref}{
	name = {\ensuremath{c_{p,ref}}},
	description = {Reference constant pressure specific heat [\si{\joule\per\kilo\gram\per\kelvin}]}
}
\newglossaryentry{lamref}{
	name = {\ensuremath{\lambda_{ref}}},
	description = {Reference mixture thermal conductivity [\si{\watt\per\meter\per\second}]}
}
\newglossaryentry{Re}{	
	name = {\ensuremath{\text{Re}}},
	description = {Reynolds Number [-] ($\gls{Re} = \frac{\gls{Uref}\gls{Lref}}{\gls{nuref}}$)}
}
\newglossaryentry{Frel}{	
name = {\ensuremath{\text{Fr}_{el}}},
description = {Electric Froude Number [-] ($\gls{Frel} = \gls{Uref} \sqrt{\frac{\gls{mmref}}{\gls{ec} \gls{Na} \gls{efref} \gls{Lref}}}$)}
}
\newglossaryentry{Pr}{	
	name = {\ensuremath{\text{Pr}}},
	description = {Prandtl Number [-] ($\gls{Pr} = \frac{\gls{nuref} \gls{lamref}}{\gls{cpref} \gls{denref}}$)}
}
\newglossaryentry{Sc}{	
	name = {\ensuremath{\text{Sc}}},
	description = {Schmidt Number [-] ($\gls{Sc} = \frac{\gls{difref}}{\gls{nuref}}$)}
}
\newglossaryentry{Mael}{	
	name = {\ensuremath{\text{Ma}_{el}}},
	description = {Electric Mach Number [-] ($\gls{Mael} = \frac{\gls{Uref}}{\gls{efref} \gls{mobref}}$)}
}
\newglossaryentry{Da}{	
	name = {\ensuremath{\text{Da}}},
	description = {Damk\"{o}hler Number [-] ($\gls{Da} = \frac{\gls{Lref}}{\gls{Uref}} \frac{\gls{denref} \gls{Ajref}}{\gls{mmref}} \exp \Big[-\frac{\gls{Eajref}}{\gls{R}\gls{delTref}}\Big]$)}
}
\newglossaryentry{Ec}{	
	name = {\ensuremath{\text{Ec}}},
	description = {Eckert Number [-] ($\gls{Ec} = \frac{\gls{Uref}^2}{\gls{cpref} \gls{delTref}}$)}
}
\newacronym[longplural={Large Eddy Simulations}]{les}{LES}{Large Eddy Simulation}
\newacronym[longplural={Left Hand Sides}]{lhs}{LHS}{Left Hand Side}
\newacronym[longplural={Right Hand Sides}]{rhs}{RHS}{Right Hand Side}
\newacronym[shortplural={EEDFs},longplural={Electron Energy Distribution Functions}]{eedf}{EEDF}{Electron Energy Distribution Function}
\newacronym{cfd}{CFD}{Computational Fluid Dynamics}
\newacronym{fpv}{FPV}{Flamelet Progress-Variable}
\newacronym{cfl}{CFL}{Courant--Friedrichs--Lewy}
\newacronym{pdf}{PDF}{Probability Density Function}

%% file: Sections/Intro.tex
%
%
\section{Introduction}
\label{sec:Introduction}
Flow control is a crucial issue to enhance the performance of modern internal combustion engines.
In particular, experimental tests have demonstrated that the extinction limit of both premixed and diffusive flames can be controlled by the application of an external electric field~\citep{Cessou2012,Duan2015,Lee2005,Hutchins2014}.
The effectiveness of this method comes from the action of the imposed electric field on the charged particles produced by the combustion process.
In this way, the fluid mixture is polarized around the flame front, where a body force is thus applied.
Moreover, such a mixture polarization determines a distortion of the electric potential field, where the flame behaves as an additional electrode, whose electric potential is comparable to the applied voltage \citep{Han2017}.
Being the flame usually contained between the two electrodes, the presence of the flame front further increases the electric field strength applied to the fluid mixture.\par
However, in spite of the experimental evidence, the models developed for the prediction of this phenomenon are computationally too expensive for practical design purposes.
For this reason, it is mandatory to design methods that reduce the additional cost associated with the adoption of these models in flame simulations.
At the same time, considering the importance of the local charge distribution, any model should correctly predict the amount of cations and anions produced by the flame, requiring a comprehensive kinetic mechanism and accurately taking into account the right transport properties of the ions~\citep{Fialkov1997}.
In particular, the present work deals with the combustion of methane in air for which, at the best of the authors' knowledge, a complete mechanism for ionized species has been proposed by \citet{Starik2002}.
It consists of 392 reactions for the production and depletion of 59 species.
Another noteworthy kinetic scheme, proposed by \citet{Prager2007}, is based on the mechanism for the prediction of lean methane-air mixtures combustion assembled by \citet{Warnatz2001} (208 reactions among 38 species) and takes into account the production and depletion of \num{11} charged species through \num{67} reactions.
The model has been validated against the experimental data of \citet{Goodings1979,Goodings1979b}.\par
A big effort has also been spent in the determination of the transport properties, especially concerning the free electrons produced by the flame.
A very sophisticated model has been proposed by \citet{Bisetti2012}, where the electron properties are computed using the momentum transfer cross-sections of the electrons with the main components of the gas mixture.
The complexity of the model has been further improved by the same authors~\citep{Bisetti2014} including non-thermal effects in the ionization process.
Interesting work has also been done by \citet{Han2015} including charges-charges and charges-neutral interaction modeling in order to study the effect of polarizability of the species.
Probably, \citet{Speelman2015a} have been the first, in this context, to employ a complete binary diffusion approach, which takes into account at the same time the molecular diffusion and the drift due to the electric field.
At the present time, it has only been possible to use these models in low dimensional flame descriptions because of their high complexity and computational cost.\par
Only a few simulations have been carried out in realistic configurations in conjunction with a \gls{cfd} approach.
One of the first attempts in modeling the interaction of the electric field with the combustion process has been done by \citet{Hu2000}.
In this model, a co-flow flame and a candle flame of methane in air have been studied using a reduced kinetic mechanism computed at run-time in a two-dimensional configuration.
This approach neglects the effect of the local charge distribution on the electric field, therefore considering it constant at each point of the domain.
This assumption can often lead to a large underestimation of the local electric field strength and, therefore, to a reduced effect of the voltage difference on the flow.
A few years later, a large improvement in modeling the phenomenon has been achieved by \citet{Yamashita2009}, who computed the capillary combustion chamber already studied experimentally and numerically by \citet{Papac2005} and \citet{Papac2008}.
The interaction between the charge produced by the flame and the local electrical potential was taken into account solving the Gauss law at each time-step of the simulation.
Since the flame was mostly confined close to the metallic surfaces, the presence of the electrons in the mixture was neglected assuming that, because of the high mobility, they would have been rapidly removed.
This assumption, in conjunction with a reaction mechanism which considers only the electrons as negatively charged species, probably leads to an over-estimation of the flame response to the voltage.\par
Although the transport of the entire set of species of the kinetic mechanism guarantees the best accuracy during the calculation, this approach is still computationally too expensive to be applied to real industrial cases.
For this reason, it is necessary to employ a reduced combustion model even in two-dimensional cases.
The first solution to this problem has been proposed by \citet{Belhi2010,Belhi2013}.
Neglecting the effect of the production of charged species on the neutral chemistry, the proposed model uses a kind of laminar \gls{fpv} approach~\citep{Fiorina2005} to simulate the combustion process.
Two equations (one for the mixture fraction \gls{mf} and one for the progress-variable \gls{prv}) are solved and a tabulated function, namely
\begin{equation}
	\phi = \mathcal{F}_\phi(\gls{mf}, \gls{prv}),
	\label{eqn:fun_manifold}
\end{equation}
is used to predict the generic thermo-chemical mixture property $\phi$.
An additional transport equation is then added to the system for each charged species considered in the mechanism.
The species properties and production rates are computed using the temperature and mass fractions stored in the \gls{fpv} chem-table (Eq. \eqref{eqn:fun_manifold}).
This approach definitely allows one to use detailed schemes for the combustion description, but still, poses limits on the number of species used in the ionization mechanism.\par
Considering the good predictive capabilities shown by \gls{fpv} models in a wide range of cases~\citep{Pierce+Moin,Ihme+Pitsch1,Ihme+Pitsch2,Pecquery2014,Moureau2011,Fiorina2010}, the aim of the present work is to develop a model for the interaction of electric field with lifted diffusive flames, completely based on and consistent with the flamelet formulation.
Such a formulation guarantees the possibility of using arbitrarily complex mechanisms for both the neutral and charged species, without any computational cost overhead.
In the next sections, after the description of the main assumptions and equations employed by the model, a one-dimensional validation test is presented to assess the capability of the \gls{fpv} approach to reproduce the results of the corresponding detailed chemi-ionization mechanism; then, a two-dimensional numerical test case using two different kinetic mechanisms is considered to assess the effectiveness of the proposed approach.

%% file: Sections/Model.tex
%
%
\section{Mathematical model}
\label{sec: mat_mod}
\subsection{Governing equations}
In order to focus our attention on modeling electro-chemical phenomena, the present work deals only with laminar flows, avoiding, in this way, the demanding task of the turbulence modeling.
Moreover, the low-Mach-number regime considered allows one to neglect flow compressibility, therefore decoupling the Navier--Stokes equations from the energy transport equation.
Following the classical derivation of low-Mach-number equations, the pressure (\gls{pre}) field is decomposed into a spatially-uniform thermodynamic pressure and a hydrodynamic component, which is retained only in the momentum equation.
In fact, under the assumption of small pressure fluctuations, the density can be computed by the mean thermodynamic pressure together with the fluid temperature and composition.
Thus, the fluid dynamics is modeled by solving the following mass and momentum conservation equations
\begin{equation}
	\tder{\gls{den}} + \divg (\gls{den} \gls{u}) = 0,
	\label{eqn:mass_conv}
\end{equation}
\begin{equation}
	\tder{\gls{den} \gls{u}} + \divg ( \gls{den} \gls{u} \gls{u}) = -\nabla \gls{pre} + \gls{den} \gls{fm} + \divg \gls{sig};
	\label{eqn:mom_conv}
\end{equation}
where 
\tim is the time,
\gls{den} is the mixture density,
\gls{u} is the velocity,
\gls{fm} is a generalized specific force field.
The local shear stress tensor \gls{sig} is modeled as
\begin{equation}
	\gls{sig} = 2  \gls{den} \gls{nu} \Bigl[ \gls{StrT} - \frac{1}{3} (\divg \gls{u}) \tens{I} \Bigr],
	\label{eqn:sigma}
\end{equation}
where 
\begin{equation}
	\gls{StrT} = \frac{1}{2} \Bigl[ \nabla \gls{u} + (\nabla \gls{u})^T \Bigr],
	\label{eqn:ShearT}
\end{equation}
and \gls{nu} is the kinematic viscosity, evaluated as a function of temperature and mixture composition.
The energy equation together with the balance equations of the chemistry model close the system of the governing equations as described in the next section.
\subsection{Chemistry model}
\label{subsec:chemMod}
The present approach is based on the flamelet model proposed by \citet{Fiorina2005}, which uses a set of one-dimensional premixed unstrained flames for solving a detailed mechanism and composing the two-dimensional manifold~(Eq.~\eqref{eqn:fun_manifold}).
Species mass fraction (\gls{imf}), temperature (\gls{T}) and mixture velocity (\gls{um}) distributions satisfy the following equations:
\begin{equation}
	\xder{\gls{den} \gls{um}} = 0,
	\label{eqn:flamelet_m_eq}
\end{equation}
\begin{equation}
	\gls{den} \gls{um} \xder{\gls{imf}} + \xder{\gls{den} \gls{ivdif} \gls{imf}} = \gls{ipr},
	\label{eqn:flamelet_Yi_eq}
\end{equation}
\begin{equation}
	\gls{den} \gls{um} \gls{cp} \xder{\gls{T}} + \isumsp \gls{den} \gls{ivdif} \gls{imf} \gls{icp} \xder{\gls{T}} = \xder{} \Bigl( \gls{lam} \xder{\gls{T}} \Bigr) + \isumsp \gls{ih} \gls{ipr}.
	\label{eqn:flamelet_T_eq}
\end{equation}
In the previous equations:
{\ns} is the number of species;
$(\cdot)_i$ refers to i$^{th}$ species quantity;
\gls{ipr} is the production rate computed using the chosen chemical mechanism;
\gls{icp} is the constant pressure heat capacity;
\gls{cp} is the mixture constant pressure heat capacity defined as $\gls{cp} = \displaystyle \isumsp \gls{imf} \gls{icp}$;
\gls{ih} is the species enthalpy.\par
The diffusion velocity of the i$^{th}$ species (\gls{ivdif}) is computed as
\begin{equation}
	\gls{ivdif} = - \frac{1}{\gls{xi}} \gls{idif} \xder{\gls{xi}} + \ksumsp \frac{\mf_k}{\xf_k} \dif_k \xder{\xf_k} + \frac{1}{\gls{xi}} \dif^T \xder{(\ln \gls{T})},
	\label{eqn:Vij_def}
\end{equation}
where:
\gls{xi} is the molar fraction;
\gls{idif} is the diffusivity;
$\dif^T$ is the thermo diffusivity due to the Soret effect.\par
The previous equations, coupled with the ideal gas law, are used to predict the distribution of all the neutral and charged species with exception to the electrons.
In fact, the mass fraction of the electrons has been calculated imposing the charge neutrality of the mixture with the equation
\begin{equation}
	\np^{e^-} = \np^{cations} - \np^{anions},
	\label{eqn:crg_neutrality}
\end{equation}
where the number of particles per unit of volume (\np) is computed as 
\begin{equation}
	\np^{i} = \gls{Na}\gls{den} \frac{\gls{imf}}{\gls{imm}}.
	\label{eqn:n_def}
\end{equation}
In the previous equation, \gls{Na} is the Avogadro number and \gls{imm} is the i$^{th}$ species molar mass.
Such an approximation is used only during the FlameMaster pre-processing calculations.
In fact, as described in details in the next sub-section, the present model does not take into account the effect of the electric field on the flamelets.
This assumption allows one to simplify the model, avoiding the numerical and theoretical complexity of a functional mapping involving the local electric field strength and direction at the expense of a limitation in the applied electric field intensity, as properly discussed in the following.\par
Defining the progress variable as the linear combination of the mass fractions of the main combustion products, it is possible to embed the entire combustion process in a functional manifold.
This manifold is populated using premixed unstrained flamelet solutions for a wide range of equivalence ratio and considering the mixture fraction and the progress variable as independent variables.
Therefore, only these two quantities are transported through the computational domain solving the following equations together with Equations~\eqref{eqn:mass_conv} and \eqref{eqn:mom_conv}:
\begin{equation}
	\tder{\gls{den} \gls{mf}} + \divg (\gls{den} \gls{u} \gls{mf}) = \divg ( \gls{den} \gls{zdif} \nabla \gls{mf} ),
	\label{eqn:mix_eq}
\end{equation}
\begin{equation}
	\tder{\gls{den} \gls{prv}} + \divg (\gls{den} \gls{u} \gls{prv}) = \divg ( \gls{den} \gls{cdif} \nabla \gls{prv} ) + \gls{den} \gls{cpr}.
	\label{eqn:C_eq}
\end{equation}
The Lewis number for these two scalars is assumed to be equal to one leading to $\gls{cdif} = \gls{zdif} = \frac{\gls{lam}}{\gls{den}\gls{cp}}$; this quantity is computed and stored in a two-dimensional chem-table along with the chemical source term of the progress-variable, \gls{cpr}, the mixture density and viscosity.
\subsection{Charged species transport model}
\label{subsec:crgMod}
The proposed model for charge transport is based on the assumption that the presence of the electric field does not affect the combustion process of the neutral species.
This assumption is valid only when the applied electric field is weak enough not to activate non-thermal phenomena due to the presence of free-electrons and when ionized species mass fractions are much smaller than combustion radical ones.
On the other hand, this hypothesis strongly simplifies the model, reducing the dimensions of the needed functional manifold, with obvious advantages in terms of computational cost and memory footprint.
Moreover, the use of the low-Mach-number formulation of the Navier--Stokes equations has forced the authors to neglect all the effects of the applied electric field on the energy of the system.
This assumption is reasonable considering the low ion currents developed in the domain, which would lead to a negligible heating due to the Joule effect.
Furthermore, the low amount of charges produced in the flame by chemi-ionization with respect to the neutral species implies that the enthalpy fluxes activated by the electric field would have only a minor effect of the total enthalpy of the system.\par
Since cations and anions move in opposite directions, when exposed to an electric field, at least two scalar quantities are necessary in order to predict the distribution of positive and negative charges in the domain.
Using an approach similar to the definition of the progress variable, we have employed the two quantities \gls{pcrg} and \gls{ncrg}, defined as
\begin{equation}
	\gls{pcrg} = \left. \gls{ec} \gls{Na} \isumsp \frac{\gls{Si} \gls{imf}}{\gls{imm}}\right\vert_{\gls{Si}>0}
	\label{eqn:P_def}
\end{equation}
and
\begin{equation}
	\gls{ncrg} = \left. - \gls{ec} \gls{Na} \isumsp \frac{\gls{Si} \gls{imf}}{\gls{imm}}\right\vert_{\gls{Si}<0};
	\label{eqn:M_def}
\end{equation}
where
\gls{Si} is the number of elementary charges (positive for cations and negative for anions)
and \gls{ec} is the elementary charge expressed in Coulomb. 
Being \gls{pcrg} and \gls{ncrg} linear combinations of species mass fractions, their transport equations read
\begin{equation}
	\tder{\gls{den} \gls{pcrg}} + \divg (\gls{den} \gls{u} \gls{pcrg} + \gls{den} \gls{pmob} \gls{ef} \gls{pcrg}) = \divg ( \gls{den} \gls{pdif} \nabla \gls{pcrg} ) + \gls{den} \, \gls{ppr}
	\label{eqn:P_eq}
\end{equation}
and
\begin{equation}
	\tder{\gls{den} \gls{ncrg}} + \divg (\gls{den} \gls{u} \gls{ncrg} - \gls{den} \gls{mmob} \gls{ef} \gls{ncrg}) = \divg ( \gls{den} \gls{mdif} \nabla \gls{ncrg} ) + \gls{den} \, \gls{mpr}.
	\label{eqn:M_eq}
\end{equation}
The additional advective term is due to the force applied by the electric field  on the charged particles and it takes into account the different mobility of the species.
The mobility of the drifting scalars is modeled using the following mass weighted average:
\begin{equation}
	\gls{pmob} = \frac{\left. \displaystyle \isumsp \frac{\gls{Si} \gls{imf} \gls{imob}}{\gls{imm}} \right\vert_{\gls{Si}>0}}
    {\left. \displaystyle \isumsp \frac{\gls{Si} \gls{imf}}{\gls{imm}} \right\vert_{\gls{Si}>0}}
	\label{eqn:Pmob_def}
\end{equation}
and
\begin{equation}
	\gls{mmob} = \frac{\left. \displaystyle \isumsp \frac{\gls{Si} \gls{imf} \gls{imob}}{\gls{imm}} \right\vert_{\gls{Si}<0}}{\left. \displaystyle \isumsp \frac{\gls{Si} \gls{imf}}{\gls{imm}} \right\vert_{\gls{Si}<0}}.
	\label{eqn:Mmob_def}
\end{equation}
These two expressions are ill-defined far from the flame-front, where the charged species mass fractions, computed in the flamelet environment, are zero; moreover, imposing weak electric fields, the modification of the ions spatial distribution is very small far from the flame front.
For this reason, when the sum of the charged species molar fractions (either positive or negative) is lower than \num{d-30}, $\mob_{\gls{pcrg}/\gls{ncrg}}$ are evaluated as the arithmetical average of the species mobility.
As proposed by \citet{Fialkov1997}, \SI{d-4}{\square\meter\per\volt\per\second} has been retained as the mobility of heavy cations and anions, whereas the electron mobility has been estimated as \SI{1.868d-2}{\square\meter\per\volt\per\second} using the formula proposed by \citet{Belhi2013}.
As already pointed out in literature, this value is less accurate than the estimate of \citet{Bisetti2012}, but it ensures a lower computational cost reducing the electron velocity through the domain.
With the aim of evaluating the prediction capability of the formulated reduced \gls{fpv} model, in the absence of detailed experimental data, this approximation appeared appropriate; in fact, the computational burden associated with the adoption of a more accurate and complex mobility model, leading to higher mobility values, would not have any significant information to the present work.
The evaluation of the result sensitivity to the accuracy of the mobility model of cations, anions and electrons is highly relevant but it is beyond the scope of the present work and is deferred to future works.\par
The same averaging procedure has been applied to predict the diffusivity of the two scalars, \gls{pcrg} and \gls{ncrg}.
The diffusivity of the heavy charged species has been set equal to that of their corresponding neutral, whereas the Einstein relationship is used to model the electrons diffusion coefficients,
\begin{equation}
	\dif_{e^-} = \frac{\gls{kb} \mob_{e^-} \gls{T}}{\gls{ec}}.
	\label{eqn:eDiff_def}
\end{equation}
When the average operator is ill defined, the drifting scalar diffusivity has been approximated with the thermal diffusivity.\par
The present model neglects the influence of the electric field on the combustion process.
This is acceptable for weak electric fields; however, particular attention must be paid to the evaluation of the production terms of \gls{pcrg} and \gls{ncrg}, being the recombination process of the charges strongly dependent on the local charge balance \citep{Han2017}.
In fact, computing these terms as a simple linear combination of the production rates of the corresponding species can lead to large errors in the prediction of the charge distributions.
This is due to the presence of the Coulomb force in equations (\ref{eqn:P_eq}) and (\ref{eqn:M_eq}) which renders the transport of these scalars very different from the conditions considered in the flamelet environment.
In fact, the ion-wind is not taken into account when generating the flamelet chem-table, since the steady flamelet equations \eqref{eqn:flamelet_m_eq}, \eqref{eqn:flamelet_Yi_eq}, \eqref{eqn:flamelet_T_eq} are solved neglecting the coupling with the electric field and its interaction with the charged species; therefore, in order to be consistent with the manifold used for all the other terms, it is necessary to define an appropriate scaling for the two production rates (\gls{ppr} and \gls{mpr}), taking into account the concentration of positive and negative charges computed at run-time.
The scaling used here is inspired by the work of \citet{Ihme2008} about the prediction of nitric oxide concentration using an \gls{fpv} model.\par
The general model reaction for the production/depletion of ions has the form:
\begin{equation}
	R'\;+\;R''\;+\;...\;\rightleftharpoons\;P^+\;+\;P^-\;+\;P'\;+\;...
	\label{eqn:C_react}
\end{equation}
where several neutral species ($R'$, $R''$, \dots) react to produce positively and negatively charged particles ($P^+$ and $P^-$) as well as other neutrals ($P'$, \dots).
From now on, the procedure will be explained for a reaction where the forward direction produces the ions and backward direction consumes them.
It is trivial to extend this procedure to reactions where only one direction is allowed.
The charge transfer reactions are not considered in the procedure because they do not lead to a change in the total positive or negative charge, but only to a change of the mixture composition.
The forward reaction rate will always depend on the fluid properties and on the concentrations of the neutral species.
Therefore, under the assumption that the electric field does not have any effect on the neutral chemistry, 
the forward reaction rate can be kept constant for all the \gls{pcrg} and \gls{ncrg} values.
A positive production term of the charges for the j-th reaction is defined as
\begin{equation}
	\pr^+_j = \gls{ec} \gls{Na} k^f_j n_c,
	\label{eqn:omegaplusPM_def}
\end{equation}
where $k^f_j$ is the forward reaction rate of the j-th reaction and $n_c$ is the number of cations or anions produced by the reaction.
For the conservation of charges, equation~\eqref{eqn:omegaplusPM_def} can be applied to cations or anions, indifferently.
The negative production term for the j-th reaction can be defined in a similar way,
\begin{equation}
	\pr^-_j = - \gls{ec} \gls{Na} k^b_j n_c,
	\label{eqn:omeganegPM_def}
\end{equation}
where $k^b_j$ is the backward reaction rate of the j-th reaction and $n_c$ is the number of cations or anions consumed by the reaction.
The backward reaction rate depends on the fluid properties and on the concentrations of the products of the reaction.
In particular, unlike the forward one, the backward reaction rate is strongly influenced by the imposition of an external electric field.
For this reason, in order to take into account the real distribution of charges in the domain, assuming a first order kinetics for the charges, a linear scaling is imposed to the negative production rate, leading to the following definition of the production terms:
\begin{equation}
	\gls{ppr} = \gls{mpr} = \jsumre \pr^+_j + \frac{ \displaystyle \jsumre \pr^-_j }{ (\gls{pcrg})_{FPV} (\gls{ncrg})_{FPV} } \gls{pcrg} \gls{ncrg}.
	\label{eqn:omegaPM_def}
\end{equation}
In the equation above, {\nr} is the number of reactions involving a production or a consumption of charges.
Moreover, the summations are precomputed and stored in the chem-table, along with: the \gls{pcrg} and \gls{ncrg} quantities obtained by the solution of the steady flamelet equation and indicated with subscript ``$(\cdot)_{FPV}$'' (to distinguish them from the run time values); \gls{pmob} and \gls{mmob}; \gls{pdif} and \gls{pdif}.
Equation \eqref{eqn:omegaPM_def} ensures, by construction, that both scalars will have exactly the same production rate, enforcing the conservation of charge principle.\par
Using the described model it is possible to evaluate the local charge of the fluid mixture as
\begin{equation}
	Charge = \gls{den} (\gls{pcrg} - \gls{ncrg}),
	\label{eqn:charge_eq}
\end{equation}
which is divided by the free-space electrical permittivity and used in the Gauss law, namely
\begin{equation}
	\lap \gls{v} = \frac{\gls{den} (\gls{ncrg} - \gls{pcrg})}{\gls{ep0}},
	\label{eqn:Gauss_law}
\end{equation}
to compute the local electric potential and then the electric field
\begin{equation}
	\gls{ef} = -\nabla \gls{v}.
	\label{eqn:E_eq}
\end{equation}
Finally, assuming that the entire force applied by the electric field on the charged particles is transferred to the fluid, the following specific force term is added to the momentum conservation equation
\begin{equation}
	\gls{fm} = (\gls{pcrg} - \gls{ncrg}) \gls{ef}.
	\label{eqn:mom_src}
\end{equation}

%% file: Sections/Num_proc.tex
%
%
\section{Numerical Procedure}
\label{sec:num_proc}
The low-Mach-number Navier--Stokes equations together with the transport equations of the combustion model are solved by the semi-implicit fractional-step method proposed by \citet{Shunn2012}. 
A Poisson equation for the pressure is obtained by a projection step to achieve a consistent discretization of the momentum equation.
The computational domain is discretized by an unstructured grid employing a linear reconstruction of the variable inside each cell to evaluate spatial gradients. 
The resulting spatial discretization is second-order accurate with low numerical dissipation.
As pointed out by \citet{Sommerer1992} and more recently by \citet{Belhi2013}, the high drift velocity developed by the ions, even when exposed to weak electric fields, increases the stiffness of the problem requiring a time-step much smaller than that needed to advance the Navier--Stokes equations in time.
In fact, the \gls{cfl} condition for the $k^{th}$ drifting scalar is reformulated to provide the integration time-step ($\Delta \tim$) as
\begin{equation}
	\gls{cfl}_k = \max_{1 \le j \le n_{cv}} \vast( \frac{\Delta \tim \displaystyle \sum_{i=1}^{n_{f,j}} | ( \gls{u}_i + \mob_k \gls{ef}_i ) \cdot \vect{n}_i S_i |}{V_j} \vast),
	\label{eqn:cfl_elec}
\end{equation}
where:
$V_j$ is the grid cell volume;
$n_{cv}$ is the number of cell in the computational domain;
$n_{f,j}$ is the number of faces for the $j^{th}$ control volume;
$\vect{n}_i$ is the unit normal to the $i^{th}$ face;
$S_i$ is the $i^{th}$ face surface area;
$\gls{u}_i$ and $\gls{ef}_i$ are the flow velocity and the electric field computed on the i-th face;
$\mob_k$ is the mobility of the $k^{th}$ drifting scalar.
Since electron mobility is orders of magnitude higher than that of other ionized species, the time-step will be always limited by the scalar that represents the negative charges.
In order to optimize the computational effort needed by the numerical procedure, a nested time discretization is combined with the fractional step for advancing the drifting scalars (\gls{pcrg} and \gls{ncrg}) and the electric potential, imposing a target \gls{cfl} condition for the drifting scalars advection.
In particular, the nested time procedure consists in subdividing the time step used to integrate the flow and chemistry equations into a number of steps calculated in order to guarantee the stability of the stiff charged species transport model, namely, equations \eqref{eqn:P_eq}, \eqref{eqn:M_eq} and \eqref{eqn:Gauss_law}.
The fluid and scalar properties needed for the solution of these equations are linearly interpolated at the solution time.
The Crank-Nicolson time integration in conjunction with the second-order conservative spatial operators, described in \citet{Shunn2012}, guarantees the second-order accuracy of both time discretizations.\par
The advancing procedure for the drifting scalars involves the solution of the \gls{pcrg} and \gls{ncrg} transport equations and of the electric potential Poisson equation using the same spatial operators used for the other equations.
This choice has been made for sake of simplicity and speed in the computational algorithm but it can cause numerical instabilities in the negative charges solution when they are exposed to intense electric fields.
In these situations, the P\'eclet number of this quantity increases and, therefore, the steepened gradients are more critical for the employed centered scheme.
Although this represents one of the major issues of the present numerical procedure, it has been verified that, for the range of applied voltages considered in this paper, no instabilities were encountered.

%% file: Sections/Chem_analysis.tex
%
%
\section{Chemical model}
\label{sec:chem_analysis}
\subsection{Description of the chemical mechanisms}
\label{sec:chem_description}
In order to test the behavior of the present model, when employed in conjunction with different chemical mechanisms, methane-air combustion simulations have been performed using two sets of reactions for the ionized species. 
The ionized species production has been modeled using either the mechanism proposed by \citet{Belhi2010} or the more sophisticated mechanism assembled by the same authors in a successive paper \citep{Belhi2013}.
Even though more accurate ionization mechanisms are provided in the literature \citep{Prager2007}, the use of these two mechanisms allows us to verify (as it will be described later) the all simplifying assumptions underlying the model formulation.
In both cases, the kinetic mechanism for the neutral species is the GriMech~3.0~\cite{GriMech3.0}.\par
\begin{figure}[tb]
	\centering
    \begin{tikzpicture}
		\node[anchor=south west,inner sep=0] (image) at (0,0) {
    		\def\svgwidth{0.9\columnwidth}
    		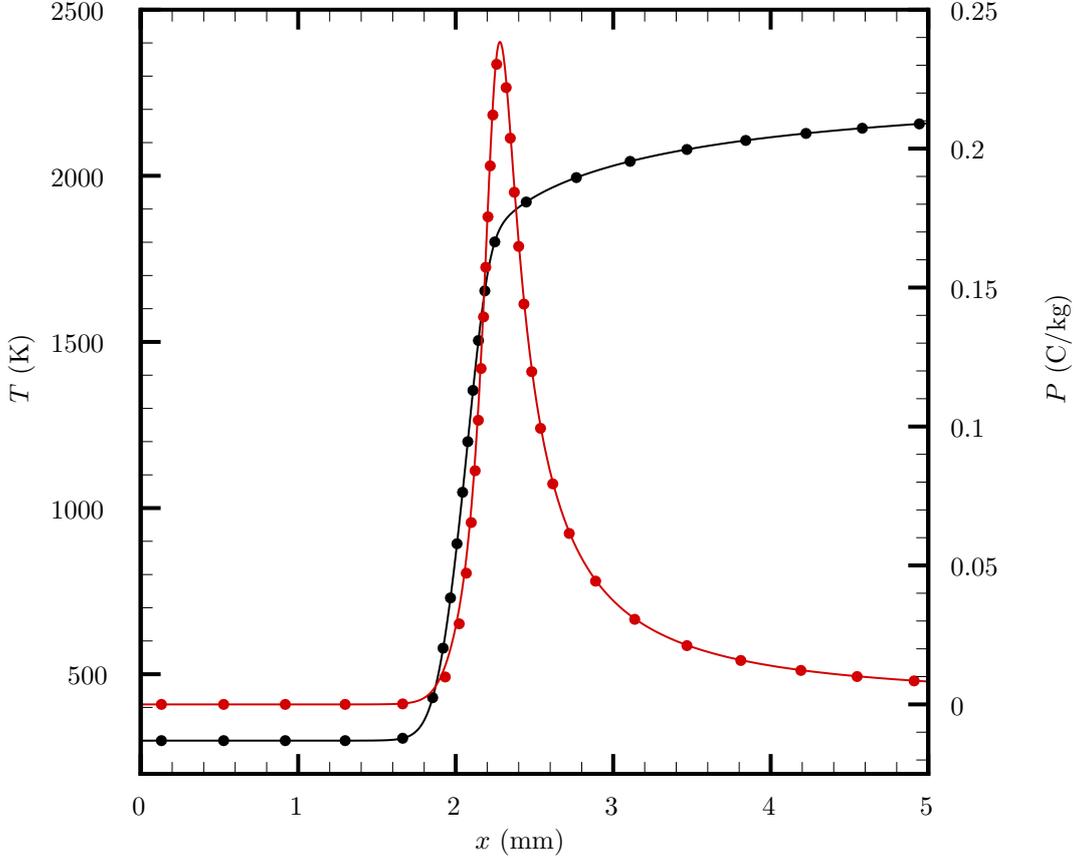
    	};
    \end{tikzpicture}
	\caption[]{Unstrained premixed flamelet solution at $\phi = 1$ of temperature (black) and \gls{pcrg} (red) for the mechanism ``A'' (symbols) and mechanism ``B'' (continuous line).}
	\label{fig:TP_comp}
\end{figure}
Using the first mechanism (later referred to as ``mechanism A''), the ionization consists of two reactions involving the production of only two charged species, namely, H$_3$O$^+$ and e$^-$.
In particular, this mechanism produces the ionized species in the reacting region of the flame through the reaction
\begin{equation}
	CH + O \rightarrow HCO^+ + e^-.
	\label{eqn:cho_reac}
\end{equation}
The produced HCO$^+$ is rapidly converted in H$_3$O$^+$ with the mechanism described by \citet{Pedersen1993}.
The originated anion and cation recombine via the reaction
\begin{equation}
	H_3O^+ + e^- \rightarrow H_2O + H,
	\label{eqn:recomb_reac}
\end{equation}
which has been assumed to be the major recombination reaction by many authors in the past \citep{Hu2000,Yamashita2009,Pedersen1993}.
Being only two charged species present in this mechanism (the presence of HCO$^+$ is negligible), our model almost recovers that proposed by \citet{Belhi2010}.
In fact, the two scalar quantities, \gls{pcrg} and \gls{ncrg}, represent the charge due to the presence of these two species and therefore modeling is not required for their transport properties.\par
Using the second mechanism (later referred to as ``mechanism B''), the electrons produced through the reaction \eqref{eqn:cho_reac} are employed in a series of electron attachment reactions which produce O$_2^-$, OH$^-$ and O$^-$.
These species have multiple effects on the mixture.
Firstly, they absorb part of the electrons preventing the recombination process of the H$_3$O$^+$ and slightly increasing the number of charged particles in the mixture.
Secondly, they strongly reduce the mobility of the negative charges being much heavier than the electrons.
This kinetic mechanism has been employed to test the capabilities of the present model to handle these phenomena without increasing the numerical cost of the computation.\par
For both mechanisms, the transport properties of the heavy ions have been assumed equal to those of the corresponding neutral species given by the GriMech~3.0~\cite{GriMech3.0} database.
The thermal properties of the ions are instead taken from the database by \citet{Burcat}.\par
Although the validation of these two mechanisms has been already provided in \citet{Belhi2010,Belhi2013}, we analyze the results for the 
stoichiometric ($\phi = 1$) unstrained flamelets computed for the conditions of the test case described in Section \ref{sec:test_case}.
The calculations have been performed using the freely distributed C++ code FlameMaster~V3.3.10~\cite{FLAMEMASTER}.\par
\begin{figure}[tb]
	\centering
	\begin{tikzpicture}
		\node[anchor=south west,inner sep=0] (image) at (0,0) {
    		\def\svgwidth{0.9\columnwidth}
    		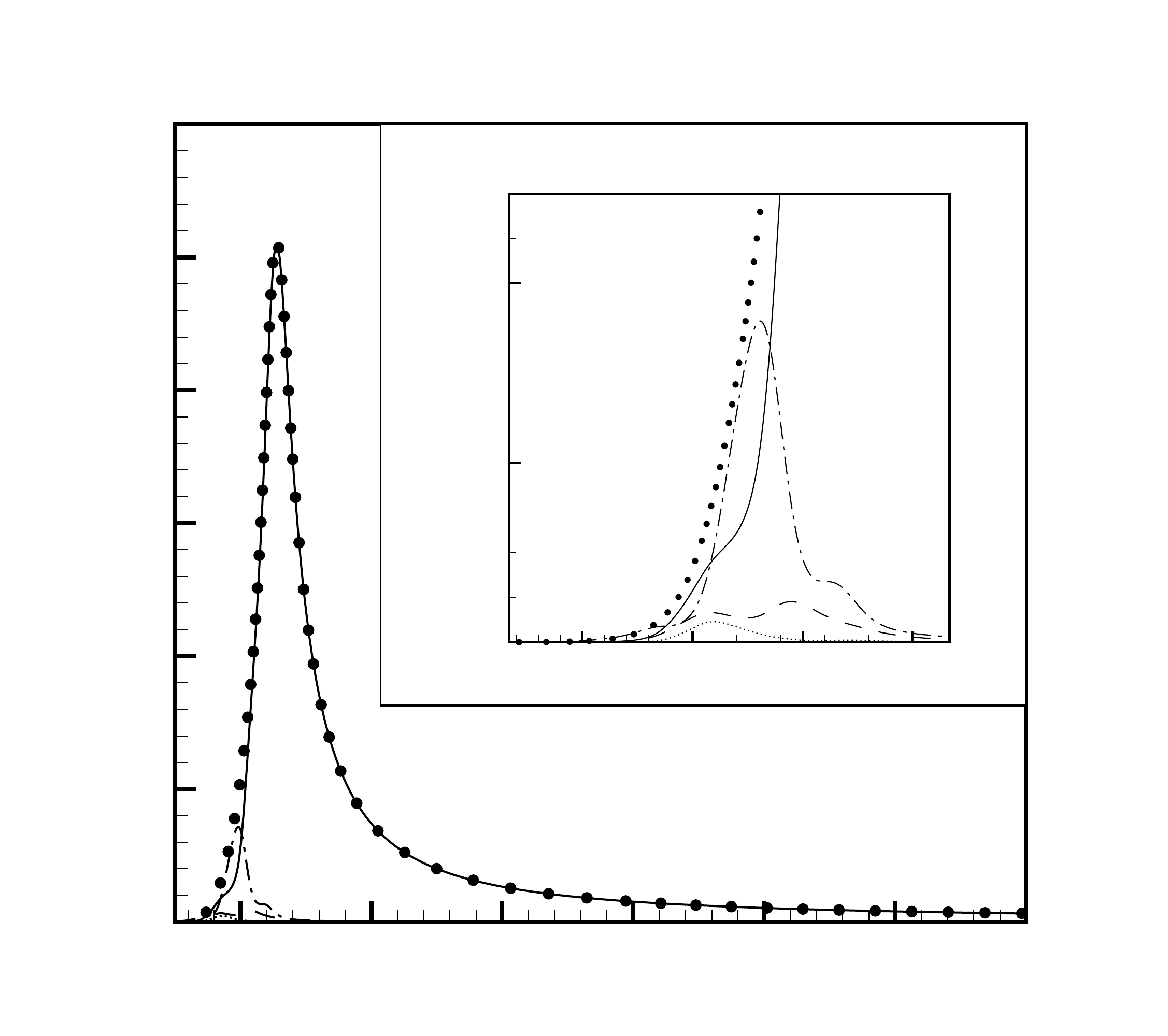
    	};
    \end{tikzpicture}
	\caption[]{Unstrained premixed flamelet solution at $\phi = 1$ for the number density of electrons obtained using mechanism ``A'' (symbols);
				number density
					of e$^-$ 		(\tikz[baseline=-0.5ex]					{ \draw (0,0) -- ++(0.6cm,0); }), 
					of O$_2^-$ 		(\tikz[baseline=-0.5ex,dashed]			{ \draw (0,0) -- ++(0.6cm,0); }), 
					of OH$^-$ 		(\tikz[baseline=-0.5ex,dashdotted]		{ \draw (0,0) -- ++(0.6cm,0); }) and 
					of O$^-$ 		(\tikz[baseline=-0.5ex,dotted]			{ \draw (0,0) -- ++(0.6cm,0); }) 
				obtained using mechanism ``B''.
				A close-up view of the lower left corner of the graph is shown by the inset.}
	\label{fig:Spec_comp}
\end{figure}
Figure~\ref{fig:TP_comp} shows temperature and \gls{pcrg} profiles for both the mechanisms.
The two temperature profiles coincide, confirming that the change of the ionization mechanism has a negligible influence on the combustion process.
Indeed, the charged species constitute only a minor part of the mixture and there are two or, in some cases, three orders of magnitude between the molar fractions of the combustion radicals and those of the anions and cations.
On the other hand, the peak value of \gls{pcrg}, which, as expected, is located close to the flame-front for both mechanisms, is slightly lower for mechanism ``A''.
The difference between the two peak values is about $1\%$ and is due to the employment of the electrons in the production of the heavy charged species.\par 
Figure~\ref{fig:Spec_comp} shows the breakdown of the negative species produced by the two mechanisms, in order to observe their influence on the transport properties of the charged particles.
As expected the main negative species for the mechanism ``B'' is the electron, whose number density has a profile almost identical to that predicted by mechanism ``A''.
The main difference between the two profiles is in the upstream part the flame front, which is shown in the inset.
In this region, the dominant negative species in the mixture is OH$^-$, whose density is in some points even higher than that of the electrons.
For this reason, the mobility of the negative species computed with Eq.~\eqref{eqn:Mmob_def} is much lower for the mechanism ``B'', entailing a different response of the mixture to an applied electric field.

%% file: Figures/TP_comp.eps_tex
\begingroup%
  \makeatletter%
  \providecommand\color[2][]{%
    \errmessage{(Inkscape) Color is used for the text in Inkscape, but the package 'color.sty' is not loaded}%
    \renewcommand\color[2][]{}%
  }%
  \providecommand\transparent[1]{%
    \errmessage{(Inkscape) Transparency is used (non-zero) for the text in Inkscape, but the package 'transparent.sty' is not loaded}%
    \renewcommand\transparent[1]{}%
  }%
  \providecommand\rotatebox[2]{#2}%
  \ifx\svgwidth\undefined%
    \setlength{\unitlength}{648bp}%
    \ifx\svgscale\undefined%
      \relax%
    \else%
      \setlength{\unitlength}{\unitlength * \real{\svgscale}}%
    \fi%
  \else%
    \setlength{\unitlength}{\svgwidth}%
  \fi%
  \global\let\svgwidth\undefined%
  \global\let\svgscale\undefined%
  \makeatother%
  \begin{picture}(1,0.88888889)%
    \put(0,0){\includegraphics[width=\unitlength]{TP_comp.pdf}}%
    \put(0.44733796,0.03047068){\color[rgb]{0,0,0}\makebox(0,0)[lb]{\smash{$x$ (\si{\milli\meter})}}}%
    \put(0.05,0.43032407){\color[rgb]{0,0,0}\rotatebox{90}{\makebox(0,0)[lb]{\smash{\gls{T} (\si{\kelvin})}}}}%
    \put(0.97175926,0.42932099){\color[rgb]{0,0,0}\rotatebox{90}{\makebox(0,0)[lb]{\smash{\gls{pcrg} (\si{\coulomb\per\kilo\gram})}}}}%
    \put(0.14251543,0.06075617){\color[rgb]{0,0,0}\makebox(0,0)[lb]{\smash{0}}}%
    \put(0.28248457,0.06075617){\color[rgb]{0,0,0}\makebox(0,0)[lb]{\smash{1}}}%
    \put(0.42249228,0.06075617){\color[rgb]{0,0,0}\makebox(0,0)[lb]{\smash{2}}}%
    \put(0.5625,0.06075617){\color[rgb]{0,0,0}\makebox(0,0)[lb]{\smash{3}}}%
    \put(0.70250772,0.06075617){\color[rgb]{0,0,0}\makebox(0,0)[lb]{\smash{4}}}%
    \put(0.84251543,0.06075617){\color[rgb]{0,0,0}\makebox(0,0)[lb]{\smash{5}}}%
    \put(0.08498457,0.17797068){\color[rgb]{0,0,0}\makebox(0,0)[lb]{\smash{500}}}%
    \put(0.07001543,0.32677469){\color[rgb]{0,0,0}\makebox(0,0)[lb]{\smash{1000}}}%
    \put(0.07001543,0.4755787){\color[rgb]{0,0,0}\makebox(0,0)[lb]{\smash{1500}}}%
    \put(0.07001543,0.62434414){\color[rgb]{0,0,0}\makebox(0,0)[lb]{\smash{2000}}}%
    \put(0.07001543,0.77314815){\color[rgb]{0,0,0}\makebox(0,0)[lb]{\smash{2500}}}%
    \put(0.86999228,0.15092593){\color[rgb]{0,0,0}\makebox(0,0)[lb]{\smash{0}}}%
    \put(0.86999228,0.2753858){\color[rgb]{0,0,0}\makebox(0,0)[lb]{\smash{0.05}}}%
    \put(0.86999228,0.3998071){\color[rgb]{0,0,0}\makebox(0,0)[lb]{\smash{0.1}}}%
    \put(0.86999228,0.52426698){\color[rgb]{0,0,0}\makebox(0,0)[lb]{\smash{0.15}}}%
    \put(0.86999228,0.64872685){\color[rgb]{0,0,0}\makebox(0,0)[lb]{\smash{0.2}}}%
    \put(0.86999228,0.77314815){\color[rgb]{0,0,0}\makebox(0,0)[lb]{\smash{0.25}}}%
  \end{picture}%
\endgroup%

%% file: Figures/Spec_comp.eps_tex
\begingroup%
  \makeatletter%
  \providecommand\color[2][]{%
    \errmessage{(Inkscape) Color is used for the text in Inkscape, but the package 'color.sty' is not loaded}%
    \renewcommand\color[2][]{}%
  }%
  \providecommand\transparent[1]{%
    \errmessage{(Inkscape) Transparency is used (non-zero) for the text in Inkscape, but the package 'transparent.sty' is not loaded}%
    \renewcommand\transparent[1]{}%
  }%
  \providecommand\rotatebox[2]{#2}%
  \ifx\svgwidth\undefined%
    \setlength{\unitlength}{648bp}%
    \ifx\svgscale\undefined%
      \relax%
    \else%
      \setlength{\unitlength}{\unitlength * \real{\svgscale}}%
    \fi%
  \else%
    \setlength{\unitlength}{\svgwidth}%
  \fi%
  \global\let\svgwidth\undefined%
  \global\let\svgscale\undefined%
  \makeatother%
  \begin{picture}(1,0.88888889)%
    \put(0,0){\includegraphics[width=\unitlength]{Spec_comp.pdf}}%
    \put(0.47731481,0.02155864){\color[rgb]{0,0,0}\makebox(0,0)[lb]{\smash{$x$ (\si{\milli\meter})}}}%
    \put(0.04,0.31733025){\color[rgb]{0,0,0}\rotatebox{90}{\makebox(0,0)[lb]{\smash{n (\SI{d17}{part\per\cubic\meter})}}}}%
    \put(0.20004167,0.06698457){\color[rgb]{0,0,0}\makebox(0,0)[lb]{\smash{2}}}%
    \put(0.31071914,0.06698457){\color[rgb]{0,0,0}\makebox(0,0)[lb]{\smash{3}}}%
    \put(0.42143519,0.06698457){\color[rgb]{0,0,0}\makebox(0,0)[lb]{\smash{4}}}%
    \put(0.53511265,0.06698457){\color[rgb]{0,0,0}\makebox(0,0)[lb]{\smash{5}}}%
    \put(0.64679012,0.06698457){\color[rgb]{0,0,0}\makebox(0,0)[lb]{\smash{6}}}%
    \put(0.75950617,0.06698457){\color[rgb]{0,0,0}\makebox(0,0)[lb]{\smash{7}}}%
    \put(0.87218364,0.06698457){\color[rgb]{0,0,0}\makebox(0,0)[lb]{\smash{8}}}%
    \put(0.10596605,0.08834877){\color[rgb]{0,0,0}\makebox(0,0)[lb]{\smash{0}}}%
    \put(0.08896605,0.20239198){\color[rgb]{0,0,0}\makebox(0,0)[lb]{\smash{\num{0.5}}}}%
    \put(0.08896605,0.31647377){\color[rgb]{0,0,0}\makebox(0,0)[lb]{\smash{\num{1.0}}}}%
    \put(0.08896605,0.43055556){\color[rgb]{0,0,0}\makebox(0,0)[lb]{\smash{\num{1.5}}}}%
    \put(0.08896605,0.54463735){\color[rgb]{0,0,0}\makebox(0,0)[lb]{\smash{\num{2.0}}}}%
    \put(0.08896605,0.65871914){\color[rgb]{0,0,0}\makebox(0,0)[lb]{\smash{\num{2.5}}}}%
    \put(0.08896605,0.77276235){\color[rgb]{0,0,0}\makebox(0,0)[lb]{\smash{\num{3.0}}}}%
    \put(0.48024691,0.29588889){\color[rgb]{0,0,0}\makebox(0,0)[lb]{\smash{1.5}}}%
    \put(0.57469136,0.29588889){\color[rgb]{0,0,0}\makebox(0,0)[lb]{\smash{1.8}}}%
    \put(0.6691358 ,0.29588889){\color[rgb]{0,0,0}\makebox(0,0)[lb]{\smash{2.1}}}%
    \put(0.76358025,0.29588889){\color[rgb]{0,0,0}\makebox(0,0)[lb]{\smash{2.4}}}%
    \put(0.40547068,0.3283179){\color[rgb]{0,0,0}\makebox(0,0)[lb]{\smash{0}}}%
    \put(0.38007716,0.48233025){\color[rgb]{0,0,0}\makebox(0,0)[lb]{\smash{0.2}}}%
    \put(0.38007716,0.63634259){\color[rgb]{0,0,0}\makebox(0,0)[lb]{\smash{0.4}}}%
  \end{picture}%
\endgroup%

%% file: Sections/1D_Verification.tex
%
%
\subsection{1D verification of the method}
\label{sec:1d_ver}
In order to verify the correct implementation of the model in our \gls{cfd} solver and to provide a first validation of the present approach, we have computed the two flamelet solutions provided in the previous subsection 
with the proposed \gls{fpv} approach.
Even though this may seem a trivial test case, it requires that the production rates, the diffusive fluxes as well as the drift induced by the electric field have to be well resolved in order to obtain a good agreement between the results of the FlameMaster code and of the \gls{cfd} solver.\par
\begin{figure}[tb]
	\centering
	\begin{tikzpicture}
		\node[anchor=south west,inner sep=0] (image) at (0,0) {
    		\def\svgwidth{0.9\columnwidth}
    		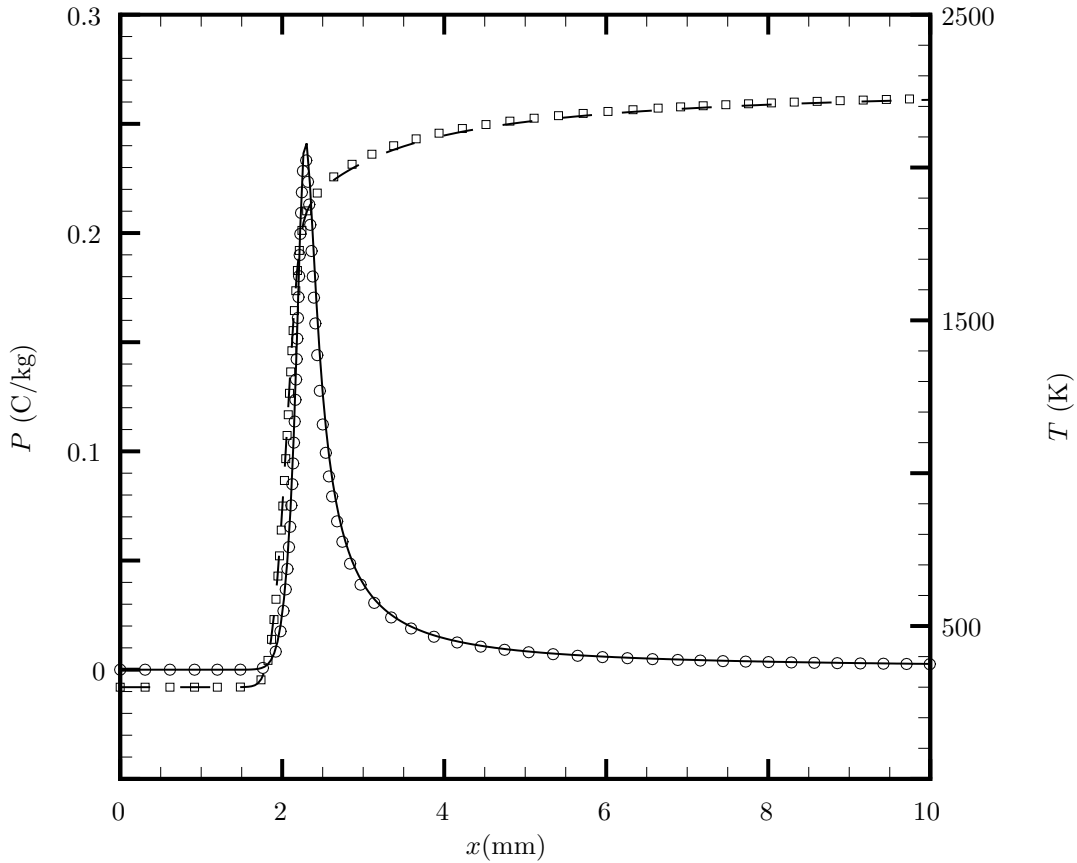
    	};
    \end{tikzpicture}
	\caption[]{Comparison of the unstrained premixed flamelet solution at $\phi = 1$ obtained using the FlameMaster code and the proposed model considering  mechanism ``A''. 
			Symbols represent the profiles of \gls{pcrg} (\tikz[baseline=-0.5ex]{ \node[circle,draw=black,inner sep=0.05cm] {};})
			 and temperature (\tikz[baseline=-0.5ex]{ \node[draw=black,inner sep=0.05cm] {};}), 
			 computed using the detailed chemistry.
			 The dashed and solid lines correspond to the solutions obtained with our CFD code.}
	\label{fig:TP_1D_belhi}
\end{figure}
\begin{figure}[tb]
	\centering
  	\begin{tikzpicture}
		\node[anchor=south west,inner sep=0] (image) at (0,0) {
    		\def\svgwidth{0.9\columnwidth}
    		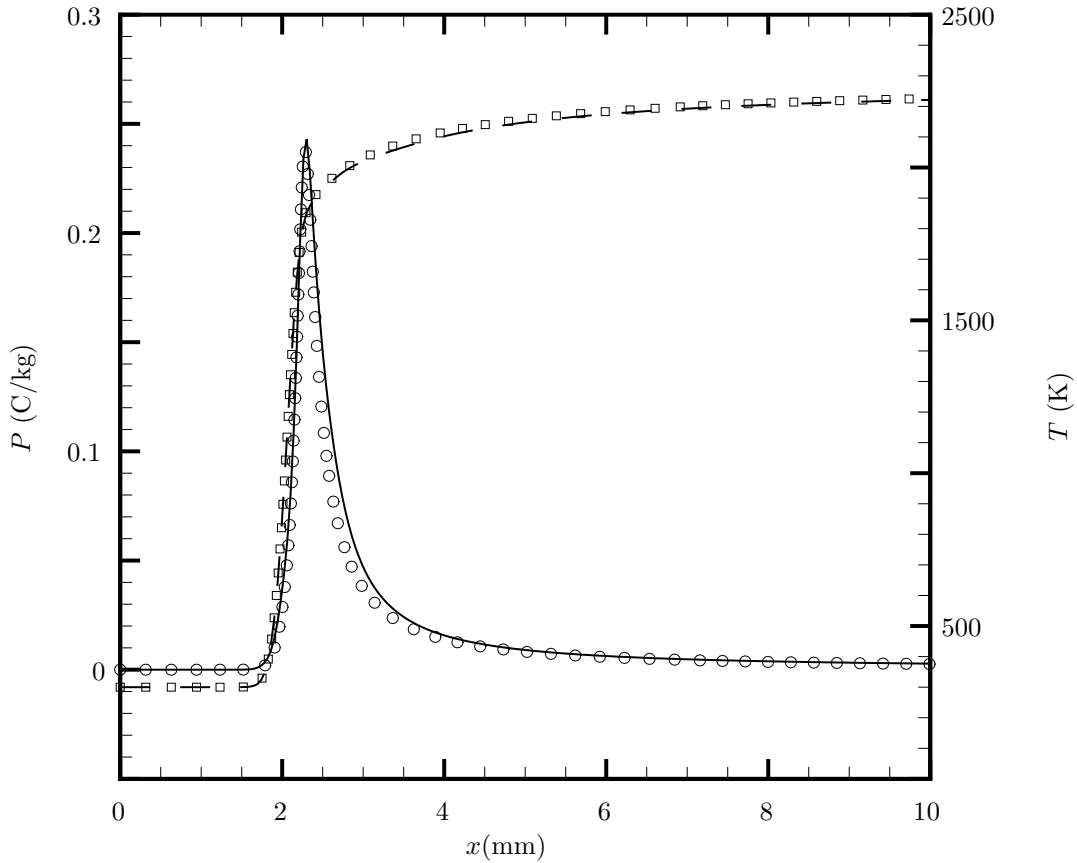
    	};
    \end{tikzpicture}
	\caption[]{Comparison of the unstrained premixed flamelet solution at $\phi = 1$ obtained using the FlameMaster code and the presented model considering the mechanism ``B''. 
		Symbols represent the profiles of \gls{pcrg} (\tikz[baseline=-0.5ex]{ \node[circle,draw=black,inner sep=0.05cm] {};})
			 and temperature (\tikz[baseline=-0.5ex]{ \node[draw=black,inner sep=0.05cm] {};}), 
			 computed using the detailed chemistry.
			 The dashed and solid lines correspond to the solutions obtained with our CFD code.}
	\label{fig:TP_1D_belhi2}
\end{figure}

The \gls{cfd} solution has been obtained using a one-dimensional grid composed of \num{3001} nodes evenly distributed over a total length of \SI{40}{\milli\meter}.
Dirichlet conditions have been imposed at the first left point of the computational domain, 
enforcing the laminar planar flame speed, the mixture fraction relative to the case at $\phi = 1$ and the progress-variable equal to zero.
The concentration of positive and negative ions has also been set to zero at this point, considering that this configuration should not produce any ion-flux.
Moreover, the large distance considered between the inlet plane and the flame-front (about \SI{1.6}{\milli\meter}) ensures the suitability of this assumption.
In fact, it has been verified that the profiles of positive and negative charges concentration reach zero at a large distance from the inlet (about \SI{1.5}{\milli\meter} as shown in Figures~\ref{fig:TP_1D_belhi}-\ref{fig:TP_1D_belhi2}), entailing that the boundary condition is not influencing their fluxes in the domain.
A convective outlet condition has been imposed at the last right point of the computational domain for the velocity and all the scalars.\par
Figure~\ref{fig:TP_1D_belhi} and Figure~\ref{fig:TP_1D_belhi2} show the results obtained for the mechanism ``A'' and ``B'', respectively. In both cases, 
the solutions obtained by the \gls{fpv} approach have been compared with those computed by the FlameMaster code.
It is noteworthy that, unlike the \gls{fpv}-\gls{cfd} solver, the FlameMaster simulations are performed considering the fully coupled detailed chemistry but enforcing the electrical neutrality of the mixture and therefore without solving the generated electric field.
There is an excellent agreement between the profiles of both temperature and positive charge density for the two simulations of each case.
Only tiny differences, which can be attributed to the charge transport computed by the \gls{fpv}-\gls{cfd} solver, can be seen on the charges profile for both cases.
This result was expected considering the small effect of the auto-generated electric field and constitutes a first validation of the proposed model.

%% file: Figures/TP_1D_belhi.eps_tex
\begingroup%
  \makeatletter%
  \providecommand\color[2][]{%
    \errmessage{(Inkscape) Color is used for the text in Inkscape, but the package 'color.sty' is not loaded}%
    \renewcommand\color[2][]{}%
  }%
  \providecommand\transparent[1]{%
    \errmessage{(Inkscape) Transparency is used (non-zero) for the text in Inkscape, but the package 'transparent.sty' is not loaded}%
    \renewcommand\transparent[1]{}%
  }%
  \providecommand\rotatebox[2]{#2}%
  \ifx\svgwidth\undefined%
    \setlength{\unitlength}{648bp}%
    \ifx\svgscale\undefined%
      \relax%
    \else%
      \setlength{\unitlength}{\unitlength * \real{\svgscale}}%
    \fi%
  \else%
    \setlength{\unitlength}{\svgwidth}%
  \fi%
  \global\let\svgwidth\undefined%
  \global\let\svgscale\undefined%
  \makeatother%
  \begin{picture}(1,0.88888889)%
    \put(0,0){\includegraphics[width=\unitlength]{TP_1D_belhi.pdf}}%
    \put(0.43734568,0.03047068){\color[rgb]{0,0,0}\makebox(0,0)[lb]{\smash{\x (\si{\milli\meter})}}}%
    \put(0.05,0.38932099){\color[rgb]{0,0,0}\rotatebox{90}{\makebox(0,0)[lb]{\smash{\gls{pcrg} (\si{\coulomb\per\kilo\gram})}}}}%
    \put(0.97175926,0.40032407){\color[rgb]{0,0,0}\rotatebox{90}{\makebox(0,0)[lb]{\smash{\gls{T} (\si{\kelvin})}}}}%
    \put(0.12249228,0.06075617){\color[rgb]{0,0,0}\makebox(0,0)[lb]{\smash{0}}}%
    \put(0.26651235,0.06075617){\color[rgb]{0,0,0}\makebox(0,0)[lb]{\smash{2}}}%
    \put(0.41049383,0.06075617){\color[rgb]{0,0,0}\makebox(0,0)[lb]{\smash{4}}}%
    \put(0.55451389,0.06075617){\color[rgb]{0,0,0}\makebox(0,0)[lb]{\smash{6}}}%
    \put(0.69849537,0.06075617){\color[rgb]{0,0,0}\makebox(0,0)[lb]{\smash{8}}}%
    \put(0.83499228,0.06075617){\color[rgb]{0,0,0}\makebox(0,0)[lb]{\smash{10}}}%
    \put(0.10501543,0.18649691){\color[rgb]{0,0,0}\makebox(0,0)[lb]{\smash{0}}}%
    \put(0.08279321,0.38206019){\color[rgb]{0,0,0}\makebox(0,0)[lb]{\smash{0.1}}}%
    \put(0.08279321,0.57758488){\color[rgb]{0,0,0}\makebox(0,0)[lb]{\smash{0.2}}}%
    \put(0.08279321,0.77314815){\color[rgb]{0,0,0}\makebox(0,0)[lb]{\smash{0.3}}}%
    \put(0.85999228,0.22561728){\color[rgb]{0,0,0}\makebox(0,0)[lb]{\smash{500}}}%
    \put(0.85999228,0.49938272){\color[rgb]{0,0,0}\makebox(0,0)[lb]{\smash{1500}}}%
    \put(0.85999228,0.77314815){\color[rgb]{0,0,0}\makebox(0,0)[lb]{\smash{2500}}}%
  \end{picture}%
\endgroup%

%% file: Figures/TP_1D_belhi2.eps_tex
\begingroup%
  \makeatletter%
  \providecommand\color[2][]{%
    \errmessage{(Inkscape) Color is used for the text in Inkscape, but the package 'color.sty' is not loaded}%
    \renewcommand\color[2][]{}%
  }%
  \providecommand\transparent[1]{%
    \errmessage{(Inkscape) Transparency is used (non-zero) for the text in Inkscape, but the package 'transparent.sty' is not loaded}%
    \renewcommand\transparent[1]{}%
  }%
  \providecommand\rotatebox[2]{#2}%
  \ifx\svgwidth\undefined%
    \setlength{\unitlength}{648.0125bp}%
    \ifx\svgscale\undefined%
      \relax%
    \else%
      \setlength{\unitlength}{\unitlength * \real{\svgscale}}%
    \fi%
  \else%
    \setlength{\unitlength}{\svgwidth}%
  \fi%
  \global\let\svgwidth\undefined%
  \global\let\svgscale\undefined%
  \makeatother%
  \begin{picture}(1,0.88889103)%
    \put(0,0){\includegraphics[width=\unitlength]{TP_1D_belhi2.pdf}}%
    \put(0.43734568,0.03047068){\color[rgb]{0,0,0}\makebox(0,0)[lb]{\smash{\x (\si{\milli\meter})}}}%
    \put(0.05,0.38932099){\color[rgb]{0,0,0}\rotatebox{90}{\makebox(0,0)[lb]{\smash{\gls{pcrg} (\si{\coulomb\per\kilo\gram})}}}}%
    \put(0.97175926,0.40032407){\color[rgb]{0,0,0}\rotatebox{90}{\makebox(0,0)[lb]{\smash{\gls{T} (\si{\kelvin})}}}}%
    \put(0.12249228,0.06075617){\color[rgb]{0,0,0}\makebox(0,0)[lb]{\smash{0}}}%
    \put(0.26651235,0.06075617){\color[rgb]{0,0,0}\makebox(0,0)[lb]{\smash{2}}}%
    \put(0.41049383,0.06075617){\color[rgb]{0,0,0}\makebox(0,0)[lb]{\smash{4}}}%
    \put(0.55451389,0.06075617){\color[rgb]{0,0,0}\makebox(0,0)[lb]{\smash{6}}}%
    \put(0.69849537,0.06075617){\color[rgb]{0,0,0}\makebox(0,0)[lb]{\smash{8}}}%
    \put(0.83499228,0.06075617){\color[rgb]{0,0,0}\makebox(0,0)[lb]{\smash{10}}}%
    \put(0.10501543,0.18649691){\color[rgb]{0,0,0}\makebox(0,0)[lb]{\smash{0}}}%
    \put(0.08279321,0.38206019){\color[rgb]{0,0,0}\makebox(0,0)[lb]{\smash{0.1}}}%
    \put(0.08279321,0.57758488){\color[rgb]{0,0,0}\makebox(0,0)[lb]{\smash{0.2}}}%
    \put(0.08279321,0.77314815){\color[rgb]{0,0,0}\makebox(0,0)[lb]{\smash{0.3}}}%
    \put(0.85999228,0.22561728){\color[rgb]{0,0,0}\makebox(0,0)[lb]{\smash{500}}}%
    \put(0.85999228,0.49938272){\color[rgb]{0,0,0}\makebox(0,0)[lb]{\smash{1500}}}%
    \put(0.85999228,0.77314815){\color[rgb]{0,0,0}\makebox(0,0)[lb]{\smash{2500}}}%
  \end{picture}%
\endgroup%

%% file: Sections/Speelman.tex
%
%
\section{Validation for burner-stabilized flame}
\label{sec:Speelman}
A further one-dimensional test case has been performed to compare the results of the present reduced order model with those obtained employing a detailed description of the phenomenon in a configuration where the flame interacts with an imposed electric field.
The test case has been carried out in the well-known configuration presented by \citet{Speelman2015b}
and consists of a premixed one-dimensional flame produced by a cylindrical heat-flux stabilized burner, whose deck area is \SI{7.069}{\square\centi\meter}.
A mixture of methane and air at $\phi = 1$ is injected with the laminar flame speed of the same mixture at \SI{298}{\kelvin} and \SI{1}{atm}, whereas the temperature of the burner is kept at \SI{350}{\kelvin}.
An external electric field is imposed using two electrodes, one positioned at the injection point of the mixture and the other \SI{1}{\centi\meter} downstream.\par
The computations employing the detailed chemi-ionization mechanism have been performed using the code FlameMaster~\citep{FLAMEMASTER} modified in order to solve the Poisson equation for the local electric potential and to account for the electric diffusion of the charged species.
The resulting computational tool is very similar to that proposed by \citet{Speelman2015b} except for the numerical discretization of the differential equations and for the molecular diffusion model employed.
In fact, FlameMaster~\citep{FLAMEMASTER} employs a central finite difference representation instead of the upwind finite-volume method used by \citet{Speelman2015b} and the diffusion model of \citet{Ern1994} is substituted with that described in the subsection \ref{subsec:chemMod}.
Since the tabulated approach, used in the present work to describe the chemistry, is not able to correctly predict heat transfer phenomena, it has been decided to perform the reduced order model simulations solving only the equations presented in the subsection \ref{subsec:crgMod}.
The data needed by the equations of the charged species have been extracted by a flamelet calculation performed in by the FlameMaster code~\citep{FLAMEMASTER} imposing the electrical neutrality of the system.
This procedure provides exactly the same data needed for the solution of the \gls{pcrg} and \gls{ncrg} equations as if they were interpolated from a table produced with the procedure described in section \ref{subsec:chemMod} but avoids the limitations imposed by the use of a tabulated chemistry approach.\par
In order to keep the formulation consistent with the rest of the paper and to reduce the computational cost of this validation, we employ the same values of the mobility described in section \ref{subsec:crgMod} for both computational approaches.
For this reason, the numerical results obtained will not be comparable with the experimental results of \citet{Speelman2015b}.\par
\begin{figure*}[tb]
	\centering
	\begin{subfigure}[b]{0.49\textwidth}
		\centering
        \begin{tikzpicture}
			\node[anchor=south west,inner sep=0] (image) at (0,0) {
    			\def\svgwidth{\columnwidth}
        		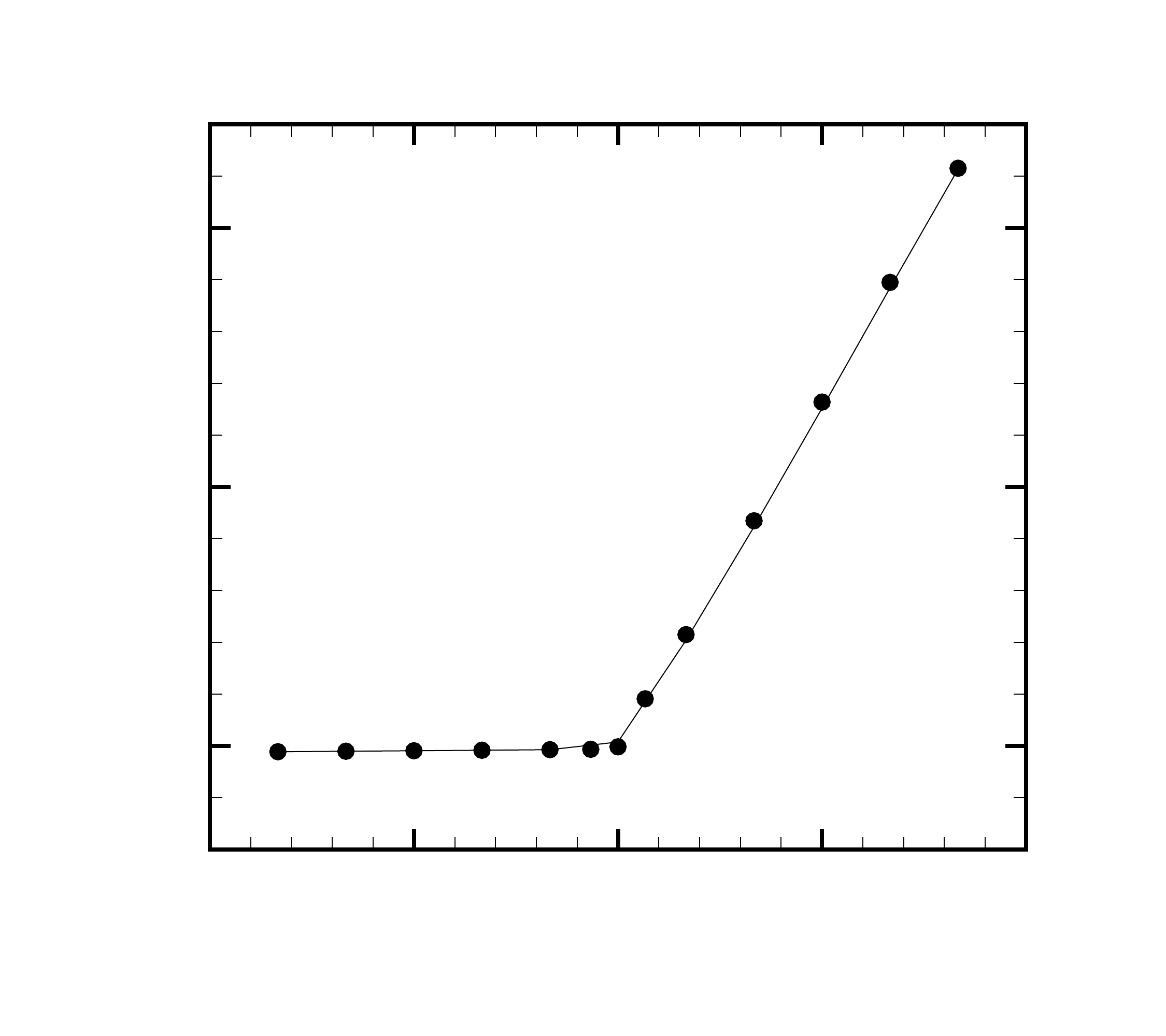
    		};
    	\end{tikzpicture}
		\caption{Mechanism ``A''}
		\label{fig:SpeelmanA}
	\end{subfigure}
	\begin{subfigure}[b]{0.49\textwidth}
		\centering
        \begin{tikzpicture}
			\node[anchor=south west,inner sep=0] (image) at (0,0) {
    			\def\svgwidth{\columnwidth}
        		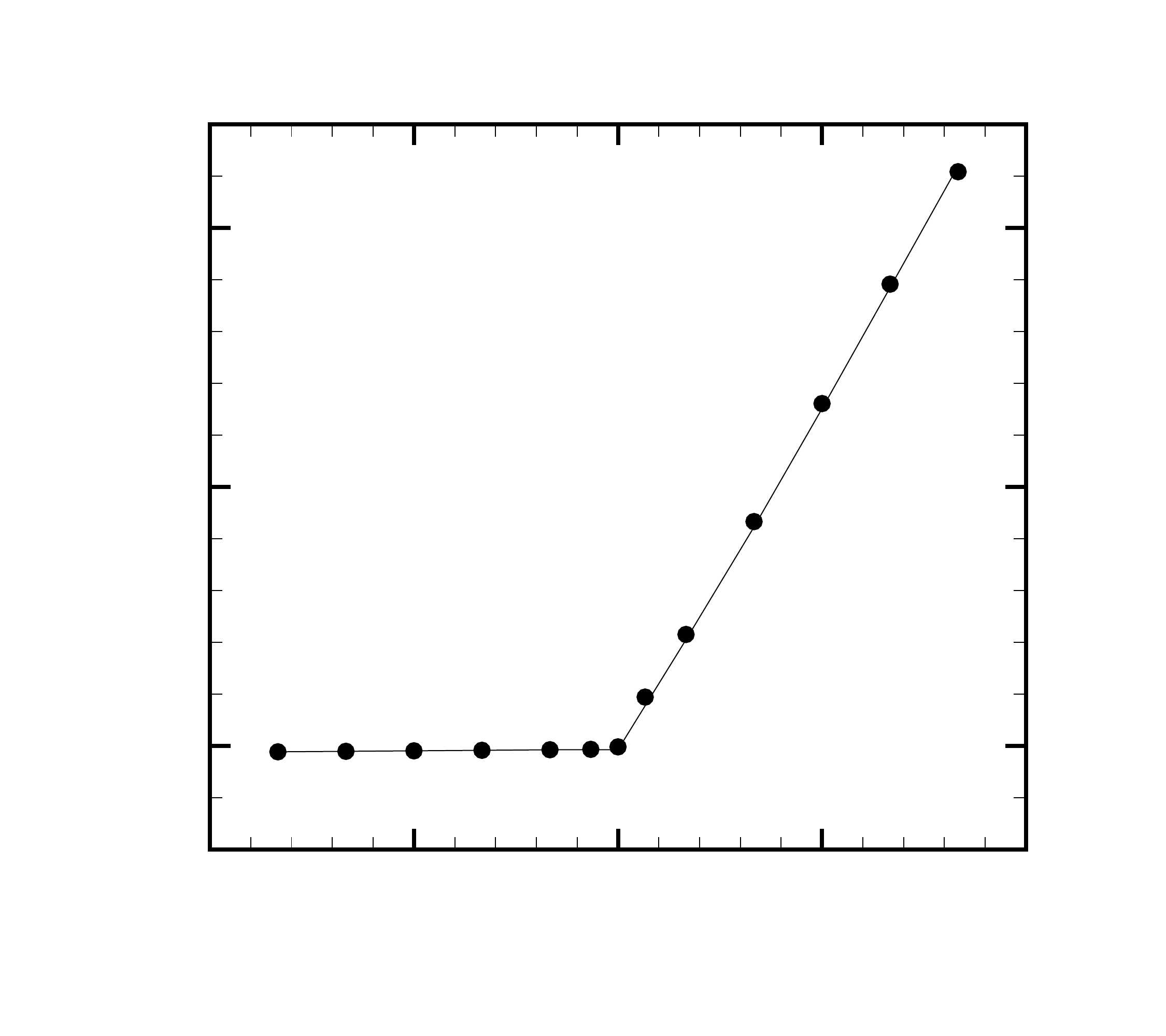
    		};
    	\end{tikzpicture}
		\caption{Mechanism ``B''}
		\label{fig:SpeelmanB}
	\end{subfigure}
	\caption[]{ Results obtained for the test case of \citet{Speelman2015b} with the detailed model (continuous line) and reduced order model (symbols).}
	\label{fig:Speelman}
\end{figure*}
Figure \ref{fig:Speelman} shows the electric current produced by the system versus the applied voltage obtained using the detailed and reduced order model in conjunction with the mechanism ``A'' and ``B''.
As expected the values of current measured with these simulations is much lower than that presented by \citet{Speelman2015b}.
Such a mismatch is only due to the reduced mobility of the ions.
Moreover, the reduced effect of the electric field on the charged species prevents the onset of the saturation on the flame.
Apart from this point, the match of the two sets of data is satisfactory for both the mechanisms and particularly relevant for the performance of the model: it ensures that the reduced model is able to reproduce the results of the detailed mechanism it has been built on.
%

%% file: Figures/SpeelmanA.eps_tex
\begingroup%
  \makeatletter%
  \providecommand\color[2][]{%
    \errmessage{(Inkscape) Color is used for the text in Inkscape, but the package 'color.sty' is not loaded}%
    \renewcommand\color[2][]{}%
  }%
  \providecommand\transparent[1]{%
    \errmessage{(Inkscape) Transparency is used (non-zero) for the text in Inkscape, but the package 'transparent.sty' is not loaded}%
    \renewcommand\transparent[1]{}%
  }%
  \providecommand\rotatebox[2]{#2}%
  \ifx\svgwidth\undefined%
    \setlength{\unitlength}{648bp}%
    \ifx\svgscale\undefined%
      \relax%
    \else%
      \setlength{\unitlength}{\unitlength * \real{\svgscale}}%
    \fi%
  \else%
    \setlength{\unitlength}{\svgwidth}%
  \fi%
  \global\let\svgwidth\undefined%
  \global\let\svgscale\undefined%
  \makeatother%
  \begin{picture}(1,0.88888889)%
    \put(0,0){\includegraphics[width=\unitlength]{SpeelmanA.pdf}}%
    \put(0.55199846,0.034375){\color[rgb]{0,0,0}\makebox(0,0)[cb]{\smash{Voltage (\si{\volt})}}}%
    \put(0.05,0.46381173){\color[rgb]{0,0,0}\rotatebox{90}{\makebox(0,0)[cb]{\smash{I (\si{\ampere}) $\cdot 10^{-4}$}}}}%
    \put(0.12600309,0.1054784){\color[rgb]{0,0,0}\makebox(0,0)[lb]{\smash{-300}}}%
    \put(0.30100309,0.1054784){\color[rgb]{0,0,0}\makebox(0,0)[lb]{\smash{-150}}}%
    \put(0.51481481,0.1054784){\color[rgb]{0,0,0}\makebox(0,0)[lb]{\smash{0}}}%
    \put(0.65941358,0.1054784){\color[rgb]{0,0,0}\makebox(0,0)[lb]{\smash{150}}}%
    \put(0.83441358,0.1054784){\color[rgb]{0,0,0}\makebox(0,0)[lb]{\smash{300}}}%
    \put(0.12962963,0.23051698){\color[rgb]{0,0,0}\makebox(0,0)[lb]{\smash{0}}}%
    \put(0.08460648,0.4527392){\color[rgb]{0,0,0}\makebox(0,0)[lb]{\smash{0.5}}}%
    \put(0.12962963,0.67496142){\color[rgb]{0,0,0}\makebox(0,0)[lb]{\smash{1}}}%
  \end{picture}%
\endgroup%

%% file: Figures/SpeelmanB.eps_tex
\begingroup%
  \makeatletter%
  \providecommand\color[2][]{%
    \errmessage{(Inkscape) Color is used for the text in Inkscape, but the package 'color.sty' is not loaded}%
    \renewcommand\color[2][]{}%
  }%
  \providecommand\transparent[1]{%
    \errmessage{(Inkscape) Transparency is used (non-zero) for the text in Inkscape, but the package 'transparent.sty' is not loaded}%
    \renewcommand\transparent[1]{}%
  }%
  \providecommand\rotatebox[2]{#2}%
  \ifx\svgwidth\undefined%
    \setlength{\unitlength}{648bp}%
    \ifx\svgscale\undefined%
      \relax%
    \else%
      \setlength{\unitlength}{\unitlength * \real{\svgscale}}%
    \fi%
  \else%
    \setlength{\unitlength}{\svgwidth}%
  \fi%
  \global\let\svgwidth\undefined%
  \global\let\svgscale\undefined%
  \makeatother%
  \begin{picture}(1,0.88888889)%
    \put(0,0){\includegraphics[width=\unitlength]{SpeelmanB.pdf}}%
    \put(0.55199846,0.034375){\color[rgb]{0,0,0}\makebox(0,0)[cb]{\smash{Voltage (\si{\volt})}}}%
    \put(0.05,0.46381173){\color[rgb]{0,0,0}\rotatebox{90}{\makebox(0,0)[cb]{\smash{I (\si{\ampere}) $\cdot 10^{-4}$}}}}%
    \put(0.12600309,0.1054784){\color[rgb]{0,0,0}\makebox(0,0)[lb]{\smash{-300}}}%
    \put(0.30100309,0.1054784){\color[rgb]{0,0,0}\makebox(0,0)[lb]{\smash{-150}}}%
    \put(0.51481481,0.1054784){\color[rgb]{0,0,0}\makebox(0,0)[lb]{\smash{0}}}%
    \put(0.65941358,0.1054784){\color[rgb]{0,0,0}\makebox(0,0)[lb]{\smash{150}}}%
    \put(0.83441358,0.1054784){\color[rgb]{0,0,0}\makebox(0,0)[lb]{\smash{300}}}%
    \put(0.12962963,0.23051698){\color[rgb]{0,0,0}\makebox(0,0)[lb]{\smash{0}}}%
    \put(0.08460648,0.4527392){\color[rgb]{0,0,0}\makebox(0,0)[lb]{\smash{0.5}}}%
    \put(0.12962963,0.67496142){\color[rgb]{0,0,0}\makebox(0,0)[lb]{\smash{1}}}%
  \end{picture}%
\endgroup%

%% file: Sections/Test_case.tex
%
%
\section{Two-dimensional test case}
\label{sec:test_case}
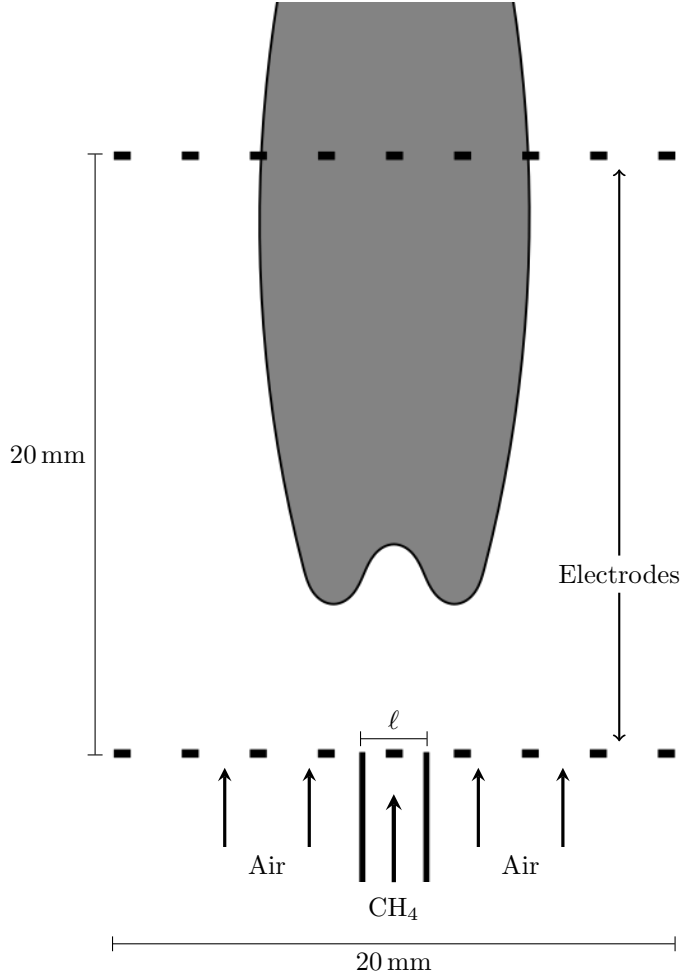
\begin{figure}[tb]
	\centering
	\begin{tikzpicture}
		\node[anchor=south west,inner sep=0] (image) at (0,0) {
			\includegraphics[trim={2 2 2 45},clip,width=0.45\textwidth]{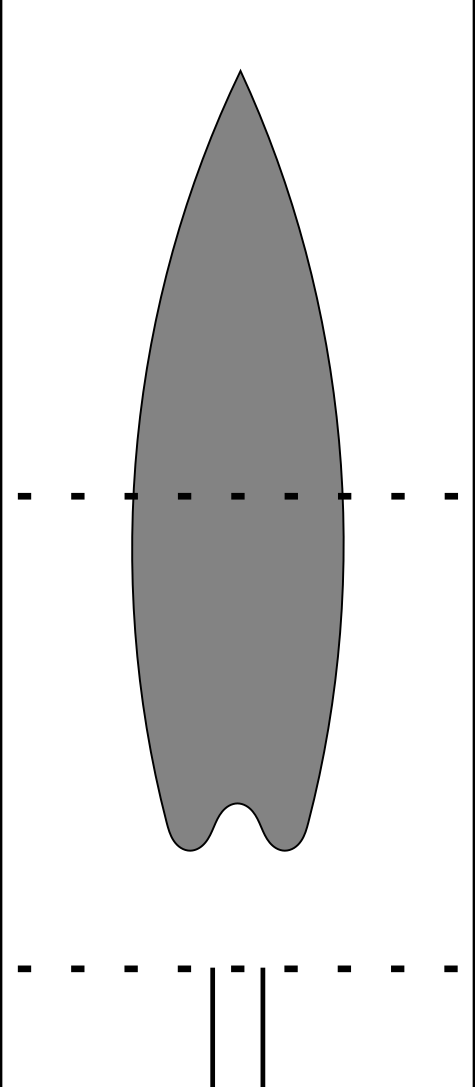}
		};
		\begin{scope}[x={(image.south east)},y={(image.north west)}]
			\draw [-stealth, black, very thick] (0.20,0.04) -- (0.20,0.13);
			\draw [-stealth, black, very thick] (0.35,0.04) -- (0.35,0.13);
			\draw [-stealth, black, very thick] (0.65,0.04) -- (0.65,0.13);
			\draw [-stealth, black, very thick] (0.80,0.04) -- (0.80,0.13);
			\node [align=center] at (0.275,0.02) {Air};
			\node [align=center] at (0.725,0.02) {Air};
			\draw [-stealth, black, ultra thick] (0.5 ,0.0) -- (0.5 ,0.10);
			\node [align=center] at (0.5,-0.03) {CH$_4$};
			\node [align=center] (electrodes) at (0.90,0.35) {Electrodes};
			\draw [->, black, thick] (electrodes) -- (0.90,0.81);
			\draw [->, black, thick] (electrodes) -- (0.90,0.16);
			\draw[arrows=|-|] (-0.03, 0.144) -- node[midway,left ]{\SI{20}{\milli\meter}} (-0.03, 0.828);
			\draw[arrows=|-|] ( 0.00,-0.070) -- node[midway,below]{\SI{20}{\milli\meter}} ( 1.00,-0.070);
			\draw[arrows=|-|] ( 0.44, 0.163) -- node[midway,above]{$\ell$}  			  ( 0.56, 0.163);
		\end{scope}
	\end{tikzpicture}
	\caption[]{Sketch of the numerical test case configuration.}
	\label{fig:Test_case}
\end{figure}
In order to test the proposed electro-chemical model in a more complex configuration, a two-dimensional configuration inspired to that of \citet{Belhi2013} has been considered.
It consists of a diffusive lifted methane-air flame burning in a two-dimensional slot burner whose geometry is shown in Figure~\ref{fig:Test_case}.
The fuel is injected in the middle of a \SI{20}{\milli\meter} wide combustion chamber with a velocity profile corresponding to a fully developed Poiseuille flow at an average velocity of \SI{4}{\meter\per\second}.
The fuel nozzle is composed of two parallel flat plates, whose thickness is \SI{0.1}{\milli\meter}, at distance of $\ell=\SI{2}{\milli\meter}$, representing the reference length.
At each side of the fuel nozzle, an air co-flow enters the combustion chamber with the Blasius velocity profile having a free-stream velocity equal to \SI{0.8}{\meter\per\second} and an upstream flat plate length of \SI{3.5}{\milli\meter}.
The fuel is pure methane and the oxidizer is standard dry air; both streams enter at \SI{300}{\kelvin}.
The combustion process, which develops at atmospheric pressure, has a stoichiometric mixture fraction $\gls{mf}_{st} =$ \num{5.4d-2} and a laminar stoichiometric flame velocity of about $\gls{um}^{fl}_{st} =$ \SI{0.38}{\meter\per\second}.\par
Two mesh electrodes are positioned at the inlet of the fuel and oxidizer flow and \SI{20}{\milli\meter} downstream of the inlet, respectively, forming a sort of capacitor configuration, whose mean electric field direction is aligned with the jet velocity.
The two electrodes are attached to an electrical generator in order to apply the electric field on the flame; the electric voltage is applied to the downstream electrode, whereas the inlet one is maintained at zero voltage.
The fluid flow is supposed not to be influenced by the presence of the electrodes.

%% file: Sections/Num_setup.tex
%
%
\section{Numerical setup}
\label{sec:num_setup}
The computational domain used for all the calculations coincides with the portion of the combustion chamber located between the electrodes (Figure~\ref{fig:Test_case}).
This choice derives from a trade-off between the accuracy of the prediction and the number of points needed for the simulation.
In fact, a longer domain would guarantee reduced influence of the outlet boundary condition on the flame, but also an increase in the number of grid points.
Moreover,the chosen configuration ensures a straightforward specification of the boundary condition at the downstream electrode.
The two-dimensional computational grid has been generated starting from a structured grid made of $128^2$ uniformly distributed points.
An homothetic refinement procedure has led to a mesh of about \num{2.4d5} quadrilateral cells clustered in the flame region.
The obtained grid spacing has a minimum value of about \SI{4d-2}{\milli\meter} next to the center line of the chamber and smoothly increases going toward the sides of the domain.\par
Dirichlet boundary conditions have been imposed at the inlet points for the fluid velocity, mixture fraction and progress-variable.
A no-slip condition has been applied along the fuel nozzle lip, whereas the sides of the domain have been considered as inviscid walls.
Standard convective boundary condition for the flow velocity has been imposed at the downstream outlet points.
The effect of buoyancy has been neglected.\par
The electrodes have been modeled by a Dirichlet boundary condition for the Gauss equation and alternatively using Neumann or Dirichlet boundary conditions for \gls{pcrg} and \gls{ncrg}.
In the particular electrical configuration considered here, for the cathode (negative pole, being this a power consuming device), the Dirichlet condition is applied to the negative charges ($\gls{ncrg} = 0$), which are repelled by the surface, and the Neumann condition is applied for the positive charges ($\nabla \gls{pcrg} \cdot \vect{n} = 0$), which are attracted by the surface.
Concerning the anode, the numerical boundary conditions are the opposite.
However, since the flow outlet section intercepts the flame, the charged particles are always present at this section and the Dirichlet condition would cause numerical instabilities.
For this reason, the zero normal gradient boundary condition for the charges has been imposed at the outlet points both for an anode or a cathode.\par
The flamelet profiles have been computed by the code FlameMaster~V3.3.10~\citep{FLAMEMASTER} using \num{1000} unevenly spaced grid points;
the output data are organized in a chemical look-up table.
The progress-variable is defined as $C = Y_{CO} + Y_{CO_2}$, as proposed by \citet{Domingo2008}.
Probably, a better description of the tabulated quantities can be achieved using an ad-hoc optimized definition of the progress-variable, obtained using the procedures described by \citet{Niu2013}, \citet{Ihme2012} or by \citet{Prufert2015}.
This optimization procedure has been omitted in the present work in order to keep a chemistry description consistent with \citet{Belhi2010,Belhi2013}.
The table, obtained in this way, is discretized using \num{301} unevenly distributed points in both \gls{mf} and \gls{prv} directions, in order to accurately interpolate the thermo-chemical variable fields needed by the flow solver.

%% file: Sections/Results.tex
%
%
\section{Results}
\label{sec:results}
This section provides the analysis of the flame structure corresponding to different applied voltages, including the configuration without any imposed electric field.
An important feature of the flame structure is the flame-tip position, which is here defined as the first point where \gls{prv} reaches the value of \num{5d-3} on the stoichiometric mixture fraction iso-line.\par
In the next sub-section, the flame configuration computed without charges transport model is discussed, being taken as baseline configuration.
Afterwards, the effect of an applied continuous voltage on the flame-tip position will be presented analyzing the new equilibrium configurations.
In order to further study the local behavior of the flame immersed in an electric field, the analysis of the electric potential and mixture charge will be performed in the steady configuration of each value of the applied voltage.
%
\subsection{Simulations without charge transport model}
\begin{figure}[tb]
	\centering
    \begin{tikzpicture}
        \node[anchor=south west,inner sep=0] (image) at (0,0) {
    		\def\svgwidth{0.7\columnwidth}
    		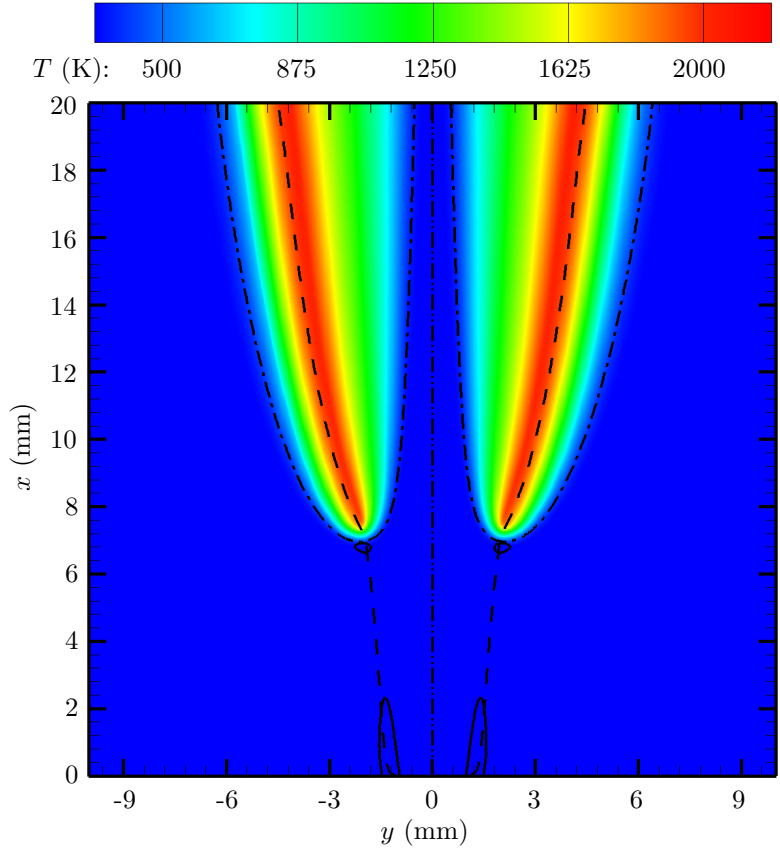
    	};
    \end{tikzpicture}
	\caption[]{Comparison of temperature fields (expressed in \si{\kelvin}) in the cases without charge transport for the mechanism ``A'' (left) and mechanism ``B'' (right). The dot-dashed line is the iso-line at $\gls{prv} =$ \num{5d-3}; the dashed line is the stoichiometric iso-line location; the continuous line is the contour at $\gls{um}_x = \gls{um}^{fl}_{st} =$ \SI{0.38}{\meter\per\second}; the dot-dot-dashed line is the symmetry axis.}
	\label{fig:T0_noemhd}
\end{figure}
For both chemical mechanisms considered in this work, these simulations have been initialized using a zero velocity flow-field.
A first computation has been performed to determine the steady solution for the non-reactive flow.
Then, the flame has been ignited imposing the maximum physical value of the progress-variable (based on the local \gls{mf} value) and a second simulation has been performed in order to reach the steady configuration.\par
For both mechanisms, the flame is symmetric from the ignition time to the stabilization, which is reached in about \SI{1.5}{\second}.
Figure~\ref{fig:T0_noemhd} shows the temperature contours computed using mechanisms ``A'' (on the left-hand side) and ``B'' (on the right-hand side).
The two mechanisms predict the same flame configuration, proving the marginal influence of the chemi-ionization mechanism on the combustion process.
The baseline lift-off height measured in both cases is $x_0 = $ \SI{6.97}{\milli\meter} (about \num{3.5} $\ell$), whereas the spanwise position of the flame tip is $y_0 = $ \SI{1.90}{\milli\meter} (about \num{0.95} $\ell$).
Both these data are in good agreement with the results of \citet{Belhi2013}.
The difference between the present test and that in literature is the shape of the injection velocity profile of the co-flow.
The position of the flame seems remarkably influenced by this boundary condition, in fact, the flame is easily blown-off or shifted toward the fuel nozzle slightly changing the upstream length of the plate for the Blasius profile.\par
It is noteworthy that both solutions show the well known reduction of the mixture velocity in the upstream region of the flame as described by many authors in the literature \citep{Ruetsch1995,Domingo1996,Plessing1998}.
This phenomenon, together with the shape of the velocity profiles imposed at the inlet sections, determines the point of stabilization of the flame in a flow with an average velocity higher than the laminar planar flame speed.
\subsection{Flame-tip steady position for voltage imposition}
Once the steady configuration of the flame has been obtained without considering charge transport phenomena, a series of simulations has been run for both mechanisms varying the voltage applied to the flame.
Two sets of twelve simulations have been performed changing the applied voltage in the range between \SI{-750}{\volt} and \SI{1250}{\volt}.
This range has been chosen considering the attachment voltages predicted by \citet{Belhi2013} and the limitations posed by the employed central finite difference scheme.
A step of \SI{125}{\volt} has been used as sample interval for the negative polarity cases, whereas the positive polarity has been sampled by a \SI{250}{\volt} interval.
The smaller sample interval employed for the negative polarity has been chosen in order to capture the sharper attachment of the flame expected in this configuration.\par
\begin{figure*}[tb]
	\centering
	\begin{subfigure}[b]{0.49\textwidth}
		\centering
        \begin{tikzpicture}
        	\node[anchor=south west,inner sep=0] (image) at (0,0) {
    			\def\svgwidth{0.95\columnwidth}
    			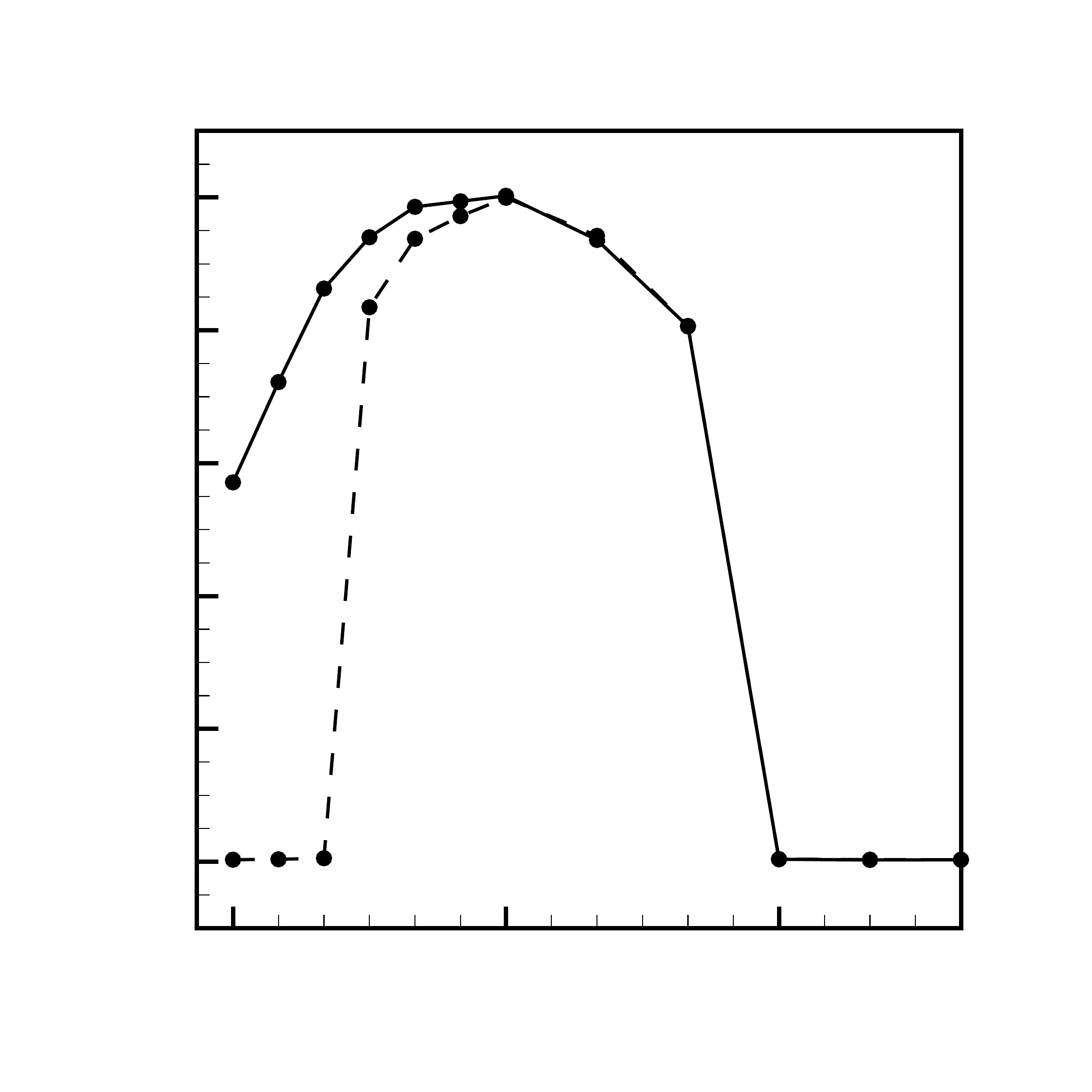
            };
		\end{tikzpicture}
		\caption{Streamwise position}
		\label{fig:XovVolt}
	\end{subfigure}
	\begin{subfigure}[b]{0.49\textwidth}
		\centering
        \begin{tikzpicture}
        	\node[anchor=south west,inner sep=0] (image) at (0,0) {
				\def\svgwidth{0.95\columnwidth}
				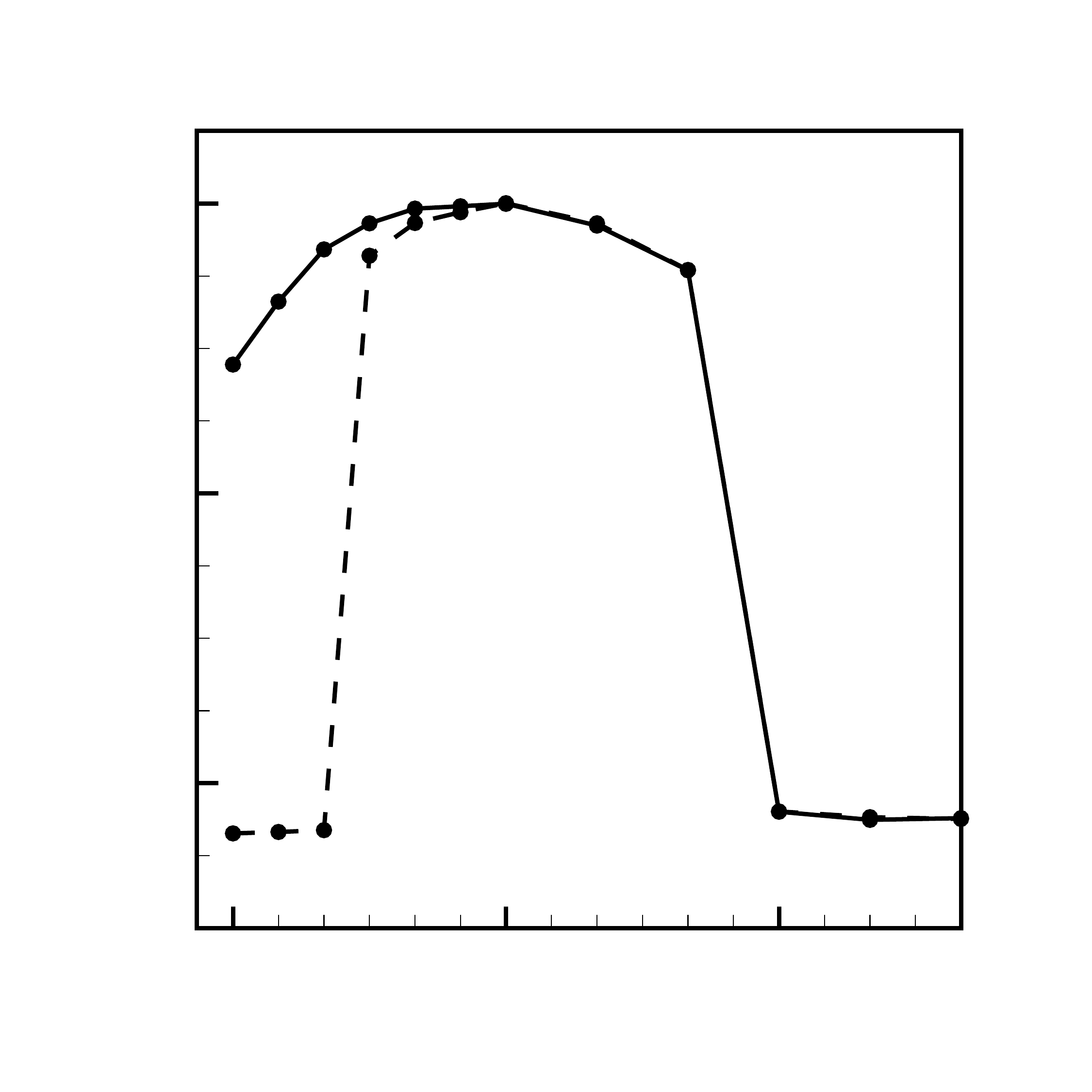
            };
        \end{tikzpicture}
		\caption{Spanwise position}
		\label{fig:YovVolt}
	\end{subfigure}
	\caption[]{Flame-tip steady position in the streamwise and spanwise directions expressed as fraction of the reference position for
	mechanism ``A'' (\tikz[baseline=-0.5ex] {
		\draw[decoration={markings, mark=at position 0.5 with {\node[circle,draw=black,fill=black,inner sep=0.04cm] {};}},postaction={decorate}]
			(0,0) -- ++(0.6cm,0);
	}) and 
	mechanism ``B'' (\tikz[baseline=-0.5ex,dashed] {
		\draw[decoration={markings, mark=at position 0.5 with {\node[circle,draw=black,fill=black,inner sep=0.04cm] {};}},postaction={decorate}] (0,0) -- ++(0.6cm,0);}).}
	\label{fig:XYovVolt}
\end{figure*}
The streamwise ($x/x_0$) and spanwise ($y/y_0$) non-dimensional coordinates of the flame-tip at steady-state are plotted in Figure~\ref{fig:XYovVolt} versus the applied voltage, $x_0$, $y_0$ being the streamwise and spanwise coordinates of the flame-tip for $\Delta \gls{v} = $ \SI{0}{\volt}.
For consistency, it has been verified that the flame-tip position does not change when the charge transport model is employed and no voltage is imposed.
On the other hand, as shown in the figure, both the positive and negative polarities lead to a reduction of the flame-tip lift-off height.
In the unattached cases, the new equilibrium point is reached when the force due to the electric field is balanced by the higher momentum of the flow next to the injection region.
Because of this mechanism, the flame will eventually attach to the upstream electrode when the flow momentum is not strong enough to counteract the electric force.
In this context, it is evident the importance of the injection velocity profiles of both the co-flow and fuel jet, as well as of the nozzle lip thickness.
It is noteworthy that the present model is not adequate to reproduce quasi-attached configurations because of the strong electric field generated between the flame and the electrode.
The electric field intensities predicted in these configurations would definitely activate non-thermal processes and probably lead to the electrical breakdown of the mixture.
Moreover, the high strain of the flow and the possible heat flux through the fuel nozzle would make the present chemical model inadequate to describe the combustion process in these configurations.\par
Figure~\ref{fig:XovVolt} shows that both the chemical set of reactions predict the attachment of the flame to the upstream electrode at $\Delta \gls{v} = $ \SI{750}{\volt} for the positive polarity.
In particular, the right-hand sides of the two graphs are almost identical because the two considered chemical mechanisms produce a similar quantity of positive ions with the same transport properties, exchanging the same momentum with the incoming flow.
The branches of the graph related to the negative polarities (left-hand side of the graphs) show a completely different behavior for each mechanism.
In fact, the flames computed with the mechanism ``A'' are less responsive to the applied voltage, if compared with the positive polarity cases, and do not achieve the attachment to the upstream electrode in the considered voltage range.
On the other hand, mechanism ``B'' provides flames which are continuously closer to the upstream electrode and predicts an attachment of the flame-tip between $\Delta \gls{v} = $ \SI{-375}{\volt} and $\Delta \gls{v} = $ \SI{-500}{\volt}.
Such a difference, commonly called ``diodic effect'', has been already observed in the literature~\citep{Won2008}.
%
The difference in the shape of the plotted graphs is due to the different nature of the ions produced by the two mechanisms.
The presence of the anions in the upstream part of the flame, predicted by mechanism ``B'' as shown in Figure~\ref{fig:Spec_comp}, is responsible for the stronger attraction of the flame by the upstream electrode in the cases with negative polarity.
Firstly, the heavy anions are more effective than the electrons in reducing the flow momentum, thanks to their lower mobility \citep{Han2017}.
Then, the formation of anions inhibits the recombination of the electrons with the cations through the mechanism described in Section~\ref{sec:chem_analysis}, generating a larger region of negatively charged mixture, as described by \citet{Belhi2013}.
On the other hand, mechanism ``A'' produces only electrons, which are lighter than the cations and, therefore, less effective in reducing the momentum of the incoming flow.\par
Figure~\ref{fig:YovVolt} shows the spanwise non-dimensional coordinate of the flame-tip for all the tested configurations.
The behavior of both mechanisms is very similar to that of the streamwise position plots, except that it appears to be less affected by the applied voltage.
The cause of this difference is that the applied electric field is mainly aligned with the flow direction with only minor deviations close to the flame-tip due to the flame curvature.
The displacement toward the center of the flame-tip is, therefore, due to the shape of the stoichiometric mixture fraction iso-line shown in Figure~\ref{fig:T0_noemhd}.\par
The results shown here are in good agreement with the findings of \citet{Belhi2013}, considering the high sensibility of this kind of system to the modeling assumptions and numerical setup.
Regarding the positive polarity, the available calculations \citep{Belhi2013} show an attachment of the flame at \SI{1250}{\volt}.
This value is rather higher than those found in the present work, but this difference is probably due to a different injection profile of the co-flow.
Furthermore, analyzing the negative polarity branches of the graph, it is possible to find a substantial agreement of the flame position computed using both the considered chemical mechanisms.
In fact, the sharp attachment achieved with the mechanism ``B'' and the high distance of the flame from the nozzle predicted with the mechanism ``A'' reduce the influence of the inlet injection profile uncertainty on the flame position.
Although the very good agreement obtained in the validation in Section \ref{sec:Speelman} suggests that the assumptions made for evaluating the transport properties and reaction terms of the drifting scalars are appropriate, another source of discrepancy between the present results and those of \citep{Belhi2013} can still be in the use of Eq.~\eqref{eqn:Pmob_def}, Eq.~\eqref{eqn:Mmob_def} and  Eq.~\eqref{eqn:omegaPM_def} when the model is employed for more complex configurations.
Unfortunately, the impossibility of isolating these two sources of error and the lack of experimental data on this case make further analyses unfeasible at the present time.
\subsection{Electric potential distribution}
\label{subsec:V}
\begin{figure}[tb]
	\centering
    \begin{tikzpicture}
      	\node[anchor=south west,inner sep=0] (image) at (0,0) {
    		\def\svgwidth{0.7\columnwidth}
    		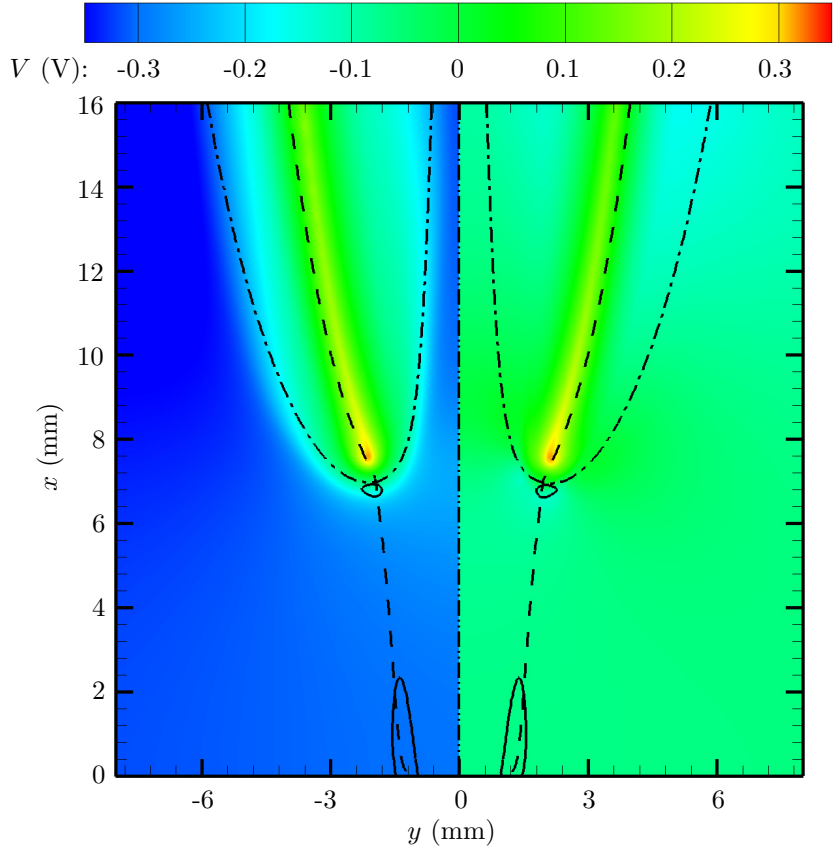
    	};
    \end{tikzpicture}
	\caption[]{Comparison of electric potential field (expressed in \si{\volt}) in the cases at $\Delta \gls{v} = $ \SI{0}{\volt} for the mechanism ``A'' (left) and mechanism ``B'' (right).
		The iso-lines are defined as in Figure~\ref{fig:T0_noemhd}.}
	\label{fig:V_DV0}
\end{figure}
As expected and described in the previous section, the introduction of the electrical model does not have any influence on the flame configuration when no electric voltage is imposed, but this case is interesting because of the ability of the present model to reproduce the electric field generated by the flame.
In fact, the large difference in diffusivity between the heavy charged particles and the electrons, which are produced in the reaction region of the flame, leads to a charge unbalance reducing the number of anions in this zone.
This mixture polarization generates an electric field pointing away from the flame and inducing a drift velocity of the electrons which counteracts their diffusive flux.
The electric potential field is shown in Figure~\ref{fig:V_DV0} for the region surrounding the flame.
This generated electric field is too weak to influence the fluid dynamics producing a negligible force, and, for this reason, it is usually neglected in combustion simulations.
On the other hand, this phenomenon is widely used in various type of combustion chambers in order to monitor the behavior of the flame with ion-sensors.\par
Mechanism ``B'' predicts a slightly lower electric potential difference, whose magnitude is \SI{0.6}{\volt}, and it is concentrated in the reacting region.
On the other hand, the high potential region computed with mechanism ``A'' is larger and more intense (the total difference of potential is about \SI{1.2}{\volt}).
Both predicted values are in good agreement with the measurements available in literature obtained by ion-sensors~\citep{Nair2005,Aithal2013,Mehresh2005}.\par
\begin{figure*}[tb]
	\centering
	\begin{subfigure}[b]{0.49\textwidth}
		\centering
		\begin{tikzpicture}
       		\node[anchor=south west,inner sep=0] (image) at (0,0) {
    			\def\svgwidth{\columnwidth}
    			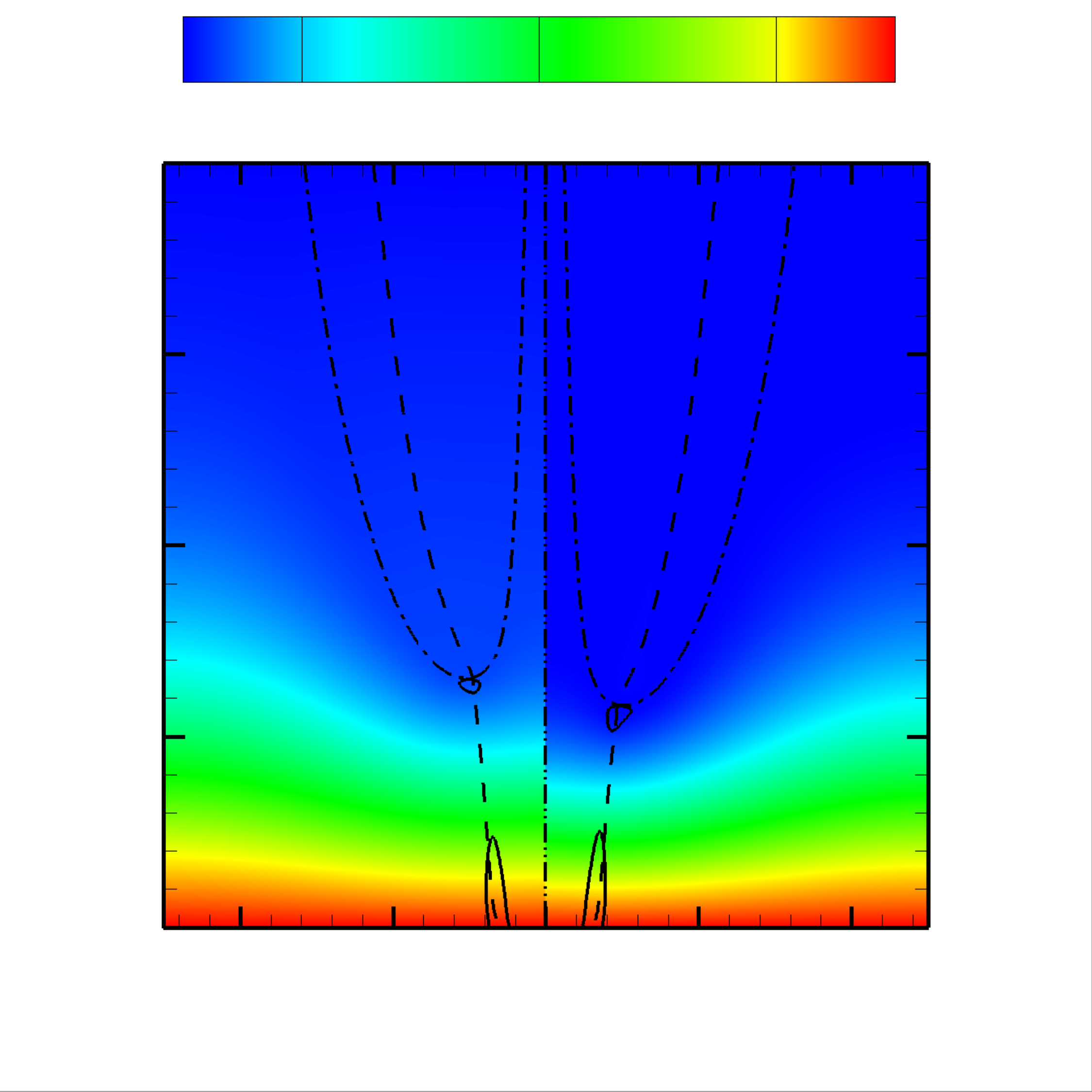
    		};
		\end{tikzpicture}
		\caption{$\Delta \gls{v}$ = \SI{-375}{\volt}}
		\label{fig:V_DV-375}
	\end{subfigure}
	\begin{subfigure}[b]{0.49\textwidth}
		\centering
        \begin{tikzpicture}
      		\node[anchor=south west,inner sep=0] (image) at (0,0) {
    			\def\svgwidth{\columnwidth}
    			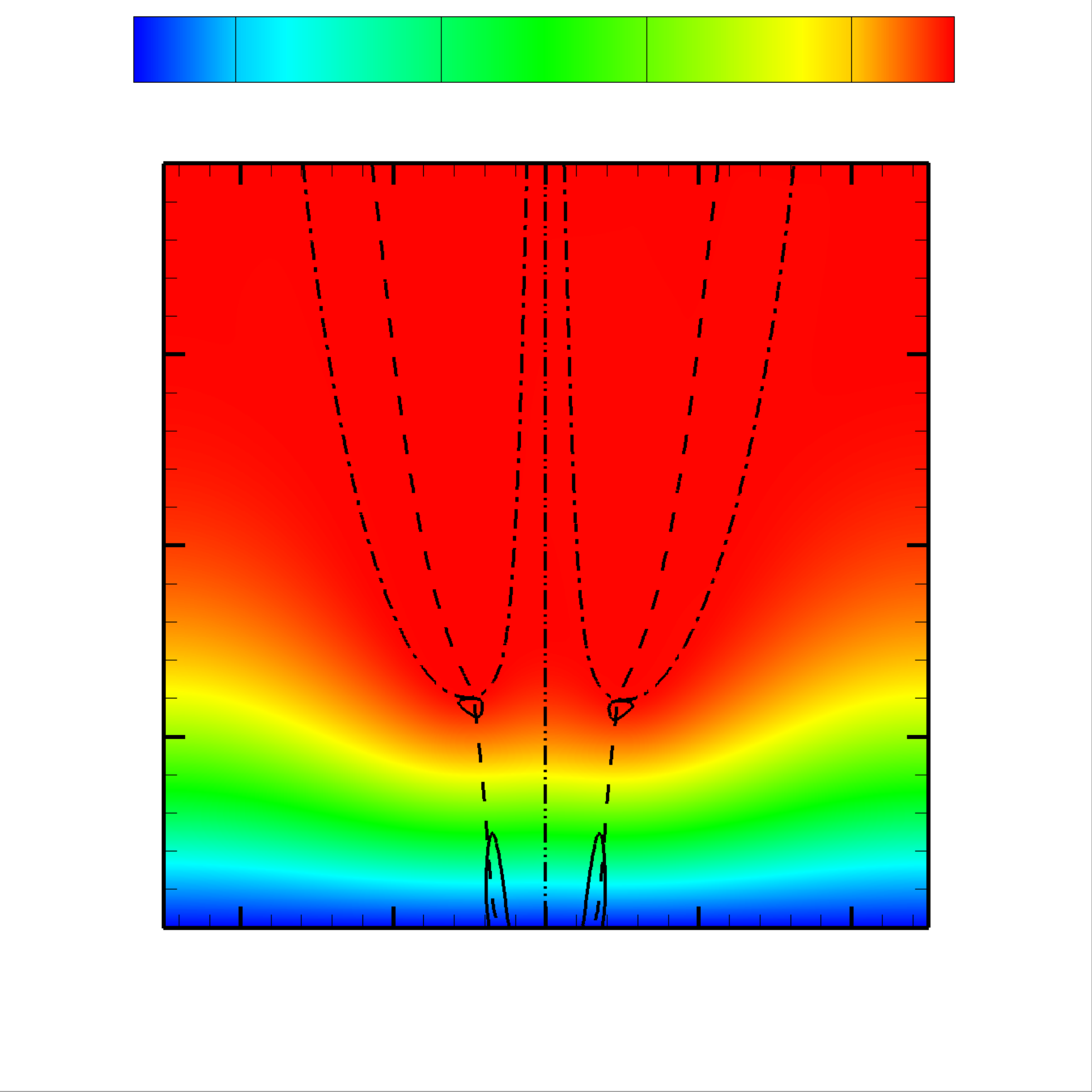
    		};
    	\end{tikzpicture}
		\caption{$\Delta \gls{v}$ = \SI{500}{\volt}}
		\label{fig:V_DV500}
	\end{subfigure}
	\caption[]{Electric potential contour plots (expressed in \si{\volt}) in the flame-front region at two different imposed voltage difference. Results for mechanism ``A'' are on the left and for mechanism ``B'' on the right of each picture. The iso-lines are defined as in Figure~\ref{fig:T0_noemhd}.}
	\label{fig:V_DV}
\end{figure*}
The contour plots provided in Figure~\ref{fig:V_DV} show the distribution of electric potential in the entire computational domain for two values of applied voltage with opposite polarity, $\SI{500}{\volt}$ and $\SI{-375}{\volt}$, for both chemical mechanisms.
In both graphs, the flame creates a region with constant electric potential between the flame-front and the downstream electrode, highlighting the importance of modeling the interaction of the flame with the electric field.
The extension of this region is determined by the ability of the flame to produce a large amount of charges in its reacting layer.
In fact, in a steady condition, the charges are usually depleted at a rate close to the production rate and the difference between the two rates is equal to the advective and diffusive fluxes.
When the electric field induces a displacement on the charges distribution, they move in opposite direction locally reducing their consumption rate.
The equilibrium point is reached either when the charge separation is sufficient to absorb the imposed voltage or when the maximum amount of charges is produced by the flame.
In both the presented cases, the steady configuration is obtained through the first configuration, being the flame sub-saturated.
%
The electric potential distribution, shown in the contour plot, entails that the electric field intensity is higher in the region between the flame-tip and the upstream electrode and highlights the need of considering the mixture charge in the Gauss equation (Eq. \eqref{eqn:Gauss_law}).
Moreover, this modification of the electric potential field by the flame strongly affects the convective movement of the charges.
In fact, their distribution is marginally affected in the diffusive region of the flame and, instead, strongly modified in the premixed part of the triple flame.\par
\begin{figure}[tb]
	\centering
    \begin{tikzpicture}
      	\node[anchor=south west,inner sep=0] (image) at (0,0) {
    		\def\svgwidth{0.7\columnwidth}
    		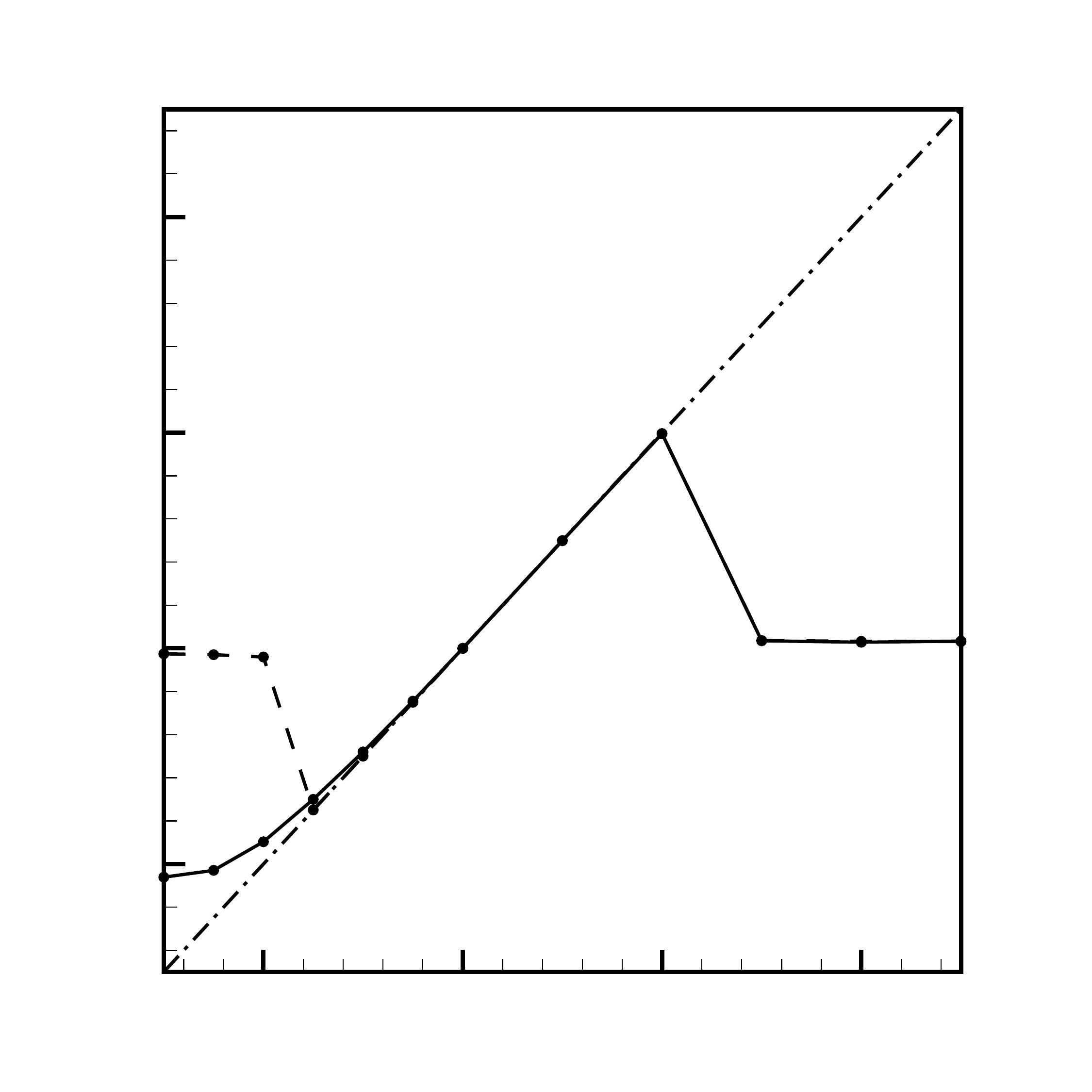
    	};
    \end{tikzpicture}
	\caption[]{Flame-tip electric potential for 
		mechanism ``A'' (\tikz[baseline=-0.5ex]					{ 
			\draw[decoration={markings, mark=at position 0.5 with {\node[circle,draw=black,fill=black,inner sep=0.04cm] {};}},postaction={decorate}] (0,0) -- ++(0.6cm,0); 
		}) and 
		mechanism ``B'' (\tikz[baseline=-0.5ex, dashed]			{ 
			\draw[decoration={markings, mark=at position 0.5 with {\node[circle,draw=black,fill=black,inner sep=0.04cm] {};}},postaction={decorate}] (0,0) -- ++(0.6cm,0);
		}). 
		The bisector is plotted as (\tikz[baseline=-0.5ex, dashdotted] { \draw (0,0) -- ++(0.6cm,0); }).}
	\label{fig:VovVolt}
\end{figure}
Figure~\ref{fig:VovVolt} shows a plot of the value of the electric potential measured at the flame-tip position versus the applied voltage for mechanisms ``A'' and ``B''.
The plotted line is almost coincident with the bisector for all the lifted flame configurations of both polarities.
The line sharply deviates from the bisector when the flame attaches to the upstream electrode and the ions motion becomes more complex and not suitable to be predicted by the present model.
The largest difference between the flame-tip voltage and the imposed one for an unattached flame is in the region between \SI{-500}{\volt} and \SI{-750}{\volt} for mechanism ``A''.
The smooth but consistent deviation from the bisector suggests an incipient electrical saturation of the flame.\par
Figure~\ref{fig:VovVolt} also explains why the streamwise and spanwise position of the flame-tip exhibit a parabolic scaling for both simulation set (see Figures~\ref{fig:XovVolt} and \ref{fig:YovVolt}).
This result is in good agreement with the results present in literature.
This non-linearity is due to the increasing polarization of the flame, which experiences a larger charge separation in order to absorb the rising difference of potential, as shown in Figure~\ref{fig:VovVolt}.
In fact, the larger charges separation over the flame-front region, in conjunction with the increased difference of potential, leads to the mentioned quadratic scaling of the equilibrium point position between the flow momentum and the electric force.
\subsection{Charge repartition and local electric field}
\label{subsec:Crg}
\begin{figure*}[tb]
	\centering
	\begin{subfigure}[b]{0.49\textwidth}
		\centering
		\begin{tikzpicture}
       		\node[anchor=south west,inner sep=0] (image) at (0,0) {
    			\def\svgwidth{\columnwidth}
    			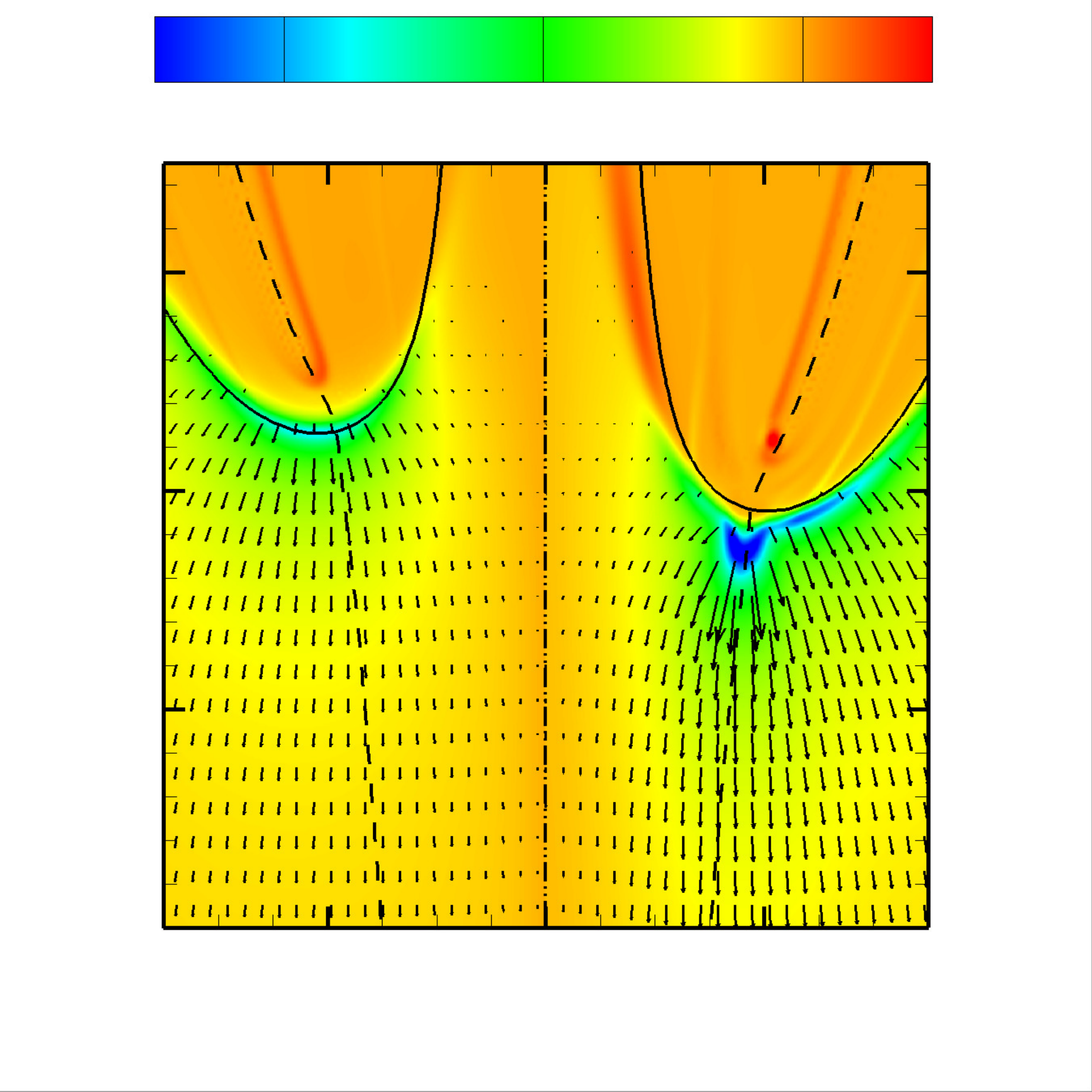
    		};
		\end{tikzpicture}
		\caption{$\Delta \gls{v}$ = \SI{-375}{\volt}}
		\label{fig:C+F_DV-375}
	\end{subfigure}
	\begin{subfigure}[b]{0.49\textwidth}
		\centering
		\begin{tikzpicture}
       		\node[anchor=south west,inner sep=0] (image) at (0,0) {
    			\def\svgwidth{\columnwidth}
    			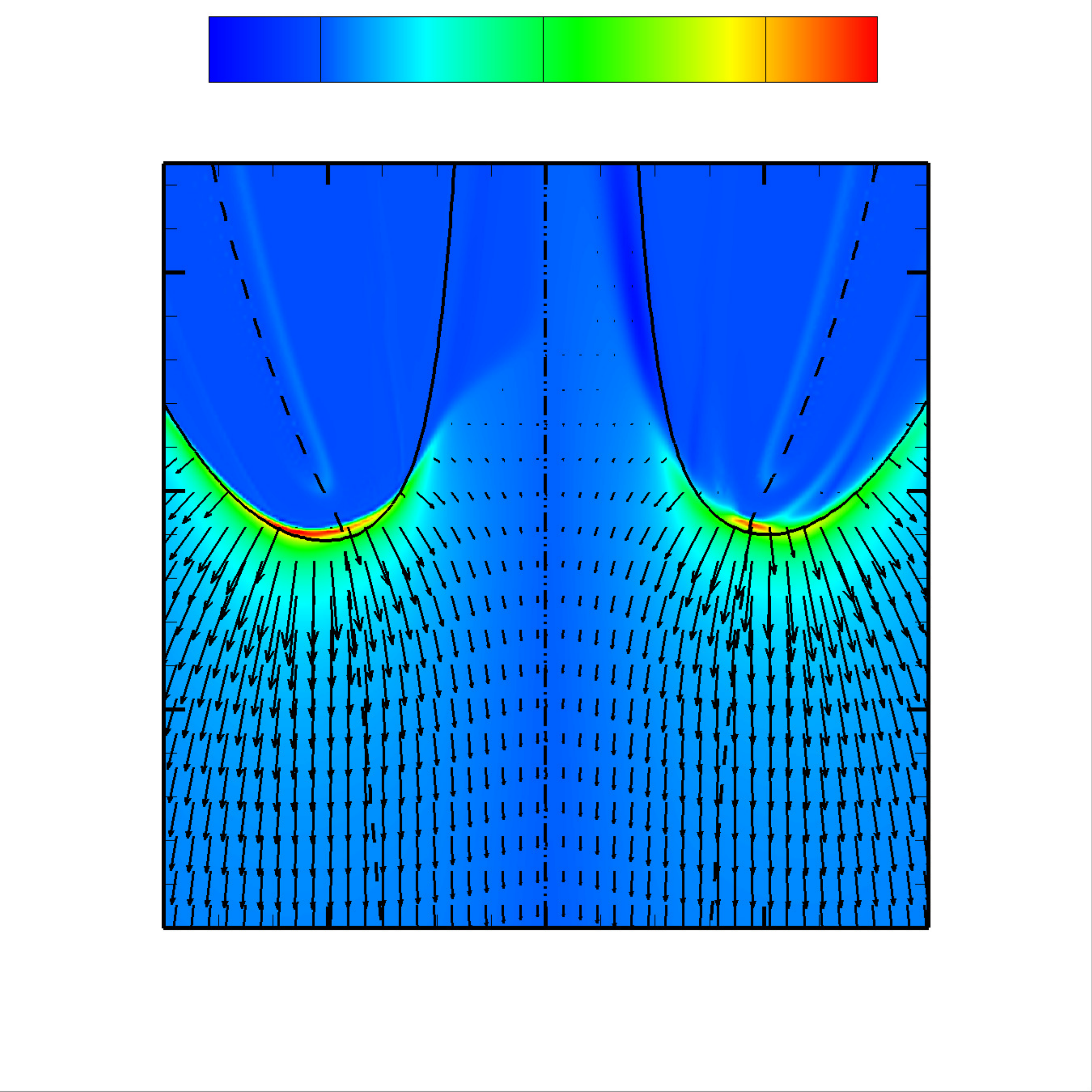
    		};
		\end{tikzpicture}
		\caption{$\Delta \gls{v}$ = \SI{500}{\volt}}
		\label{fig:C+F_DV500}
	\end{subfigure}
	\caption[]{Electric charge density contour plots (expressed in \si{\milli\coulomb\per\cubic\meter}) for two imposed voltage difference. The vectors represent the electric force exchanged with the fluid mixture. Results for mechanism ``A'' are on the left panel and for mechanism ``B'' on the right panel. The iso-lines are defined as in Figure~\ref{fig:T0_noemhd}.}
	\label{fig:C+F_DV}
\end{figure*}
Figure~\ref{fig:C+F_DV} shows the charge distribution in the surroundings of the flame tip at both \SI{-375}{\volt} and \SI{500}{\volt}.
The four combinations of mechanisms and polarities predict two charge concentration regions, one next to the \gls{prv} iso-line and the other located further downstream in the reaction layer.
The first charge peak is due to modification of the ionized species distribution caused by the presence of the electric field.
In fact, the sign of this peak is a direct consequence of the applied field polarity, being positive in the positive polarity case and negative in the other.
The second peak of charge, which is positive in all the four cases examined here, coincides with the region of the flame where the charges are produced and it is due to the high diffusivity of the electrons.
In fact, such a charge peak is responsible for the electric field generated by the flame and described in the previous sub-sections (Figure~\ref{fig:V_DV0}).
Its effect on the electric potential field is not visible in the contour plots relative to the present cases (Figure~\ref{fig:V_DV}), being the magnitude of the generated potential much weaker than the applied voltage.\par
In the positive polarity case, both the fields of the Figure~\ref{fig:V_DV500} show similar charged areas along the \gls{prv} iso-line, but the calculations performed with the mechanisms ``A'' are characterized by a higher and more elongated peak.
On the other hand, the solution obtained with the mechanisms ``B'' has two additional small charge concentration regions, located between the \gls{prv} iso-line and the reacting region.
These differences between the two mechanisms are ascribable to the higher charges mobility of the mechanism ``A'', which contains only electrons as negatively charged species.
The higher negative mobility induces a higher mixture polarization localized in the upstream part of the flame-front.
Regarding mechanism ``B'', the presence of heavy ions in the mixture, which modify the local transport properties of the scalar \gls{ncrg}, entails more complicated equilibrium points for the distribution of the electrical quantities.\par
In the negative polarity case (Figure~\ref{fig:V_DV-375}), mechanism ``B'' predicts a higher peak due to the electric field induced charges transport.
The diametrical behavior of the two mechanisms with respect to the positive polarity is due to the already described effect of the heavy anions (present in the upstream part of the flame) that inhibit the recombination of the electrons and therefore cause an anion concentration.
The shape of this concentration and its location are in good agreement with the contour plots shown by \citet{Belhi2013} in his Figure 11.
This phenomenon is responsible for the stronger influence of the imposed electric field on the flame position for this type of polarity.
\begin{figure*}[htb]
	\centering
	\begin{subfigure}[b]{0.49\textwidth}
		\centering
		\begin{tikzpicture}
       		\node[anchor=south west,inner sep=0] (image) at (0,0) {
    			\def\svgwidth{\columnwidth}
    			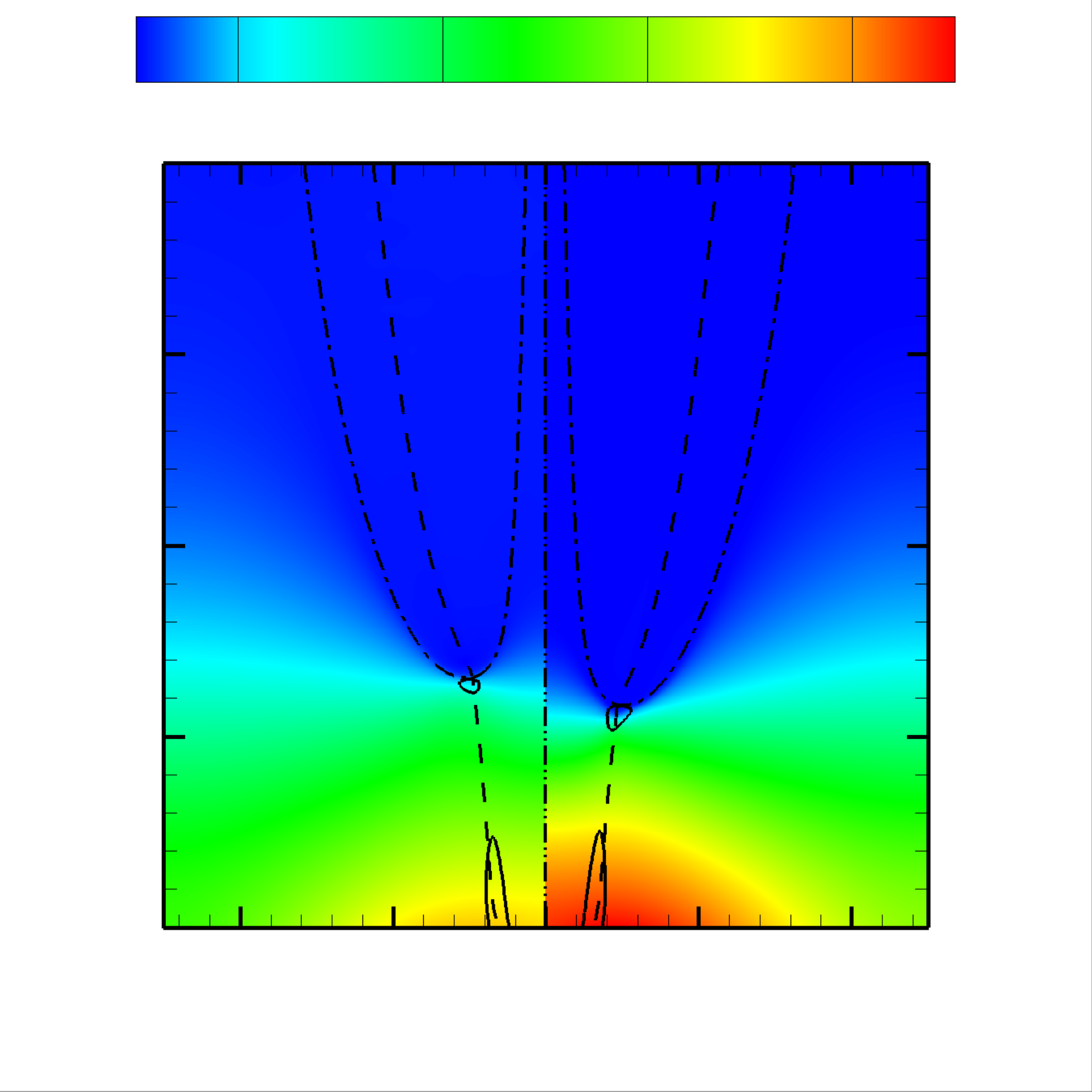
    		};
		\end{tikzpicture}
		\caption{$\Delta \gls{v}$ = \SI{-375}{\volt} $x$ direction}
		\label{fig:EF-X_DV-375}
	\end{subfigure}
	\begin{subfigure}[b]{0.49\textwidth}
		\centering
		\begin{tikzpicture}
       		\node[anchor=south west,inner sep=0] (image) at (0,0) {
    			\def\svgwidth{\columnwidth}
    			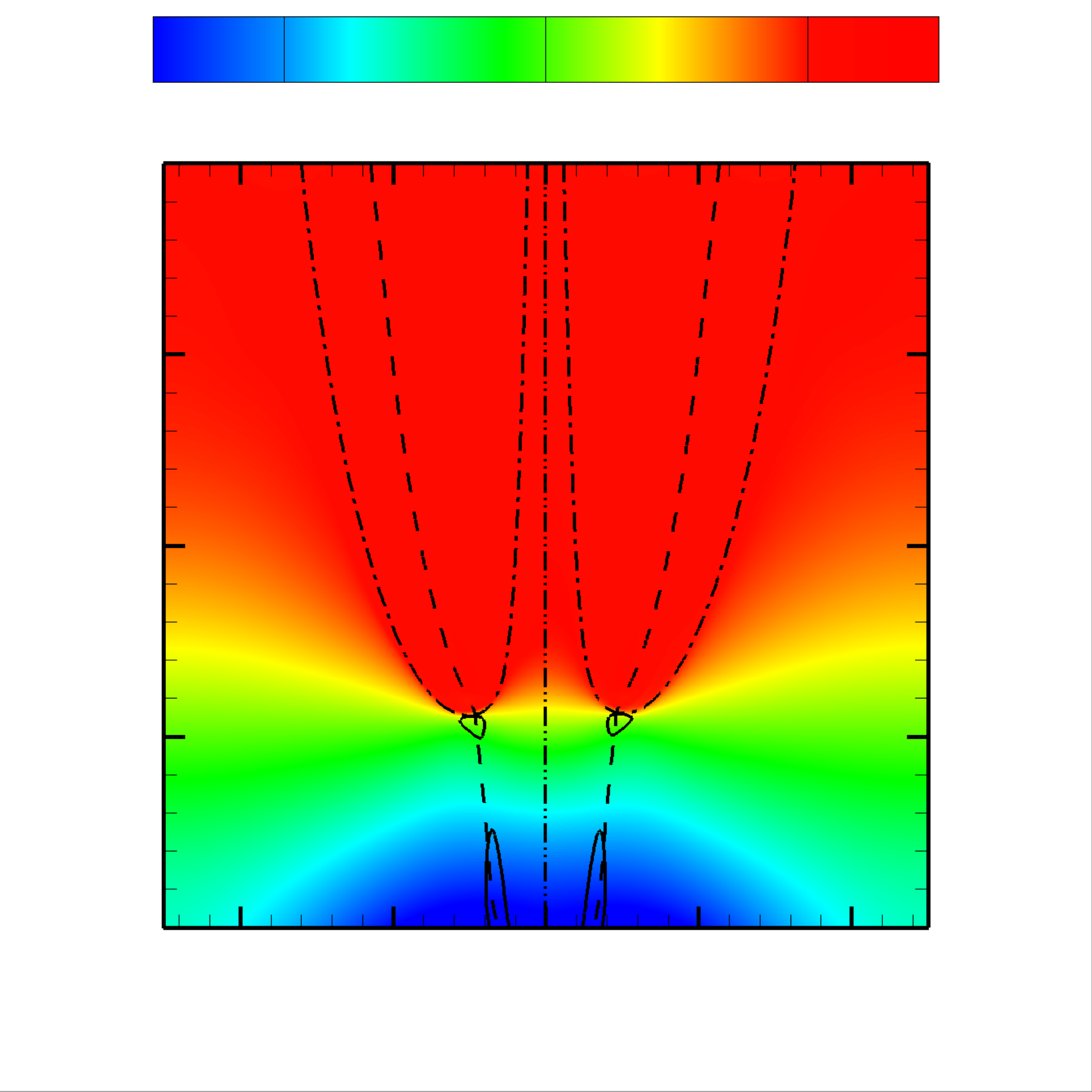
    		};
		\end{tikzpicture}
		\caption{$\Delta \gls{v}$ = \SI{500}{\volt} $x$ direction}
		\label{fig:EF-X_DV500}
	\end{subfigure}
	\begin{subfigure}[b]{0.49\textwidth}
		\centering
		\begin{tikzpicture}
      		\node[anchor=south west,inner sep=0] (image) at (0,0) {
    			\def\svgwidth{\columnwidth}
    			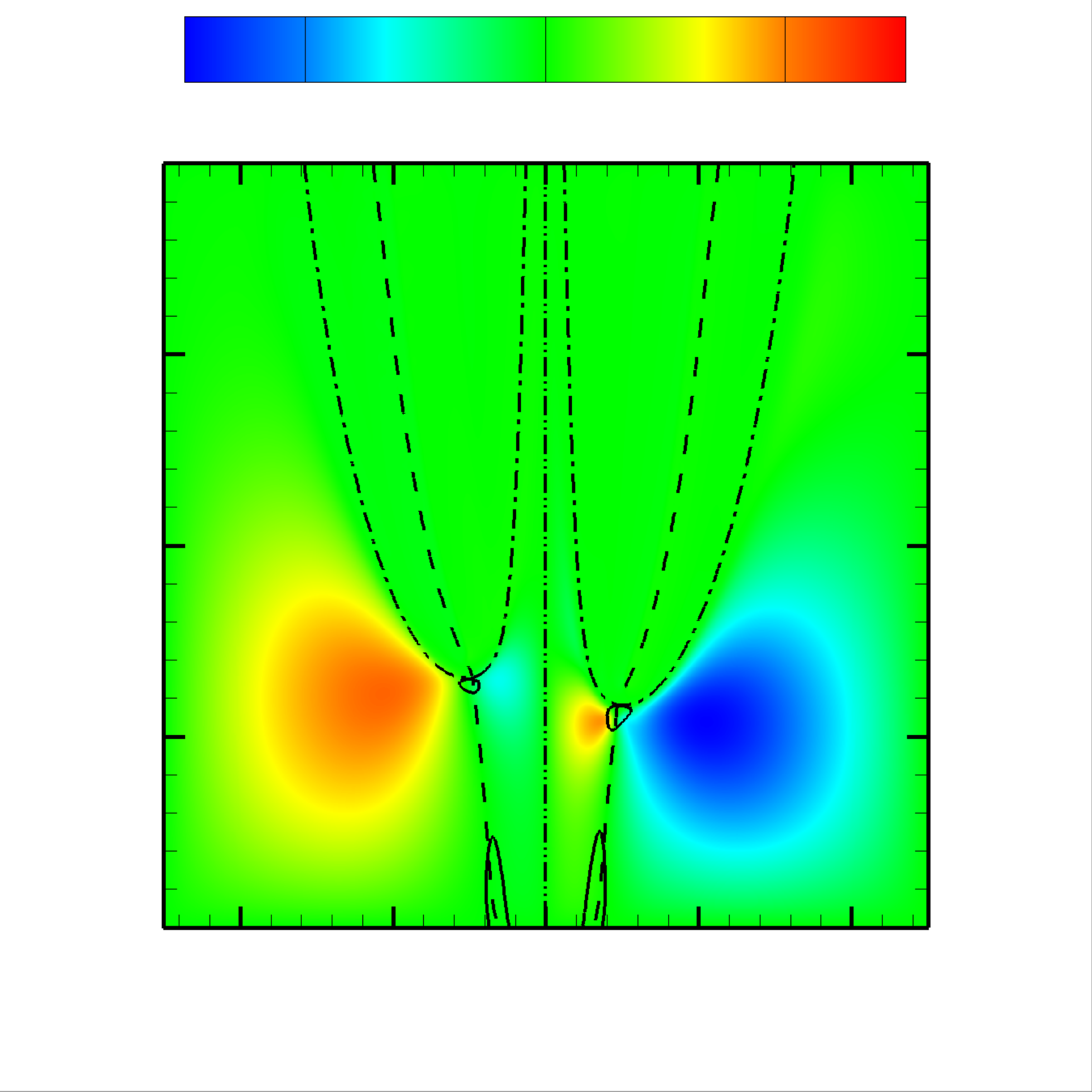
    		};
		\end{tikzpicture}
		\caption{$\Delta \gls{v}$ = \SI{-375}{\volt} $y$ direction}
		\label{fig:EF-Y_DV-375}
	\end{subfigure}
	\begin{subfigure}[b]{0.49\textwidth}
    	\vspace{0.5cm}
		\centering
		\begin{tikzpicture}
       		\node[anchor=south west,inner sep=0] (image) at (0,0) {
    			\def\svgwidth{\columnwidth}
    			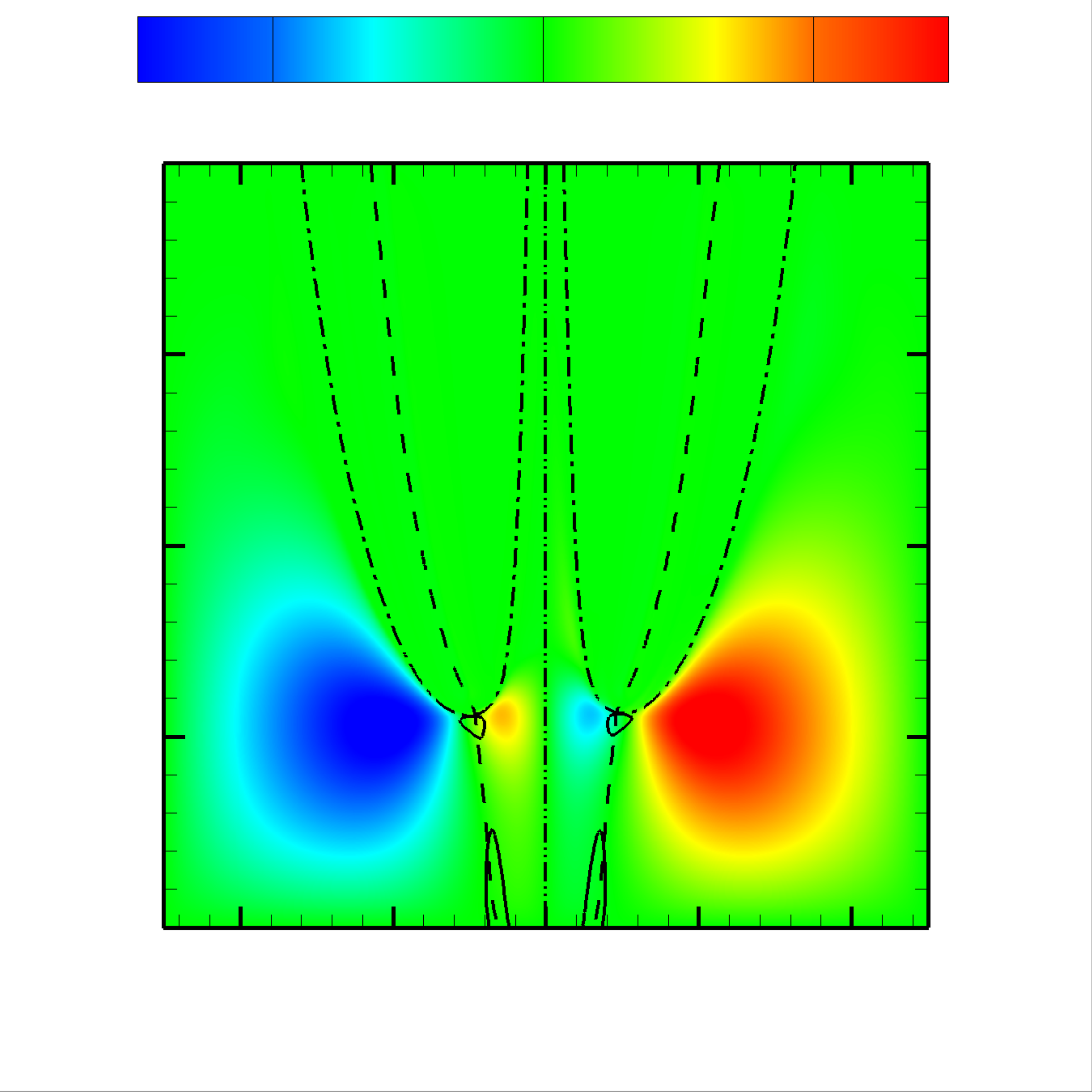
    		};
		\end{tikzpicture}
		\caption{$\Delta \gls{v}$ = \SI{500}{\volt} $y$ direction}
		\label{fig:EF-Y_DV500}
	\end{subfigure}
	\caption[]{Electric field intensity projected along the $x$ (a-b figures) and $y$ (c-d figures) directions (expressed in \si{\kilo\volt\per\meter}) at two different imposed voltage. Results for mechanism ``A'' and for mechanism ``B'' are shown on the left and right hand-side of each picture, respectively. The iso-lines are defined as in Figure~\ref{fig:T0_noemhd}.}
	\label{fig:EF}
\end{figure*}
%

%% file: Figures/T0_noemhd.eps_tex
\begingroup%
  \makeatletter%
  \providecommand\color[2][]{%
    \errmessage{(Inkscape) Color is used for the text in Inkscape, but the package 'color.sty' is not loaded}%
    \renewcommand\color[2][]{}%
  }%
  \providecommand\transparent[1]{%
    \errmessage{(Inkscape) Transparency is used (non-zero) for the text in Inkscape, but the package 'transparent.sty' is not loaded}%
    \renewcommand\transparent[1]{}%
  }%
  \providecommand\rotatebox[2]{#2}%
  \ifx\svgwidth\undefined%
    \setlength{\unitlength}{648bp}%
    \ifx\svgscale\undefined%
      \relax%
    \else%
      \setlength{\unitlength}{\unitlength * \real{\svgscale}}%
    \fi%
  \else%
    \setlength{\unitlength}{\svgwidth}%
  \fi%
  \global\let\svgwidth\undefined%
  \global\let\svgscale\undefined%
  \makeatother%
  \begin{picture}(1,1)%
    \put(0,0){\includegraphics[trim={2 2 2 2},clip,width=\unitlength]{T0_noemhd.pdf}}%
    \put(0.44074074,0.01550926){\color[rgb]{0,0,0}\makebox(0,0)[lb]{\smash{$y$ (\si{\milli\meter})}}}%
    \put(0.04,0.41574074){\color[rgb]{0,0,0}\rotatebox{90}{\makebox(0,0)[lb]{\smash{$x$ (\si{\milli\meter})}}}}%
    \put(0.13587963,0.05060648){\color[rgb]{0,0,0}\makebox(0,0)[lb]{\smash{-9}}}%
    \put(0.25285494,0.05060648){\color[rgb]{0,0,0}\makebox(0,0)[lb]{\smash{-6}}}%
    \put(0.36983025,0.05060648){\color[rgb]{0,0,0}\makebox(0,0)[lb]{\smash{-3}}}%
    \put(0.49151235,0.05060648){\color[rgb]{0,0,0}\makebox(0,0)[lb]{\smash{0}}}%
    \put(0.60852623,0.05060648){\color[rgb]{0,0,0}\makebox(0,0)[lb]{\smash{3}}}%
    \put(0.72550154,0.05060648){\color[rgb]{0,0,0}\makebox(0,0)[lb]{\smash{6}}}%
    \put(0.84247685,0.05060648){\color[rgb]{0,0,0}\makebox(0,0)[lb]{\smash{9}}}%
    \put(0.08014043,0.07978395){\color[rgb]{0,0,0}\makebox(0,0)[lb]{\smash{0}}}%
    \put(0.08014043,0.15679012){\color[rgb]{0,0,0}\makebox(0,0)[lb]{\smash{2}}}%
    \put(0.08014043,0.2337963){\color[rgb]{0,0,0}\makebox(0,0)[lb]{\smash{4}}}%
    \put(0.08014043,0.31076389){\color[rgb]{0,0,0}\makebox(0,0)[lb]{\smash{6}}}%
    \put(0.08014043,0.38777006){\color[rgb]{0,0,0}\makebox(0,0)[lb]{\smash{8}}}%
    \put(0.06324228,0.46477623){\color[rgb]{0,0,0}\makebox(0,0)[lb]{\smash{10}}}%
    \put(0.06324228,0.54174383){\color[rgb]{0,0,0}\makebox(0,0)[lb]{\smash{12}}}%
    \put(0.06324228,0.61875){\color[rgb]{0,0,0}\makebox(0,0)[lb]{\smash{14}}}%
    \put(0.06324228,0.69575617){\color[rgb]{0,0,0}\makebox(0,0)[lb]{\smash{16}}}%
    \put(0.06324228,0.77272377){\color[rgb]{0,0,0}\makebox(0,0)[lb]{\smash{18}}}%
    \put(0.06324228,0.84972994){\color[rgb]{0,0,0}\makebox(0,0)[lb]{\smash{20}}}%
    \put(0.04324228,0.89290123){\color[rgb]{0,0,0}\makebox(0,0)[lb]{\smash{\gls{T} (\si{\kelvin}):}}}%
    \put(0.1679784,0.89290123){\color[rgb]{0,0,0}\makebox(0,0)[lb]{\smash{500}}}%
    \put(0.32172068,0.89290123){\color[rgb]{0,0,0}\makebox(0,0)[lb]{\smash{875}}}%
    \put(0.46701389,0.89290123){\color[rgb]{0,0,0}\makebox(0,0)[lb]{\smash{1250}}}%
    \put(0.62079475,0.89290123){\color[rgb]{0,0,0}\makebox(0,0)[lb]{\smash{1625}}}%
    \put(0.77453704,0.89290123){\color[rgb]{0,0,0}\makebox(0,0)[lb]{\smash{2000}}}%
  \end{picture}%
\endgroup%

%% file: Figures/XovVolt.eps_tex
\begingroup%
  \makeatletter%
  \providecommand\color[2][]{%
    \errmessage{(Inkscape) Color is used for the text in Inkscape, but the package 'color.sty' is not loaded}%
    \renewcommand\color[2][]{}%
  }%
  \providecommand\transparent[1]{%
    \errmessage{(Inkscape) Transparency is used (non-zero) for the text in Inkscape, but the package 'transparent.sty' is not loaded}%
    \renewcommand\transparent[1]{}%
  }%
  \providecommand\rotatebox[2]{#2}%
  \ifx\svgwidth\undefined%
    \setlength{\unitlength}{648bp}%
    \ifx\svgscale\undefined%
      \relax%
    \else%
      \setlength{\unitlength}{\unitlength * \real{\svgscale}}%
    \fi%
  \else%
    \setlength{\unitlength}{\svgwidth}%
  \fi%
  \global\let\svgwidth\undefined%
  \global\let\svgscale\undefined%
  \makeatother%
  \begin{picture}(1,1)%
    \put(0,0){\includegraphics[width=\unitlength]{XovVolt.pdf}}%
    \put(0.37376543,0.02020525){\color[rgb]{0,0,0}\makebox(0,0)[lb]{\smash{Voltage (\si{\volt})}}}%
    \put(0.04,0.40){\color[rgb]{0,0,0}\rotatebox{90}{\makebox(0,0)[lb]{\smash{$x/x_0$ (-)}}}}%
    \put(0.1533179,0.08918981){\color[rgb]{0,0,0}\makebox(0,0)[lb]{\smash{-750}}}%
    \put(0.44605062,0.08918981){\color[rgb]{0,0,0}\makebox(0,0)[lb]{\smash{0}}}%
    \put(0.6626929,0.08918981){\color[rgb]{0,0,0}\makebox(0,0)[lb]{\smash{750}}}%
    \put(0.13626543,0.1904321){\color[rgb]{0,0,0}\makebox(0,0)[lb]{\smash{0}}}%
    \put(0.08626543,0.3121142){\color[rgb]{0,0,0}\makebox(0,0)[lb]{\smash{0.2}}}%
    \put(0.08626543,0.43375772){\color[rgb]{0,0,0}\makebox(0,0)[lb]{\smash{0.4}}}%
    \put(0.08626543,0.55543981){\color[rgb]{0,0,0}\makebox(0,0)[lb]{\smash{0.6}}}%
    \put(0.08626543,0.67708333){\color[rgb]{0,0,0}\makebox(0,0)[lb]{\smash{0.8}}}%
    \put(0.13626543,0.79876543){\color[rgb]{0,0,0}\makebox(0,0)[lb]{\smash{1}}}%
  \end{picture}%
\endgroup%

%% file: Figures/YovVolt.eps_tex
\begingroup%
  \makeatletter%
  \providecommand\color[2][]{%
    \errmessage{(Inkscape) Color is used for the text in Inkscape, but the package 'color.sty' is not loaded}%
    \renewcommand\color[2][]{}%
  }%
  \providecommand\transparent[1]{%
    \errmessage{(Inkscape) Transparency is used (non-zero) for the text in Inkscape, but the package 'transparent.sty' is not loaded}%
    \renewcommand\transparent[1]{}%
  }%
  \providecommand\rotatebox[2]{#2}%
  \ifx\svgwidth\undefined%
    \setlength{\unitlength}{648bp}%
    \ifx\svgscale\undefined%
      \relax%
    \else%
      \setlength{\unitlength}{\unitlength * \real{\svgscale}}%
    \fi%
  \else%
    \setlength{\unitlength}{\svgwidth}%
  \fi%
  \global\let\svgwidth\undefined%
  \global\let\svgscale\undefined%
  \makeatother%
  \begin{picture}(1,1)%
    \put(0,0){\includegraphics[width=\unitlength]{YovVolt.pdf}}%
    \put(0.37376543,0.02020525){\color[rgb]{0,0,0}\makebox(0,0)[lb]{\smash{Voltage (\si{\volt})}}}%
    \put(0.04,0.40){\color[rgb]{0,0,0}\rotatebox{90}{\makebox(0,0)[lb]{\smash{$y/y_0$ (-)}}}}%
    \put(0.1533179,0.08918981){\color[rgb]{0,0,0}\makebox(0,0)[lb]{\smash{-750}}}%
    \put(0.44605062,0.08918981){\color[rgb]{0,0,0}\makebox(0,0)[lb]{\smash{0}}}%
    \put(0.6626929,0.08918981){\color[rgb]{0,0,0}\makebox(0,0)[lb]{\smash{750}}}%
    \put(0.08626543,0.26234568){\color[rgb]{0,0,0}\makebox(0,0)[lb]{\smash{0.6}}}%
    \put(0.08626543,0.52777778){\color[rgb]{0,0,0}\makebox(0,0)[lb]{\smash{0.8}}}%
    \put(0.13626543,0.79324846){\color[rgb]{0,0,0}\makebox(0,0)[lb]{\smash{1}}}%
  \end{picture}%
\endgroup%

%% file: Figures/V_DV0.eps_tex
\begingroup%
  \makeatletter%
  \providecommand\color[2][]{%
    \errmessage{(Inkscape) Color is used for the text in Inkscape, but the package 'color.sty' is not loaded}%
    \renewcommand\color[2][]{}%
  }%
  \providecommand\transparent[1]{%
    \errmessage{(Inkscape) Transparency is used (non-zero) for the text in Inkscape, but the package 'transparent.sty' is not loaded}%
    \renewcommand\transparent[1]{}%
  }%
  \providecommand\rotatebox[2]{#2}%
  \ifx\svgwidth\undefined%
    \setlength{\unitlength}{648bp}%
    \ifx\svgscale\undefined%
      \relax%
    \else%
      \setlength{\unitlength}{\unitlength * \real{\svgscale}}%
    \fi%
  \else%
    \setlength{\unitlength}{\svgwidth}%
  \fi%
  \global\let\svgwidth\undefined%
  \global\let\svgscale\undefined%
  \makeatother%
  \begin{picture}(1,1)%
    \put(0,0){\includegraphics[trim={2 2 2 2},clip,width=\unitlength]{V_DV0.pdf}}%
    \put(0.44074074,0.01550926){\color[rgb]{0,0,0}\makebox(0,0)[lb]{\smash{$y$ (\si{\milli\meter})}}}%
    \put(0.04,0.41574074){\color[rgb]{0,0,0}\rotatebox{90}{\makebox(0,0)[lb]{\smash{$x$ (\si{\milli\meter})}}}}%
    \put(0.19436728,0.05060648){\color[rgb]{0,0,0}\makebox(0,0)[lb]{\smash{-6}}}%
    \put(0.33436728,0.05060648){\color[rgb]{0,0,0}\makebox(0,0)[lb]{\smash{-3}}}%
    \put(0.48436728,0.05060648){\color[rgb]{0,0,0}\makebox(0,0)[lb]{\smash{ 0}}}%
    \put(0.63036728,0.05060648){\color[rgb]{0,0,0}\makebox(0,0)[lb]{\smash{ 3}}}%
    \put(0.77836728,0.05060648){\color[rgb]{0,0,0}\makebox(0,0)[lb]{\smash{ 6}}}%
    \put(0.08014043,0.07978395){\color[rgb]{0,0,0}\makebox(0,0)[lb]{\smash{0}}}%
    \put(0.08014043,0.17604167){\color[rgb]{0,0,0}\makebox(0,0)[lb]{\smash{2}}}%
    \put(0.08014043,0.27229938){\color[rgb]{0,0,0}\makebox(0,0)[lb]{\smash{4}}}%
    \put(0.08014043,0.3685571){\color[rgb]{0,0,0}\makebox(0,0)[lb]{\smash{6}}}%
    \put(0.08014043,0.46481481){\color[rgb]{0,0,0}\makebox(0,0)[lb]{\smash{8}}}%
    \put(0.06324228,0.56103395){\color[rgb]{0,0,0}\makebox(0,0)[lb]{\smash{10}}}%
    \put(0.06324228,0.65729167){\color[rgb]{0,0,0}\makebox(0,0)[lb]{\smash{12}}}%
    \put(0.06324228,0.75354938){\color[rgb]{0,0,0}\makebox(0,0)[lb]{\smash{14}}}%
    \put(0.06324228,0.8498071){\color[rgb]{0,0,0}\makebox(0,0)[lb]{\smash{16}}}%
    \put(0.07624228,0.89290123){\color[rgb]{0,0,0}\makebox(0,0)[rb]{\smash{\gls{v} (\si{\volt}):}}}%
    \put(0.10945216,0.89290123){\color[rgb]{0,0,0}\makebox(0,0)[lb]{\smash{-0.3}}}%
    \put(0.23070988,0.89290123){\color[rgb]{0,0,0}\makebox(0,0)[lb]{\smash{-0.2}}}%
    \put(0.35196759,0.89290123){\color[rgb]{0,0,0}\makebox(0,0)[lb]{\smash{-0.1}}}%
    \put(0.49039352,0.89290123){\color[rgb]{0,0,0}\makebox(0,0)[lb]{\smash{0}}}%
    \put(0.59915123,0.89290123){\color[rgb]{0,0,0}\makebox(0,0)[lb]{\smash{0.1}}}%
    \put(0.72040895,0.89290123){\color[rgb]{0,0,0}\makebox(0,0)[lb]{\smash{0.2}}}%
    \put(0.84166667,0.89290123){\color[rgb]{0,0,0}\makebox(0,0)[lb]{\smash{0.3}}}%
  \end{picture}%
\endgroup%

%% file: Figures/V_DV-375.eps_tex
\begingroup%
  \makeatletter%
  \providecommand\color[2][]{%
    \errmessage{(Inkscape) Color is used for the text in Inkscape, but the package 'color.sty' is not loaded}%
    \renewcommand\color[2][]{}%
  }%
  \providecommand\transparent[1]{%
    \errmessage{(Inkscape) Transparency is used (non-zero) for the text in Inkscape, but the package 'transparent.sty' is not loaded}%
    \renewcommand\transparent[1]{}%
  }%
  \providecommand\rotatebox[2]{#2}%
  \ifx\svgwidth\undefined%
    \setlength{\unitlength}{648bp}%
    \ifx\svgscale\undefined%
      \relax%
    \else%
      \setlength{\unitlength}{\unitlength * \real{\svgscale}}%
    \fi%
  \else%
    \setlength{\unitlength}{\svgwidth}%
  \fi%
  \global\let\svgwidth\undefined%
  \global\let\svgscale\undefined%
  \makeatother%
  \begin{picture}(1,1)%
    \put(0,0){\includegraphics[width=\unitlength]{V_DV-375.pdf}}%
        \put(0.41512346,0.03598765){\color[rgb]{0,0,0}\makebox(0,0)[lb]{\smash{$y$ (\si{\milli\meter})}}}%
    \put(0.04,0.41508488){\color[rgb]{0,0,0}\rotatebox{90}{\makebox(0,0)[lb]{\smash{$x$ (\si{\milli\meter})}}}}%
    \put(0.19810957,0.09601852){\color[rgb]{0,0,0}\makebox(0,0)[lb]{\smash{-8}}}%
    \put(0.33811728,0.09601852){\color[rgb]{0,0,0}\makebox(0,0)[lb]{\smash{-4}}}%
    \put(0.48595679,0.09601852){\color[rgb]{0,0,0}\makebox(0,0)[lb]{\smash{0}}}%
    \put(0.62592593,0.09601852){\color[rgb]{0,0,0}\makebox(0,0)[lb]{\smash{4}}}%
    \put(0.76593364,0.09601852){\color[rgb]{0,0,0}\makebox(0,0)[lb]{\smash{8}}}%
    \put(0.11188272,0.13298611){\color[rgb]{0,0,0}\makebox(0,0)[lb]{\smash{0}}}%
    \put(0.11188272,0.30798611){\color[rgb]{0,0,0}\makebox(0,0)[lb]{\smash{5}}}%
    \put(0.08375772,0.48298611){\color[rgb]{0,0,0}\makebox(0,0)[lb]{\smash{10}}}%
    \put(0.08375772,0.65798611){\color[rgb]{0,0,0}\makebox(0,0)[lb]{\smash{15}}}%
    \put(0.08375772,0.83298611){\color[rgb]{0,0,0}\makebox(0,0)[lb]{\smash{20}}}%
    \put(0.23641975,0.87619599){\color[rgb]{0,0,0}\makebox(0,0)[lb]{\smash{-300}}}%
    \put(0.45378086,0.87619599){\color[rgb]{0,0,0}\makebox(0,0)[lb]{\smash{-200}}}%
    \put(0.67114198,0.87619599){\color[rgb]{0,0,0}\makebox(0,0)[lb]{\smash{-100}}}%
  \end{picture}%
\endgroup%

%% file: Figures/V_DV500.eps_tex
\begingroup%
  \makeatletter%
  \providecommand\color[2][]{%
    \errmessage{(Inkscape) Color is used for the text in Inkscape, but the package 'color.sty' is not loaded}%
    \renewcommand\color[2][]{}%
  }%
  \providecommand\transparent[1]{%
    \errmessage{(Inkscape) Transparency is used (non-zero) for the text in Inkscape, but the package 'transparent.sty' is not loaded}%
    \renewcommand\transparent[1]{}%
  }%
  \providecommand\rotatebox[2]{#2}%
  \ifx\svgwidth\undefined%
    \setlength{\unitlength}{648bp}%
    \ifx\svgscale\undefined%
      \relax%
    \else%
      \setlength{\unitlength}{\unitlength * \real{\svgscale}}%
    \fi%
  \else%
    \setlength{\unitlength}{\svgwidth}%
  \fi%
  \global\let\svgwidth\undefined%
  \global\let\svgscale\undefined%
  \makeatother%
  \begin{picture}(1,1)%
    \put(0,0){\includegraphics[trim={2 2 2 2},clip,width=\unitlength]{V_DV500.pdf}}%
    \put(0.41512346,0.03598765){\color[rgb]{0,0,0}\makebox(0,0)[lb]{\smash{$y$ (\si{\milli\meter})}}}%
    \put(0.04,0.41508488){\color[rgb]{0,0,0}\rotatebox{90}{\makebox(0,0)[lb]{\smash{$x$ (\si{\milli\meter})}}}}%
    \put(0.19810957,0.09601852){\color[rgb]{0,0,0}\makebox(0,0)[lb]{\smash{-8}}}%
    \put(0.33811728,0.09601852){\color[rgb]{0,0,0}\makebox(0,0)[lb]{\smash{-4}}}%
    \put(0.48595679,0.09601852){\color[rgb]{0,0,0}\makebox(0,0)[lb]{\smash{0}}}%
    \put(0.62592593,0.09601852){\color[rgb]{0,0,0}\makebox(0,0)[lb]{\smash{4}}}%
    \put(0.76593364,0.09601852){\color[rgb]{0,0,0}\makebox(0,0)[lb]{\smash{8}}}%
    \put(0.11188272,0.13298611){\color[rgb]{0,0,0}\makebox(0,0)[lb]{\smash{0}}}%
    \put(0.11188272,0.30798611){\color[rgb]{0,0,0}\makebox(0,0)[lb]{\smash{5}}}%
    \put(0.08375772,0.48298611){\color[rgb]{0,0,0}\makebox(0,0)[lb]{\smash{10}}}%
    \put(0.08375772,0.65798611){\color[rgb]{0,0,0}\makebox(0,0)[lb]{\smash{15}}}%
    \put(0.08375772,0.83298611){\color[rgb]{0,0,0}\makebox(0,0)[lb]{\smash{20}}}%
    \put(0.18225309,0.87619599){\color[rgb]{0,0,0}\makebox(0,0)[lb]{\smash{100}}}%
    \put(0.37025463,0.87619599){\color[rgb]{0,0,0}\makebox(0,0)[lb]{\smash{200}}}%
    \put(0.55825617,0.87619599){\color[rgb]{0,0,0}\makebox(0,0)[lb]{\smash{300}}}%
    \put(0.74625772,0.87619599){\color[rgb]{0,0,0}\makebox(0,0)[lb]{\smash{400}}}%
  \end{picture}%
\endgroup%

%% file: Figures/VovVolt.eps_tex
\begingroup%
  \makeatletter%
  \providecommand\color[2][]{%
    \errmessage{(Inkscape) Color is used for the text in Inkscape, but the package 'color.sty' is not loaded}%
    \renewcommand\color[2][]{}%
  }%
  \providecommand\transparent[1]{%
    \errmessage{(Inkscape) Transparency is used (non-zero) for the text in Inkscape, but the package 'transparent.sty' is not loaded}%
    \renewcommand\transparent[1]{}%
  }%
  \providecommand\rotatebox[2]{#2}%
  \ifx\svgwidth\undefined%
    \setlength{\unitlength}{648bp}%
    \ifx\svgscale\undefined%
      \relax%
    \else%
      \setlength{\unitlength}{\unitlength * \real{\svgscale}}%
    \fi%
  \else%
    \setlength{\unitlength}{\svgwidth}%
  \fi%
  \global\let\svgwidth\undefined%
  \global\let\svgscale\undefined%
  \makeatother%
  \begin{picture}(1,1)%
    \put(0,0){\includegraphics[width=\unitlength]{VovVolt.pdf}}%
    \put(0.50300926,0.03050154){\color[rgb]{0,0,0}\makebox(0,0)[cb]{\smash{Imposed voltage (\si{\volt})}}}%
    \put(0.05999228,0.49301698){\color[rgb]{0,0,0}\rotatebox{90}{\makebox(0,0)[cb]{\smash{Flame-tip voltage (\si{\volt})}}}}%
    \put(0.21126543,0.07959105){\color[rgb]{0,0,0}\makebox(0,0)[lb]{\smash{-500}}}%
    \put(0.41531636,0.07959105){\color[rgb]{0,0,0}\makebox(0,0)[lb]{\smash{0}}}%
    \put(0.58094136,0.07959105){\color[rgb]{0,0,0}\makebox(0,0)[lb]{\smash{500}}}%
    \put(0.75501543,0.07959105){\color[rgb]{0,0,0}\makebox(0,0)[lb]{\smash{1000}}}%
    \put(0.08001543,0.19853395){\color[rgb]{0,0,0}\makebox(0,0)[lb]{\smash{-500}}}%
    \put(0.12310957,0.39606481){\color[rgb]{0,0,0}\makebox(0,0)[lb]{\smash{0}}}%
    \put(0.08939043,0.5935571){\color[rgb]{0,0,0}\makebox(0,0)[lb]{\smash{500}}}%
    \put(0.07249228,0.79104938){\color[rgb]{0,0,0}\makebox(0,0)[lb]{\smash{1000}}}%
  \end{picture}%
\endgroup%

%% file: Figures/C+F_DV-375.eps_tex
\begingroup%
  \makeatletter%
  \providecommand\color[2][]{%
    \errmessage{(Inkscape) Color is used for the text in Inkscape, but the package 'color.sty' is not loaded}%
    \renewcommand\color[2][]{}%
  }%
  \providecommand\transparent[1]{%
    \errmessage{(Inkscape) Transparency is used (non-zero) for the text in Inkscape, but the package 'transparent.sty' is not loaded}%
    \renewcommand\transparent[1]{}%
  }%
  \providecommand\rotatebox[2]{#2}%
  \ifx\svgwidth\undefined%
    \setlength{\unitlength}{648bp}%
    \ifx\svgscale\undefined%
      \relax%
    \else%
      \setlength{\unitlength}{\unitlength * \real{\svgscale}}%
    \fi%
  \else%
    \setlength{\unitlength}{\svgwidth}%
  \fi%
  \global\let\svgwidth\undefined%
  \global\let\svgscale\undefined%
  \makeatother%
  \begin{picture}(1,1)%
    \put(0,0){\includegraphics[trim={2 2 2 2},clip,width=\unitlength]{C+F_DV-375.pdf}}%
    \put(0.41512346,0.03598765){\color[rgb]{0,0,0}\makebox(0,0)[lb]{\smash{$y$ (\si{\milli\meter})}}}%
    \put(0.04,0.41508488){\color[rgb]{0,0,0}\rotatebox{90}{\makebox(0,0)[lb]{\smash{$x$ (\si{\milli\meter})}}}}%
    \put(0.278125  ,0.09601852){\color[rgb]{0,0,0}\makebox(0,0)[lb]{\smash{-2}}}%
    \put(0.48595679,0.09601852){\color[rgb]{0,0,0}\makebox(0,0)[lb]{\smash{0}}}%
    \put(0.68595679,0.09601852){\color[rgb]{0,0,0}\makebox(0,0)[lb]{\smash{2}}}%
    \put(0.11188272,0.13298611){\color[rgb]{0,0,0}\makebox(0,0)[lb]{\smash{2}}}%
    \put(0.11188272,0.33298611){\color[rgb]{0,0,0}\makebox(0,0)[lb]{\smash{4}}}%
    \put(0.11188272,0.53298611){\color[rgb]{0,0,0}\makebox(0,0)[lb]{\smash{6}}}%
    \put(0.11188272,0.73298611){\color[rgb]{0,0,0}\makebox(0,0)[lb]{\smash{8}}}%
    \put(0.23374228,0.87619599){\color[rgb]{0,0,0}\makebox(0,0)[lb]{\smash{-1}}}%
    \put(0.44982253,0.87619599){\color[rgb]{0,0,0}\makebox(0,0)[lb]{\smash{-0.5}}}%
    \put(0.72384259,0.87619599){\color[rgb]{0,0,0}\makebox(0,0)[lb]{\smash{0}}}%
  \end{picture}%
\endgroup%

%% file: Figures/C+F_DV500.eps_tex
\begingroup%
  \makeatletter%
  \providecommand\color[2][]{%
    \errmessage{(Inkscape) Color is used for the text in Inkscape, but the package 'color.sty' is not loaded}%
    \renewcommand\color[2][]{}%
  }%
  \providecommand\transparent[1]{%
    \errmessage{(Inkscape) Transparency is used (non-zero) for the text in Inkscape, but the package 'transparent.sty' is not loaded}%
    \renewcommand\transparent[1]{}%
  }%
  \providecommand\rotatebox[2]{#2}%
  \ifx\svgwidth\undefined%
    \setlength{\unitlength}{648bp}%
    \ifx\svgscale\undefined%
      \relax%
    \else%
      \setlength{\unitlength}{\unitlength * \real{\svgscale}}%
    \fi%
  \else%
    \setlength{\unitlength}{\svgwidth}%
  \fi%
  \global\let\svgwidth\undefined%
  \global\let\svgscale\undefined%
  \makeatother%
  \begin{picture}(1,1)%
    \put(0,0){\includegraphics[trim={2 2 2 2},clip,width=\unitlength]{C+F_DV500.pdf}}%
    \put(0.41512346,0.03598765){\color[rgb]{0,0,0}\makebox(0,0)[lb]{\smash{$y$ (\si{\milli\meter})}}}%
    \put(0.04,0.41508488){\color[rgb]{0,0,0}\rotatebox{90}{\makebox(0,0)[lb]{\smash{$x$ (\si{\milli\meter})}}}}%
    \put(0.278125  ,0.09601852){\color[rgb]{0,0,0}\makebox(0,0)[lb]{\smash{-2}}}%
    \put(0.48595679,0.09601852){\color[rgb]{0,0,0}\makebox(0,0)[lb]{\smash{0}}}%
    \put(0.68595679,0.09601852){\color[rgb]{0,0,0}\makebox(0,0)[lb]{\smash{2}}}%
    \put(0.11188272,0.13298611){\color[rgb]{0,0,0}\makebox(0,0)[lb]{\smash{2}}}%
    \put(0.11188272,0.33298611){\color[rgb]{0,0,0}\makebox(0,0)[lb]{\smash{4}}}%
    \put(0.11188272,0.53298611){\color[rgb]{0,0,0}\makebox(0,0)[lb]{\smash{6}}}%
    \put(0.11188272,0.73298611){\color[rgb]{0,0,0}\makebox(0,0)[lb]{\smash{8}}}%
    \put(0.28209877,0.87619599){\color[rgb]{0,0,0}\makebox(0,0)[lb]{\smash{0}}}%
    \put(0.48691358,0.87619599){\color[rgb]{0,0,0}\makebox(0,0)[lb]{\smash{1}}}%
    \put(0.69092593,0.87619599){\color[rgb]{0,0,0}\makebox(0,0)[lb]{\smash{2}}}%
  \end{picture}%
\endgroup%

%% file: Figures/EF-X_DV-375.eps_tex
\begingroup%
  \makeatletter%
  \providecommand\color[2][]{%
    \errmessage{(Inkscape) Color is used for the text in Inkscape, but the package 'color.sty' is not loaded}%
    \renewcommand\color[2][]{}%
  }%
  \providecommand\transparent[1]{%
    \errmessage{(Inkscape) Transparency is used (non-zero) for the text in Inkscape, but the package 'transparent.sty' is not loaded}%
    \renewcommand\transparent[1]{}%
  }%
  \providecommand\rotatebox[2]{#2}%
  \ifx\svgwidth\undefined%
    \setlength{\unitlength}{648bp}%
    \ifx\svgscale\undefined%
      \relax%
    \else%
      \setlength{\unitlength}{\unitlength * \real{\svgscale}}%
    \fi%
  \else%
    \setlength{\unitlength}{\svgwidth}%
  \fi%
  \global\let\svgwidth\undefined%
  \global\let\svgscale\undefined%
  \makeatother%
  \begin{picture}(1,1)%
    \put(0,0){\includegraphics[trim={2 2 2 2},clip,width=\unitlength]{EF-X_DV-375.pdf}}%
    \put(0.41512346,0.03598765){\color[rgb]{0,0,0}\makebox(0,0)[lb]{\smash{$y$ (\si{\milli\meter})}}}%
    \put(0.04,0.41508488){\color[rgb]{0,0,0}\rotatebox{90}{\makebox(0,0)[lb]{\smash{$x$ (\si{\milli\meter})}}}}%
    \put(0.19810957,0.09601852){\color[rgb]{0,0,0}\makebox(0,0)[lb]{\smash{-8}}}%
    \put(0.33811728,0.09601852){\color[rgb]{0,0,0}\makebox(0,0)[lb]{\smash{-4}}}%
    \put(0.48591821,0.09601852){\color[rgb]{0,0,0}\makebox(0,0)[lb]{\smash{0}}}%
    \put(0.62588735,0.09601852){\color[rgb]{0,0,0}\makebox(0,0)[lb]{\smash{4}}}%
    \put(0.76585648,0.09601852){\color[rgb]{0,0,0}\makebox(0,0)[lb]{\smash{8}}}%
    \put(0.11188272,0.13298611){\color[rgb]{0,0,0}\makebox(0,0)[lb]{\smash{0}}}%
    \put(0.11188272,0.30798611){\color[rgb]{0,0,0}\makebox(0,0)[lb]{\smash{5}}}%
    \put(0.08375772,0.48294753){\color[rgb]{0,0,0}\makebox(0,0)[lb]{\smash{10}}}%
    \put(0.08375772,0.65794753){\color[rgb]{0,0,0}\makebox(0,0)[lb]{\smash{15}}}%
    \put(0.08375772,0.83294753){\color[rgb]{0,0,0}\makebox(0,0)[lb]{\smash{20}}}%
    \put(0.18188272,0.87619599){\color[rgb]{0,0,0}\makebox(0,0)[lb]{\smash{20}}}%
    \put(0.37438272,0.87619599){\color[rgb]{0,0,0}\makebox(0,0)[lb]{\smash{40}}}%
    \put(0.56688272,0.87619599){\color[rgb]{0,0,0}\makebox(0,0)[lb]{\smash{60}}}%
    \put(0.75438272,0.87619599){\color[rgb]{0,0,0}\makebox(0,0)[lb]{\smash{80}}}%
  \end{picture}%
\endgroup%

%% file: Figures/EF-X_DV500.eps_tex
\begingroup%
  \makeatletter%
  \providecommand\color[2][]{%
    \errmessage{(Inkscape) Color is used for the text in Inkscape, but the package 'color.sty' is not loaded}%
    \renewcommand\color[2][]{}%
  }%
  \providecommand\transparent[1]{%
    \errmessage{(Inkscape) Transparency is used (non-zero) for the text in Inkscape, but the package 'transparent.sty' is not loaded}%
    \renewcommand\transparent[1]{}%
  }%
  \providecommand\rotatebox[2]{#2}%
  \ifx\svgwidth\undefined%
    \setlength{\unitlength}{648bp}%
    \ifx\svgscale\undefined%
      \relax%
    \else%
      \setlength{\unitlength}{\unitlength * \real{\svgscale}}%
    \fi%
  \else%
    \setlength{\unitlength}{\svgwidth}%
  \fi%
  \global\let\svgwidth\undefined%
  \global\let\svgscale\undefined%
  \makeatother%
  \begin{picture}(1,1)%
    \put(0,0){\includegraphics[trim={2 2 2 2},clip,width=\unitlength]{EF-X_DV500.pdf}}%
    \put(0.41512346,0.03598765){\color[rgb]{0,0,0}\makebox(0,0)[lb]{\smash{$y$ (\si{\milli\meter})}}}%
    \put(0.04,0.41508488){\color[rgb]{0,0,0}\rotatebox{90}{\makebox(0,0)[lb]{\smash{$x$ (\si{\milli\meter})}}}}%
    \put(0.19810957,0.09601852){\color[rgb]{0,0,0}\makebox(0,0)[lb]{\smash{-8}}}%
    \put(0.33811728,0.09601852){\color[rgb]{0,0,0}\makebox(0,0)[lb]{\smash{-4}}}%
    \put(0.48591821,0.09601852){\color[rgb]{0,0,0}\makebox(0,0)[lb]{\smash{0}}}%
    \put(0.62588735,0.09601852){\color[rgb]{0,0,0}\makebox(0,0)[lb]{\smash{4}}}%
    \put(0.76585648,0.09601852){\color[rgb]{0,0,0}\makebox(0,0)[lb]{\smash{8}}}%
    \put(0.11188272,0.13298611){\color[rgb]{0,0,0}\makebox(0,0)[lb]{\smash{0}}}%
    \put(0.11188272,0.30798611){\color[rgb]{0,0,0}\makebox(0,0)[lb]{\smash{5}}}%
    \put(0.08375772,0.48294753){\color[rgb]{0,0,0}\makebox(0,0)[lb]{\smash{10}}}%
    \put(0.08375772,0.65794753){\color[rgb]{0,0,0}\makebox(0,0)[lb]{\smash{15}}}%
    \put(0.08375772,0.83294753){\color[rgb]{0,0,0}\makebox(0,0)[lb]{\smash{20}}}%
    \put(0.21626543,0.87619599){\color[rgb]{0,0,0}\makebox(0,0)[lb]{\smash{-100}}}%
    \put(0.4575    ,0.87619599){\color[rgb]{0,0,0}\makebox(0,0)[lb]{\smash{-50}}}%
    \put(0.72874228,0.87619599){\color[rgb]{0,0,0}\makebox(0,0)[lb]{\smash{0}}}%
  \end{picture}%
\endgroup%

%% file: Figures/EF-Y_DV-375.eps_tex
\begingroup%
  \makeatletter%
  \providecommand\color[2][]{%
    \errmessage{(Inkscape) Color is used for the text in Inkscape, but the package 'color.sty' is not loaded}%
    \renewcommand\color[2][]{}%
  }%
  \providecommand\transparent[1]{%
    \errmessage{(Inkscape) Transparency is used (non-zero) for the text in Inkscape, but the package 'transparent.sty' is not loaded}%
    \renewcommand\transparent[1]{}%
  }%
  \providecommand\rotatebox[2]{#2}%
  \ifx\svgwidth\undefined%
    \setlength{\unitlength}{648bp}%
    \ifx\svgscale\undefined%
      \relax%
    \else%
      \setlength{\unitlength}{\unitlength * \real{\svgscale}}%
    \fi%
  \else%
    \setlength{\unitlength}{\svgwidth}%
  \fi%
  \global\let\svgwidth\undefined%
  \global\let\svgscale\undefined%
  \makeatother%
  \begin{picture}(1,1)%
    \put(0,0){\includegraphics[trim={2 2 2 2},clip,width=\unitlength]{EF-Y_DV-375.pdf}}%
    \put(0.41512346,0.03598765){\color[rgb]{0,0,0}\makebox(0,0)[lb]{\smash{$y$ (\si{\milli\meter})}}}%
    \put(0.04,0.41508488){\color[rgb]{0,0,0}\rotatebox{90}{\makebox(0,0)[lb]{\smash{$x$ (\si{\milli\meter})}}}}%
    \put(0.19810957,0.09601852){\color[rgb]{0,0,0}\makebox(0,0)[lb]{\smash{-8}}}%
    \put(0.33811728,0.09601852){\color[rgb]{0,0,0}\makebox(0,0)[lb]{\smash{-4}}}%
    \put(0.48591821,0.09601852){\color[rgb]{0,0,0}\makebox(0,0)[lb]{\smash{0}}}%
    \put(0.62588735,0.09601852){\color[rgb]{0,0,0}\makebox(0,0)[lb]{\smash{4}}}%
    \put(0.76585648,0.09601852){\color[rgb]{0,0,0}\makebox(0,0)[lb]{\smash{8}}}%
    \put(0.11188272,0.13298611){\color[rgb]{0,0,0}\makebox(0,0)[lb]{\smash{0}}}%
    \put(0.11188272,0.30798611){\color[rgb]{0,0,0}\makebox(0,0)[lb]{\smash{5}}}%
    \put(0.08375772,0.48294753){\color[rgb]{0,0,0}\makebox(0,0)[lb]{\smash{10}}}%
    \put(0.08375772,0.65794753){\color[rgb]{0,0,0}\makebox(0,0)[lb]{\smash{15}}}%
    \put(0.08375772,0.83294753){\color[rgb]{0,0,0}\makebox(0,0)[lb]{\smash{20}}}%
    \put(0.24685957,0.87619599){\color[rgb]{0,0,0}\makebox(0,0)[lb]{\smash{-15}}}%
    \put(0.48811728,0.87619599){\color[rgb]{0,0,0}\makebox(0,0)[lb]{\smash{0}}}%
    \put(0.69311728,0.87619599){\color[rgb]{0,0,0}\makebox(0,0)[lb]{\smash{15}}}%
  \end{picture}%
\endgroup%

%% file: Figures/EF-Y_DV500.eps_tex
\begingroup%
  \makeatletter%
  \providecommand\color[2][]{%
    \errmessage{(Inkscape) Color is used for the text in Inkscape, but the package 'color.sty' is not loaded}%
    \renewcommand\color[2][]{}%
  }%
  \providecommand\transparent[1]{%
    \errmessage{(Inkscape) Transparency is used (non-zero) for the text in Inkscape, but the package 'transparent.sty' is not loaded}%
    \renewcommand\transparent[1]{}%
  }%
  \providecommand\rotatebox[2]{#2}%
  \ifx\svgwidth\undefined%
    \setlength{\unitlength}{648bp}%
    \ifx\svgscale\undefined%
      \relax%
    \else%
      \setlength{\unitlength}{\unitlength * \real{\svgscale}}%
    \fi%
  \else%
    \setlength{\unitlength}{\svgwidth}%
  \fi%
  \global\let\svgwidth\undefined%
  \global\let\svgscale\undefined%
  \makeatother%
  \begin{picture}(1,1)%
    \put(0,0){\includegraphics[trim={2 2 2 2},clip,width=\unitlength]{EF-Y_DV500.pdf}}%
    \put(0.41512346,0.03598765){\color[rgb]{0,0,0}\makebox(0,0)[lb]{\smash{$y$ (\si{\milli\meter})}}}%
    \put(0.04,0.41508488){\color[rgb]{0,0,0}\rotatebox{90}{\makebox(0,0)[lb]{\smash{$x$ (\si{\milli\meter})}}}}%
    \put(0.19810957,0.09601852){\color[rgb]{0,0,0}\makebox(0,0)[lb]{\smash{-8}}}%
    \put(0.33811728,0.09601852){\color[rgb]{0,0,0}\makebox(0,0)[lb]{\smash{-4}}}%
    \put(0.48591821,0.09601852){\color[rgb]{0,0,0}\makebox(0,0)[lb]{\smash{0}}}%
    \put(0.62588735,0.09601852){\color[rgb]{0,0,0}\makebox(0,0)[lb]{\smash{4}}}%
    \put(0.76585648,0.09601852){\color[rgb]{0,0,0}\makebox(0,0)[lb]{\smash{8}}}%
    \put(0.11188272,0.13298611){\color[rgb]{0,0,0}\makebox(0,0)[lb]{\smash{0}}}%
    \put(0.11188272,0.30798611){\color[rgb]{0,0,0}\makebox(0,0)[lb]{\smash{5}}}%
    \put(0.08375772,0.48294753){\color[rgb]{0,0,0}\makebox(0,0)[lb]{\smash{10}}}%
    \put(0.08375772,0.65794753){\color[rgb]{0,0,0}\makebox(0,0)[lb]{\smash{15}}}%
    \put(0.08375772,0.83294753){\color[rgb]{0,0,0}\makebox(0,0)[lb]{\smash{20}}}%
    \put(0.2175    ,0.87619599){\color[rgb]{0,0,0}\makebox(0,0)[lb]{\smash{-20}}}%
    \put(0.48626543,0.87619599){\color[rgb]{0,0,0}\makebox(0,0)[lb]{\smash{0}}}%
    \put(0.70873457,0.87619599){\color[rgb]{0,0,0}\makebox(0,0)[lb]{\smash{20}}}%
  \end{picture}%
\endgroup%

%% file: Sections/Comp_cost.tex
%
%
\section{Computational cost of the simulations}
\label{sec:comp_cost}
For the formulation of a good model, its efficiency and its affordability are not of secondary importance with respect to its accuracy.
In fact, the reduction of the computational cost of this kind of simulations is a mandatory requirement for analyzing complex flows in the future.
For this reason, a comparison of the average wall-time needed per time-step has been carried-out for each simulation done in this work.
When mechanism ``A'' is used, the computational cost of the present approach is comparable with the performance of the model proposed by \citet{Belhi2010}.
Mechanism ``B'' instead provides a completely different behavior of the model because of the introduction of the averaged properties of the scalars.\par
All the simulations were run on two Xeon 10-core E5-2660v3 (\SI{2.6}{\giga\hertz}) processors without hyper-trading arranged on a single node.
The time-step for the inner iterations has been evaluated using the \gls{cfl} number based on the advection of the negative charges ($\gls{cfl}_M$), which has been set to \num{3.5} for all the simulations.
The outer advancement step has been kept constant and equal to \SI{5E-6}{\second} for all the simulations (corresponding to a \gls{cfl} number of about \num{2.5} based on the flow velocity).
The results of the comparison are reported in Table~\ref{tab:CompTime} for each combination of applied voltage and kinetic mechanisms.\par
\begin{table}[tb]
	\begin{center}
		\begin{tabular}{|c|c|c|c|c|}
			\hline Voltage			& Mech. ``A''			& Mech. ``B''			& Speedup		\\
			\hline \SI{-750}{\volt}	& \SI{111.26}{\second}	& \SI{17.12}{\second}	& \num{6.50}	\\
			\hline \SI{-625}{\volt}	& \SI{93.15} {\second}	& \SI{14.91}{\second}	& \num{6.25}	\\
			\hline \SI{-500}{\volt}	& \SI{71.86} {\second}	& \SI{7.69} {\second}	& \num{9.34}	\\
			\hline \SI{-375}{\volt}	& \SI{55.90} {\second}	& \SI{1.36} {\second}	& \num{41.16}	\\
			\hline \SI{-250}{\volt}	& \SI{39.72} {\second}	& \SI{1.36} {\second}	& \num{29.08}	\\
			\hline \SI{-125}{\volt}	& \SI{26.96} {\second}	& \SI{0.95} {\second}	& \num{28.25}	\\
			\hline \SI{0}{\volt}	& \SI{5.29}  {\second}	& \SI{5.26} {\second}	& \num{1.01}	\\
			\hline \SI{250}{\volt}	& \SI{37.16} {\second}	& \SI{1.33} {\second}	& \num{28.01}	\\
			\hline \SI{500}{\volt}	& \SI{79.90} {\second}	& \SI{1.79} {\second}	& \num{44.61}	\\
			\hline \SI{750}{\volt}	& \SI{123.75}{\second}	& \SI{15.25}{\second}	& \num{8.11}	\\
			\hline \SI{1000}{\volt}	& \SI{177.23}{\second}	& \SI{22.67}{\second}	& \num{7.82}	\\
			\hline \SI{1250}{\volt}	& \SI{256.91}{\second}	& \SI{32.67}{\second}	& \num{7.86}	\\
		\hline
		\end{tabular}
	\end{center}
	\caption[]{Computational cost of the simulations per flow time-step.}
	\label{tab:CompTime}
\end{table}
Analyzing the data in Table~\ref{tab:CompTime}, it is clear the large speedup obtained employing the present model in conjunction with mechanism ``B''.
The best cases (where the speedup is close to 40) are those where an external voltage is applied and where the flame is not collapsed on the upstream electrode.
Such a large difference is due to the reduction of the advection velocity of the scalar \gls{ncrg} in the regions far from the flame.
This reduction of drift velocity is entailed by the averaging process on the scalar mobility.\par
For this reason, the time per step measured in the case without any applied voltage is almost identical for the two mechanisms.
In this case, the region where the electric field reaches its higher intensity coincides with the flame, where the mobilities of \gls{pcrg} and \gls{ncrg} are computed using the species distributions obtained by the flamelet equations.
Being the distribution predicted by the flamelets very similar to those predicted by the \gls{cfd} simulation, the drift velocity computed for the scalar \gls{ncrg} does not change with the ionization mechanism.
As consequence, the number of inner time-steps needed by both mechanisms is almost the same (about \num{5}).\par
On the other hand, when an external voltage is applied, the maximum electric field intensity is going to be placed next to the upstream electrode, as shown in Figure~\ref{fig:EF}, where no charges are present especially in the weak voltage cases.
In this region, mechanism ``A'' computes the drift velocity of the scalar \gls{ncrg} using the mobility of the electrons.
On the other hand, mechanism ``B'' uses an arithmetically averaged mobility which is two orders of magnitude lower than the electrons mobility.
In the second case, the drift velocity computed for the anions is much lower and, therefore, the set-up employing mechanism ``A'' experience a most restrictive time advancement condition.
Accordingly, the solver is forced to perform a larger number of inner time-steps (for a $\Delta\gls{v} = \SI{500}{\volt}$ the simulations with the mechanism ``A'' and ``B'' require about \num{200} and \num{5} inner time-steps, respectively).
Since the number of operations performed per inner time-steps is the same regardless the complexity of the chemistry model, a large variation of wall-time needed for the solution is registered.\par
The speedup effect of mechanism ``B'' vanishes in the cases with attached flame because the reacting region reaches the zone of the computational domain where the electric field is maximum and therefore the computed velocity of the anions becomes again comparable between the two mechanisms.\par
Such a reduction of the numerical cost is not a mere numerical artifact produced by the present model, but it relies on the fact that the electrons, produced in the flame region, are very likely to collide with molecular oxygen present in the mixture upstream the flame, producing O$_{2}^-$.
This phenomenon, which tracking each species does not make any impact on the stiffness of the problem, strongly enhances the numerical efficiency of the system in this configuration where the maximum electric field is reached well outside the reacting region (see Figure~\ref{fig:EF}).

%% file: Sections/Conclusions.tex
%
%
\section{Conclusions}
\label{sec:conclusions}
This work provides a model for the prediction of the effect of weak electric fields on diffusive laminar flames.
The proposed formulation solves the same number of governing equations independently of the chemical mechanism used to predict the chemi-ionization of the flame.
This model, based on the \acrfull{fpv} approach, only introduces two scalar quantities, which are used to predict the local presence of positive and negative charges in the mixture.
Therefore, the proposed approach allows one to use detailed and more accurate chemistry models, improving the accuracy of the numerical simulations without any increase of computational cost.
The model has been tested using two numerical test cases and employing two different chemical mechanisms for the ionization of the mixture.\par
A one-dimensional test case has been first considered in order to validate the accuracy of the proposed reduced-order model to reproduce the results of a detailed chemi-ionization model.
Then, the results of a two-dimensional simulations, performed with a wide range of constant imposed voltage, have shown a good agreement with the calculations available in literature regarding the flame position and its electrical response.
It has also been shown that it is possible to reduce the computational cost of each simulation by a factor of 40.
This aspect, together with the low dimensionality of the chem-tables required by the model, can be very useful for the extension of the present formulation to turbulent flows.\par
The two scalar quantities defined to describe the behavior of the positive and negative charges, which are the key feature of the proposed \gls{fpv} approach, render on one hand its use very efficient and, on the other hand, introduce some approximations in modeling the charge transport phenomena.  
This aspect should be definitely investigated in the future taking advantage of experimental test case specifically produced for model validation.
Moreover, the behavior of the present formulation with more complex species transport models and configuration is a current field of investigation.